\documentclass[12pt]{article}
\usepackage{times,latexsym,amssymb,calc,alg,ifthen}
\usepackage{graphicx,color,geometry}
\newcommand{\ignore}[1]{}
\ignore{

latex ecprhtml
bibtex ecprhtml
latex ecprhtml
latex ecprhtml

latex2html -reuse 1 ecprhtml

rsync -v -a --delete -e ssh ecprhtml traal:public_html/qip
ssh traal 'chmod -R a+rX public_html/qip/ecprhtml'

pdflatex ecprpdf
bibtex ecprpdf
pdflatex ecprpdf
pdflatex ecprpdf

scp ecprpdf.pdf traal:public_html/drafts
ssh traal 'chmod -R a+rX public_html/drafts'

latex2html -reuse 1 -dir ecprhtml1 -split 0 -no_navigation ecprhtml

rsync -v -a --delete -e ssh ecprhtml1 traal:public_html/drafts
ssh traal 'chmod -R a+rX public_html/drafts'

rsync -v -a --delete -e ssh ecprhtml traal:/n/u1/www-external/users/knill/public_html/qip
scp ecprpdf.pdf traal:/n/u1/www-external/users/knill/public_html/qip/ecprhtml
ssh traal 'cdwww; chmod -R a+rX qip/ecprhtml'

ssh traal
cdwww; cd ..
rsync -av --delete -e ssh public_html www.c3.lanl.gov:/u2/users/knill 

mkdir /tmp/ecprhtml
cp `texfls ecprhtml.log` /tmp/ecprhtml
(cd /tmp/ecprhtml; tar czvf ecprhtml.tar.gz *)

}
\definecolor{snow}{rgb}{0.99609375,0.9765625,0.9765625}
\definecolor{ghost}{rgb}{0.96875,0.96875,0.99609375}
\definecolor{GhostWhite}{rgb}{0.96875,0.96875,0.99609375}
\definecolor{WhiteSmoke}{rgb}{0.95703125,0.95703125,0.95703125}
\definecolor{gainsboro}{rgb}{0.859375,0.859375,0.859375}
\definecolor{floral}{rgb}{0.99609375,0.9765625,0.9375}
\definecolor{FloralWhite}{rgb}{0.99609375,0.9765625,0.9375}
\definecolor{old}{rgb}{0.98828125,0.95703125,0.8984375}
\definecolor{OldLace}{rgb}{0.98828125,0.95703125,0.8984375}
\definecolor{linen}{rgb}{0.9765625,0.9375,0.8984375}
\definecolor{antique}{rgb}{0.9765625,0.91796875,0.83984375}
\definecolor{AntiqueWhite}{rgb}{0.9765625,0.91796875,0.83984375}
\definecolor{papaya}{rgb}{0.99609375,0.93359375,0.83203125}
\definecolor{PapayaWhip}{rgb}{0.99609375,0.93359375,0.83203125}
\definecolor{blanched}{rgb}{0.99609375,0.91796875,0.80078125}
\definecolor{BlanchedAlmond}{rgb}{0.99609375,0.91796875,0.80078125}
\definecolor{bisque}{rgb}{0.99609375,0.890625,0.765625}
\definecolor{peach}{rgb}{0.99609375,0.8515625,0.72265625}
\definecolor{PeachPuff}{rgb}{0.99609375,0.8515625,0.72265625}
\definecolor{navajo}{rgb}{0.99609375,0.8671875,0.67578125}
\definecolor{NavajoWhite}{rgb}{0.99609375,0.8671875,0.67578125}
\definecolor{moccasin}{rgb}{0.99609375,0.890625,0.70703125}
\definecolor{cornsilk}{rgb}{0.99609375,0.96875,0.859375}
\definecolor{ivory}{rgb}{0.99609375,0.99609375,0.9375}
\definecolor{lemon}{rgb}{0.99609375,0.9765625,0.80078125}
\definecolor{LemonChiffon}{rgb}{0.99609375,0.9765625,0.80078125}
\definecolor{seashell}{rgb}{0.99609375,0.95703125,0.9296875}
\definecolor{honeydew}{rgb}{0.9375,0.99609375,0.9375}
\definecolor{mint}{rgb}{0.95703125,0.99609375,0.9765625}
\definecolor{MintCream}{rgb}{0.95703125,0.99609375,0.9765625}
\definecolor{azure}{rgb}{0.9375,0.99609375,0.99609375}
\definecolor{alice}{rgb}{0.9375,0.96875,0.99609375}
\definecolor{AliceBlue}{rgb}{0.9375,0.96875,0.99609375}
\definecolor{lavender}{rgb}{0.99609375,0.9375,0.95703125}
\definecolor{LavenderBlush}{rgb}{0.99609375,0.9375,0.95703125}
\definecolor{misty}{rgb}{0.99609375,0.890625,0.87890625}
\definecolor{MistyRose}{rgb}{0.99609375,0.890625,0.87890625}
\definecolor{DarkSlateGray}{rgb}{0.18359375,0.30859375,0.30859375}
\definecolor{dim}{rgb}{0.41015625,0.41015625,0.41015625}
\definecolor{DimGray}{rgb}{0.41015625,0.41015625,0.41015625}
\definecolor{dim}{rgb}{0.41015625,0.41015625,0.41015625}
\definecolor{DimGrey}{rgb}{0.41015625,0.41015625,0.41015625}
\definecolor{SlateGray}{rgb}{0.4375,0.5,0.5625}
\definecolor{SlateGrey}{rgb}{0.4375,0.5,0.5625}
\definecolor{LightSlateGray}{rgb}{0.46484375,0.53125,0.59765625}
\definecolor{LightSlateGrey}{rgb}{0.46484375,0.53125,0.59765625}
\definecolor{gray}{rgb}{0.7421875,0.7421875,0.7421875}
\definecolor{grey}{rgb}{0.7421875,0.7421875,0.7421875}
\definecolor{LightGrey}{rgb}{0.82421875,0.82421875,0.82421875}
\definecolor{LightGray}{rgb}{0.82421875,0.82421875,0.82421875}
\definecolor{midnight}{rgb}{0.09765625,0.09765625,0.4375}
\definecolor{MidnightBlue}{rgb}{0.09765625,0.09765625,0.4375}
\definecolor{NavyBlue}{rgb}{0,0,0.5}
\definecolor{cornflower}{rgb}{0.390625,0.58203125,0.92578125}
\definecolor{CornflowerBlue}{rgb}{0.390625,0.58203125,0.92578125}
\definecolor{DarkSlateBlue}{rgb}{0.28125,0.23828125,0.54296875}
\definecolor{SlateBlue}{rgb}{0.4140625,0.3515625,0.80078125}
\definecolor{MediumSlateBlue}{rgb}{0.48046875,0.40625,0.9296875}
\definecolor{light}{rgb}{0.515625,0.4375,0.99609375}
\definecolor{LightSlateBlue}{rgb}{0.515625,0.4375,0.99609375}
\definecolor{MediumBlue}{rgb}{0,0,0.80078125}
\definecolor{royal}{rgb}{0.25390625,0.41015625,0.87890625}
\definecolor{RoyalBlue}{rgb}{0.25390625,0.41015625,0.87890625}
\definecolor{dodger}{rgb}{0.1171875,0.5625,0.99609375}
\definecolor{DodgerBlue}{rgb}{0.1171875,0.5625,0.99609375}
\definecolor{deep}{rgb}{0,0.74609375,0.99609375}
\definecolor{DeepSkyBlue}{rgb}{0,0.74609375,0.99609375}
\definecolor{sky}{rgb}{0.52734375,0.8046875,0.91796875}
\definecolor{SkyBlue}{rgb}{0.52734375,0.8046875,0.91796875}
\definecolor{LightSkyBlue}{rgb}{0.52734375,0.8046875,0.9765625}
\definecolor{steel}{rgb}{0.2734375,0.5078125,0.703125}
\definecolor{SteelBlue}{rgb}{0.2734375,0.5078125,0.703125}
\definecolor{LightSteelBlue}{rgb}{0.6875,0.765625,0.8671875}
\definecolor{LightBlue}{rgb}{0.67578125,0.84375,0.8984375}
\definecolor{powder}{rgb}{0.6875,0.875,0.8984375}
\definecolor{PowderBlue}{rgb}{0.6875,0.875,0.8984375}
\definecolor{PaleTurquoise}{rgb}{0.68359375,0.9296875,0.9296875}
\definecolor{DarkTurquoise}{rgb}{0,0.8046875,0.81640625}
\definecolor{MediumTurquoise}{rgb}{0.28125,0.81640625,0.796875}
\definecolor{turquoise}{rgb}{0.25,0.875,0.8125}
\definecolor{LightCyan}{rgb}{0.875,0.99609375,0.99609375}
\definecolor{cadet}{rgb}{0.37109375,0.6171875,0.625}
\definecolor{CadetBlue}{rgb}{0.37109375,0.6171875,0.625}
\definecolor{MediumAquamarine}{rgb}{0.3984375,0.80078125,0.6640625}
\definecolor{aquamarine}{rgb}{0.49609375,0.99609375,0.828125}
\definecolor{DarkGreen}{rgb}{0,0.390625,0}
\definecolor{DarkOliveGreen}{rgb}{0.33203125,0.41796875,0.18359375}
\definecolor{DarkSeaGreen}{rgb}{0.55859375,0.734375,0.55859375}
\definecolor{sea}{rgb}{0.1796875,0.54296875,0.33984375}
\definecolor{SeaGreen}{rgb}{0.1796875,0.54296875,0.33984375}
\definecolor{MediumSeaGreen}{rgb}{0.234375,0.69921875,0.44140625}
\definecolor{LightSeaGreen}{rgb}{0.125,0.6953125,0.6640625}
\definecolor{PaleGreen}{rgb}{0.59375,0.98046875,0.59375}
\definecolor{spring}{rgb}{0,0.99609375,0.49609375}
\definecolor{SpringGreen}{rgb}{0,0.99609375,0.49609375}
\definecolor{lawn}{rgb}{0.484375,0.984375,0}
\definecolor{LawnGreen}{rgb}{0.484375,0.984375,0}
\definecolor{chartreuse}{rgb}{0.49609375,0.99609375,0}
\definecolor{MediumSpringGreen}{rgb}{0,0.9765625,0.6015625}
\definecolor{GreenYellow}{rgb}{0.67578125,0.99609375,0.18359375}
\definecolor{lime}{rgb}{0.1953125,0.80078125,0.1953125}
\definecolor{LimeGreen}{rgb}{0.1953125,0.80078125,0.1953125}
\definecolor{YellowGreen}{rgb}{0.6015625,0.80078125,0.1953125}
\definecolor{forest}{rgb}{0.1328125,0.54296875,0.1328125}
\definecolor{ForestGreen}{rgb}{0.1328125,0.54296875,0.1328125}
\definecolor{olive}{rgb}{0.41796875,0.5546875,0.13671875}
\definecolor{OliveDrab}{rgb}{0.41796875,0.5546875,0.13671875}
\definecolor{DarkKhaki}{rgb}{0.73828125,0.71484375,0.41796875}
\definecolor{khaki}{rgb}{0.9375,0.8984375,0.546875}
\definecolor{PaleGoldenrod}{rgb}{0.9296875,0.90625,0.6640625}
\definecolor{LightGoldenrodYellow}{rgb}{0.9765625,0.9765625,0.8203125}
\definecolor{LightYellow}{rgb}{0.99609375,0.99609375,0.875}
\definecolor{gold}{rgb}{0.99609375,0.83984375,0}
\definecolor{LightGoldenrod}{rgb}{0.9296875,0.86328125,0.5078125}
\definecolor{goldenrod}{rgb}{0.8515625,0.64453125,0.125}
\definecolor{DarkGoldenrod}{rgb}{0.71875,0.5234375,0.04296875}
\definecolor{rosy}{rgb}{0.734375,0.55859375,0.55859375}
\definecolor{RosyBrown}{rgb}{0.734375,0.55859375,0.55859375}
\definecolor{indian}{rgb}{0.80078125,0.359375,0.359375}
\definecolor{IndianRed}{rgb}{0.80078125,0.359375,0.359375}
\definecolor{saddle}{rgb}{0.54296875,0.26953125,0.07421875}
\definecolor{SaddleBrown}{rgb}{0.54296875,0.26953125,0.07421875}
\definecolor{sienna}{rgb}{0.625,0.3203125,0.17578125}
\definecolor{peru}{rgb}{0.80078125,0.51953125,0.24609375}
\definecolor{burlywood}{rgb}{0.8671875,0.71875,0.52734375}
\definecolor{beige}{rgb}{0.95703125,0.95703125,0.859375}
\definecolor{wheat}{rgb}{0.95703125,0.8671875,0.69921875}
\definecolor{sandy}{rgb}{0.953125,0.640625,0.375}
\definecolor{SandyBrown}{rgb}{0.953125,0.640625,0.375}
\definecolor{tan}{rgb}{0.8203125,0.703125,0.546875}
\definecolor{chocolate}{rgb}{0.8203125,0.41015625,0.1171875}
\definecolor{firebrick}{rgb}{0.6953125,0.1328125,0.1328125}
\definecolor{brown}{rgb}{0.64453125,0.1640625,0.1640625}
\definecolor{DarkSalmon}{rgb}{0.91015625,0.5859375,0.4765625}
\definecolor{salmon}{rgb}{0.9765625,0.5,0.4453125}
\definecolor{LightSalmon}{rgb}{0.99609375,0.625,0.4765625}
\definecolor{orange}{rgb}{0.99609375,0.64453125,0}
\definecolor{DarkOrange}{rgb}{0.99609375,0.546875,0}
\definecolor{coral}{rgb}{0.99609375,0.49609375,0.3125}
\definecolor{LightCoral}{rgb}{0.9375,0.5,0.5}
\definecolor{tomato}{rgb}{0.99609375,0.38671875,0.27734375}
\definecolor{OrangeRed}{rgb}{0.99609375,0.26953125,0}
\definecolor{HotPink}{rgb}{0.99609375,0.41015625,0.703125}
\definecolor{DeepPink}{rgb}{0.99609375,0.078125,0.57421875}
\definecolor{pink}{rgb}{0.99609375,0.75,0.79296875}
\definecolor{LightPink}{rgb}{0.99609375,0.7109375,0.75390625}
\definecolor{PaleVioletRed}{rgb}{0.85546875,0.4375,0.57421875}
\definecolor{maroon}{rgb}{0.6875,0.1875,0.375}
\definecolor{MediumVioletRed}{rgb}{0.77734375,0.08203125,0.51953125}
\definecolor{violet}{rgb}{0.8125,0.125,0.5625}
\definecolor{VioletRed}{rgb}{0.8125,0.125,0.5625}
\definecolor{plum}{rgb}{0.86328125,0.625,0.86328125}
\definecolor{orchid}{rgb}{0.8515625,0.4375,0.8359375}
\definecolor{MediumOrchid}{rgb}{0.7265625,0.33203125,0.82421875}
\definecolor{DarkOrchid}{rgb}{0.59765625,0.1953125,0.796875}
\definecolor{DarkViolet}{rgb}{0.578125,0,0.82421875}
\definecolor{blue}{rgb}{0.5390625,0.16796875,0.8828125}
\definecolor{BlueViolet}{rgb}{0.5390625,0.16796875,0.8828125}
\definecolor{purple}{rgb}{0.625,0.125,0.9375}
\definecolor{MediumPurple}{rgb}{0.57421875,0.4375,0.85546875}
\definecolor{thistle}{rgb}{0.84375,0.74609375,0.84375}
\definecolor{snow1}{rgb}{0.99609375,0.9765625,0.9765625}
\definecolor{snow2}{rgb}{0.9296875,0.91015625,0.91015625}
\definecolor{snow3}{rgb}{0.80078125,0.78515625,0.78515625}
\definecolor{snow4}{rgb}{0.54296875,0.53515625,0.53515625}
\definecolor{seashell1}{rgb}{0.99609375,0.95703125,0.9296875}
\definecolor{seashell2}{rgb}{0.9296875,0.89453125,0.8671875}
\definecolor{seashell3}{rgb}{0.80078125,0.76953125,0.74609375}
\definecolor{seashell4}{rgb}{0.54296875,0.5234375,0.5078125}
\definecolor{AntiqueWhite1}{rgb}{0.99609375,0.93359375,0.85546875}
\definecolor{AntiqueWhite2}{rgb}{0.9296875,0.87109375,0.796875}
\definecolor{AntiqueWhite3}{rgb}{0.80078125,0.75,0.6875}
\definecolor{AntiqueWhite4}{rgb}{0.54296875,0.51171875,0.46875}
\definecolor{bisque1}{rgb}{0.99609375,0.890625,0.765625}
\definecolor{bisque2}{rgb}{0.9296875,0.83203125,0.71484375}
\definecolor{bisque3}{rgb}{0.80078125,0.71484375,0.6171875}
\definecolor{bisque4}{rgb}{0.54296875,0.48828125,0.41796875}
\definecolor{PeachPuff1}{rgb}{0.99609375,0.8515625,0.72265625}
\definecolor{PeachPuff2}{rgb}{0.9296875,0.79296875,0.67578125}
\definecolor{PeachPuff3}{rgb}{0.80078125,0.68359375,0.58203125}
\definecolor{PeachPuff4}{rgb}{0.54296875,0.46484375,0.39453125}
\definecolor{NavajoWhite1}{rgb}{0.99609375,0.8671875,0.67578125}
\definecolor{NavajoWhite2}{rgb}{0.9296875,0.80859375,0.62890625}
\definecolor{NavajoWhite3}{rgb}{0.80078125,0.69921875,0.54296875}
\definecolor{NavajoWhite4}{rgb}{0.54296875,0.47265625,0.3671875}
\definecolor{LemonChiffon1}{rgb}{0.99609375,0.9765625,0.80078125}
\definecolor{LemonChiffon2}{rgb}{0.9296875,0.91015625,0.74609375}
\definecolor{LemonChiffon3}{rgb}{0.80078125,0.78515625,0.64453125}
\definecolor{LemonChiffon4}{rgb}{0.54296875,0.53515625,0.4375}
\definecolor{cornsilk1}{rgb}{0.99609375,0.96875,0.859375}
\definecolor{cornsilk2}{rgb}{0.9296875,0.90625,0.80078125}
\definecolor{cornsilk3}{rgb}{0.80078125,0.78125,0.69140625}
\definecolor{cornsilk4}{rgb}{0.54296875,0.53125,0.46875}
\definecolor{ivory1}{rgb}{0.99609375,0.99609375,0.9375}
\definecolor{ivory2}{rgb}{0.9296875,0.9296875,0.875}
\definecolor{ivory3}{rgb}{0.80078125,0.80078125,0.75390625}
\definecolor{ivory4}{rgb}{0.54296875,0.54296875,0.51171875}
\definecolor{honeydew1}{rgb}{0.9375,0.99609375,0.9375}
\definecolor{honeydew2}{rgb}{0.875,0.9296875,0.875}
\definecolor{honeydew3}{rgb}{0.75390625,0.80078125,0.75390625}
\definecolor{honeydew4}{rgb}{0.51171875,0.54296875,0.51171875}
\definecolor{LavenderBlush1}{rgb}{0.99609375,0.9375,0.95703125}
\definecolor{LavenderBlush2}{rgb}{0.9296875,0.875,0.89453125}
\definecolor{LavenderBlush3}{rgb}{0.80078125,0.75390625,0.76953125}
\definecolor{LavenderBlush4}{rgb}{0.54296875,0.51171875,0.5234375}
\definecolor{MistyRose1}{rgb}{0.99609375,0.890625,0.87890625}
\definecolor{MistyRose2}{rgb}{0.9296875,0.83203125,0.8203125}
\definecolor{MistyRose3}{rgb}{0.80078125,0.71484375,0.70703125}
\definecolor{MistyRose4}{rgb}{0.54296875,0.48828125,0.48046875}
\definecolor{azure1}{rgb}{0.9375,0.99609375,0.99609375}
\definecolor{azure2}{rgb}{0.875,0.9296875,0.9296875}
\definecolor{azure3}{rgb}{0.75390625,0.80078125,0.80078125}
\definecolor{azure4}{rgb}{0.51171875,0.54296875,0.54296875}
\definecolor{SlateBlue1}{rgb}{0.51171875,0.43359375,0.99609375}
\definecolor{SlateBlue2}{rgb}{0.4765625,0.40234375,0.9296875}
\definecolor{SlateBlue3}{rgb}{0.41015625,0.34765625,0.80078125}
\definecolor{SlateBlue4}{rgb}{0.27734375,0.234375,0.54296875}
\definecolor{RoyalBlue1}{rgb}{0.28125,0.4609375,0.99609375}
\definecolor{RoyalBlue2}{rgb}{0.26171875,0.4296875,0.9296875}
\definecolor{RoyalBlue3}{rgb}{0.2265625,0.37109375,0.80078125}
\definecolor{RoyalBlue4}{rgb}{0.15234375,0.25,0.54296875}
\definecolor{blue1}{rgb}{0,0,0.99609375}
\definecolor{blue2}{rgb}{0,0,0.9296875}
\definecolor{blue3}{rgb}{0,0,0.80078125}
\definecolor{blue4}{rgb}{0,0,0.54296875}
\definecolor{DodgerBlue1}{rgb}{0.1171875,0.5625,0.99609375}
\definecolor{DodgerBlue2}{rgb}{0.109375,0.5234375,0.9296875}
\definecolor{DodgerBlue3}{rgb}{0.09375,0.453125,0.80078125}
\definecolor{DodgerBlue4}{rgb}{0.0625,0.3046875,0.54296875}
\definecolor{SteelBlue1}{rgb}{0.38671875,0.71875,0.99609375}
\definecolor{SteelBlue2}{rgb}{0.359375,0.671875,0.9296875}
\definecolor{SteelBlue3}{rgb}{0.30859375,0.578125,0.80078125}
\definecolor{SteelBlue4}{rgb}{0.2109375,0.390625,0.54296875}
\definecolor{DeepSkyBlue1}{rgb}{0,0.74609375,0.99609375}
\definecolor{DeepSkyBlue2}{rgb}{0,0.6953125,0.9296875}
\definecolor{DeepSkyBlue3}{rgb}{0,0.6015625,0.80078125}
\definecolor{DeepSkyBlue4}{rgb}{0,0.40625,0.54296875}
\definecolor{SkyBlue1}{rgb}{0.52734375,0.8046875,0.99609375}
\definecolor{SkyBlue2}{rgb}{0.4921875,0.75,0.9296875}
\definecolor{SkyBlue3}{rgb}{0.421875,0.6484375,0.80078125}
\definecolor{SkyBlue4}{rgb}{0.2890625,0.4375,0.54296875}
\definecolor{LightSkyBlue1}{rgb}{0.6875,0.8828125,0.99609375}
\definecolor{LightSkyBlue2}{rgb}{0.640625,0.82421875,0.9296875}
\definecolor{LightSkyBlue3}{rgb}{0.55078125,0.7109375,0.80078125}
\definecolor{LightSkyBlue4}{rgb}{0.375,0.48046875,0.54296875}
\definecolor{SlateGray1}{rgb}{0.7734375,0.8828125,0.99609375}
\definecolor{SlateGray2}{rgb}{0.72265625,0.82421875,0.9296875}
\definecolor{SlateGray3}{rgb}{0.62109375,0.7109375,0.80078125}
\definecolor{SlateGray4}{rgb}{0.421875,0.48046875,0.54296875}
\definecolor{LightSteelBlue1}{rgb}{0.7890625,0.87890625,0.99609375}
\definecolor{LightSteelBlue2}{rgb}{0.734375,0.8203125,0.9296875}
\definecolor{LightSteelBlue3}{rgb}{0.6328125,0.70703125,0.80078125}
\definecolor{LightSteelBlue4}{rgb}{0.4296875,0.48046875,0.54296875}
\definecolor{LightBlue1}{rgb}{0.74609375,0.93359375,0.99609375}
\definecolor{LightBlue2}{rgb}{0.6953125,0.87109375,0.9296875}
\definecolor{LightBlue3}{rgb}{0.6015625,0.75,0.80078125}
\definecolor{LightBlue4}{rgb}{0.40625,0.51171875,0.54296875}
\definecolor{LightCyan1}{rgb}{0.875,0.99609375,0.99609375}
\definecolor{LightCyan2}{rgb}{0.81640625,0.9296875,0.9296875}
\definecolor{LightCyan3}{rgb}{0.703125,0.80078125,0.80078125}
\definecolor{LightCyan4}{rgb}{0.4765625,0.54296875,0.54296875}
\definecolor{PaleTurquoise1}{rgb}{0.73046875,0.99609375,0.99609375}
\definecolor{PaleTurquoise2}{rgb}{0.6796875,0.9296875,0.9296875}
\definecolor{PaleTurquoise3}{rgb}{0.5859375,0.80078125,0.80078125}
\definecolor{PaleTurquoise4}{rgb}{0.3984375,0.54296875,0.54296875}
\definecolor{CadetBlue1}{rgb}{0.59375,0.95703125,0.99609375}
\definecolor{CadetBlue2}{rgb}{0.5546875,0.89453125,0.9296875}
\definecolor{CadetBlue3}{rgb}{0.4765625,0.76953125,0.80078125}
\definecolor{CadetBlue4}{rgb}{0.32421875,0.5234375,0.54296875}
\definecolor{turquoise1}{rgb}{0,0.95703125,0.99609375}
\definecolor{turquoise2}{rgb}{0,0.89453125,0.9296875}
\definecolor{turquoise3}{rgb}{0,0.76953125,0.80078125}
\definecolor{turquoise4}{rgb}{0,0.5234375,0.54296875}
\definecolor{cyan1}{rgb}{0,0.99609375,0.99609375}
\definecolor{cyan2}{rgb}{0,0.9296875,0.9296875}
\definecolor{cyan3}{rgb}{0,0.80078125,0.80078125}
\definecolor{cyan4}{rgb}{0,0.54296875,0.54296875}
\definecolor{DarkSlateGray1}{rgb}{0.58984375,0.99609375,0.99609375}
\definecolor{DarkSlateGray2}{rgb}{0.55078125,0.9296875,0.9296875}
\definecolor{DarkSlateGray3}{rgb}{0.47265625,0.80078125,0.80078125}
\definecolor{DarkSlateGray4}{rgb}{0.3203125,0.54296875,0.54296875}
\definecolor{aquamarine1}{rgb}{0.49609375,0.99609375,0.828125}
\definecolor{aquamarine2}{rgb}{0.4609375,0.9296875,0.7734375}
\definecolor{aquamarine3}{rgb}{0.3984375,0.80078125,0.6640625}
\definecolor{aquamarine4}{rgb}{0.26953125,0.54296875,0.453125}
\definecolor{DarkSeaGreen1}{rgb}{0.75390625,0.99609375,0.75390625}
\definecolor{DarkSeaGreen2}{rgb}{0.703125,0.9296875,0.703125}
\definecolor{DarkSeaGreen3}{rgb}{0.60546875,0.80078125,0.60546875}
\definecolor{DarkSeaGreen4}{rgb}{0.41015625,0.54296875,0.41015625}
\definecolor{SeaGreen1}{rgb}{0.328125,0.99609375,0.62109375}
\definecolor{SeaGreen2}{rgb}{0.3046875,0.9296875,0.578125}
\definecolor{SeaGreen3}{rgb}{0.26171875,0.80078125,0.5}
\definecolor{SeaGreen4}{rgb}{0.1796875,0.54296875,0.33984375}
\definecolor{PaleGreen1}{rgb}{0.6015625,0.99609375,0.6015625}
\definecolor{PaleGreen2}{rgb}{0.5625,0.9296875,0.5625}
\definecolor{PaleGreen3}{rgb}{0.484375,0.80078125,0.484375}
\definecolor{PaleGreen4}{rgb}{0.328125,0.54296875,0.328125}
\definecolor{SpringGreen1}{rgb}{0,0.99609375,0.49609375}
\definecolor{SpringGreen2}{rgb}{0,0.9296875,0.4609375}
\definecolor{SpringGreen3}{rgb}{0,0.80078125,0.3984375}
\definecolor{SpringGreen4}{rgb}{0,0.54296875,0.26953125}
\definecolor{green1}{rgb}{0,0.99609375,0}
\definecolor{green2}{rgb}{0,0.9296875,0}
\definecolor{green3}{rgb}{0,0.80078125,0}
\definecolor{green4}{rgb}{0,0.54296875,0}
\definecolor{chartreuse1}{rgb}{0.49609375,0.99609375,0}
\definecolor{chartreuse2}{rgb}{0.4609375,0.9296875,0}
\definecolor{chartreuse3}{rgb}{0.3984375,0.80078125,0}
\definecolor{chartreuse4}{rgb}{0.26953125,0.54296875,0}
\definecolor{OliveDrab1}{rgb}{0.75,0.99609375,0.2421875}
\definecolor{OliveDrab2}{rgb}{0.69921875,0.9296875,0.2265625}
\definecolor{OliveDrab3}{rgb}{0.6015625,0.80078125,0.1953125}
\definecolor{OliveDrab4}{rgb}{0.41015625,0.54296875,0.1328125}
\definecolor{DarkOliveGreen1}{rgb}{0.7890625,0.99609375,0.4375}
\definecolor{DarkOliveGreen2}{rgb}{0.734375,0.9296875,0.40625}
\definecolor{DarkOliveGreen3}{rgb}{0.6328125,0.80078125,0.3515625}
\definecolor{DarkOliveGreen4}{rgb}{0.4296875,0.54296875,0.23828125}
\definecolor{khaki1}{rgb}{0.99609375,0.9609375,0.55859375}
\definecolor{khaki2}{rgb}{0.9296875,0.8984375,0.51953125}
\definecolor{khaki3}{rgb}{0.80078125,0.7734375,0.44921875}
\definecolor{khaki4}{rgb}{0.54296875,0.5234375,0.3046875}
\definecolor{LightGoldenrod1}{rgb}{0.99609375,0.921875,0.54296875}
\definecolor{LightGoldenrod2}{rgb}{0.9296875,0.859375,0.5078125}
\definecolor{LightGoldenrod3}{rgb}{0.80078125,0.7421875,0.4375}
\definecolor{LightGoldenrod4}{rgb}{0.54296875,0.50390625,0.296875}
\definecolor{LightYellow1}{rgb}{0.99609375,0.99609375,0.875}
\definecolor{LightYellow2}{rgb}{0.9296875,0.9296875,0.81640625}
\definecolor{LightYellow3}{rgb}{0.80078125,0.80078125,0.703125}
\definecolor{LightYellow4}{rgb}{0.54296875,0.54296875,0.4765625}
\definecolor{yellow1}{rgb}{0.99609375,0.99609375,0}
\definecolor{yellow2}{rgb}{0.9296875,0.9296875,0}
\definecolor{yellow3}{rgb}{0.80078125,0.80078125,0}
\definecolor{yellow4}{rgb}{0.54296875,0.54296875,0}
\definecolor{gold1}{rgb}{0.99609375,0.83984375,0}
\definecolor{gold2}{rgb}{0.9296875,0.78515625,0}
\definecolor{gold3}{rgb}{0.80078125,0.67578125,0}
\definecolor{gold4}{rgb}{0.54296875,0.45703125,0}
\definecolor{goldenrod1}{rgb}{0.99609375,0.75390625,0.14453125}
\definecolor{goldenrod2}{rgb}{0.9296875,0.703125,0.1328125}
\definecolor{goldenrod3}{rgb}{0.80078125,0.60546875,0.11328125}
\definecolor{goldenrod4}{rgb}{0.54296875,0.41015625,0.078125}
\definecolor{DarkGoldenrod1}{rgb}{0.99609375,0.72265625,0.05859375}
\definecolor{DarkGoldenrod2}{rgb}{0.9296875,0.67578125,0.0546875}
\definecolor{DarkGoldenrod3}{rgb}{0.80078125,0.58203125,0.046875}
\definecolor{DarkGoldenrod4}{rgb}{0.54296875,0.39453125,0.03125}
\definecolor{RosyBrown1}{rgb}{0.99609375,0.75390625,0.75390625}
\definecolor{RosyBrown2}{rgb}{0.9296875,0.703125,0.703125}
\definecolor{RosyBrown3}{rgb}{0.80078125,0.60546875,0.60546875}
\definecolor{RosyBrown4}{rgb}{0.54296875,0.41015625,0.41015625}
\definecolor{IndianRed1}{rgb}{0.99609375,0.4140625,0.4140625}
\definecolor{IndianRed2}{rgb}{0.9296875,0.38671875,0.38671875}
\definecolor{IndianRed3}{rgb}{0.80078125,0.33203125,0.33203125}
\definecolor{IndianRed4}{rgb}{0.54296875,0.2265625,0.2265625}
\definecolor{sienna1}{rgb}{0.99609375,0.5078125,0.27734375}
\definecolor{sienna2}{rgb}{0.9296875,0.47265625,0.2578125}
\definecolor{sienna3}{rgb}{0.80078125,0.40625,0.22265625}
\definecolor{sienna4}{rgb}{0.54296875,0.27734375,0.1484375}
\definecolor{burlywood1}{rgb}{0.99609375,0.82421875,0.60546875}
\definecolor{burlywood2}{rgb}{0.9296875,0.76953125,0.56640625}
\definecolor{burlywood3}{rgb}{0.80078125,0.6640625,0.48828125}
\definecolor{burlywood4}{rgb}{0.54296875,0.44921875,0.33203125}
\definecolor{wheat1}{rgb}{0.99609375,0.90234375,0.7265625}
\definecolor{wheat2}{rgb}{0.9296875,0.84375,0.6796875}
\definecolor{wheat3}{rgb}{0.80078125,0.7265625,0.5859375}
\definecolor{wheat4}{rgb}{0.54296875,0.4921875,0.3984375}
\definecolor{tan1}{rgb}{0.99609375,0.64453125,0.30859375}
\definecolor{tan2}{rgb}{0.9296875,0.6015625,0.28515625}
\definecolor{tan3}{rgb}{0.80078125,0.51953125,0.24609375}
\definecolor{tan4}{rgb}{0.54296875,0.3515625,0.16796875}
\definecolor{chocolate1}{rgb}{0.99609375,0.49609375,0.140625}
\definecolor{chocolate2}{rgb}{0.9296875,0.4609375,0.12890625}
\definecolor{chocolate3}{rgb}{0.80078125,0.3984375,0.11328125}
\definecolor{chocolate4}{rgb}{0.54296875,0.26953125,0.07421875}
\definecolor{firebrick1}{rgb}{0.99609375,0.1875,0.1875}
\definecolor{firebrick2}{rgb}{0.9296875,0.171875,0.171875}
\definecolor{firebrick3}{rgb}{0.80078125,0.1484375,0.1484375}
\definecolor{firebrick4}{rgb}{0.54296875,0.1015625,0.1015625}
\definecolor{brown1}{rgb}{0.99609375,0.25,0.25}
\definecolor{brown2}{rgb}{0.9296875,0.23046875,0.23046875}
\definecolor{brown3}{rgb}{0.80078125,0.19921875,0.19921875}
\definecolor{brown4}{rgb}{0.54296875,0.13671875,0.13671875}
\definecolor{salmon1}{rgb}{0.99609375,0.546875,0.41015625}
\definecolor{salmon2}{rgb}{0.9296875,0.5078125,0.3828125}
\definecolor{salmon3}{rgb}{0.80078125,0.4375,0.328125}
\definecolor{salmon4}{rgb}{0.54296875,0.296875,0.22265625}
\definecolor{LightSalmon1}{rgb}{0.99609375,0.625,0.4765625}
\definecolor{LightSalmon2}{rgb}{0.9296875,0.58203125,0.4453125}
\definecolor{LightSalmon3}{rgb}{0.80078125,0.50390625,0.3828125}
\definecolor{LightSalmon4}{rgb}{0.54296875,0.33984375,0.2578125}
\definecolor{orange1}{rgb}{0.99609375,0.64453125,0}
\definecolor{orange2}{rgb}{0.9296875,0.6015625,0}
\definecolor{orange3}{rgb}{0.80078125,0.51953125,0}
\definecolor{orange4}{rgb}{0.54296875,0.3515625,0}
\definecolor{DarkOrange1}{rgb}{0.99609375,0.49609375,0}
\definecolor{DarkOrange2}{rgb}{0.9296875,0.4609375,0}
\definecolor{DarkOrange3}{rgb}{0.80078125,0.3984375,0}
\definecolor{DarkOrange4}{rgb}{0.54296875,0.26953125,0}
\definecolor{coral1}{rgb}{0.99609375,0.4453125,0.3359375}
\definecolor{coral2}{rgb}{0.9296875,0.4140625,0.3125}
\definecolor{coral3}{rgb}{0.80078125,0.35546875,0.26953125}
\definecolor{coral4}{rgb}{0.54296875,0.2421875,0.18359375}
\definecolor{tomato1}{rgb}{0.99609375,0.38671875,0.27734375}
\definecolor{tomato2}{rgb}{0.9296875,0.359375,0.2578125}
\definecolor{tomato3}{rgb}{0.80078125,0.30859375,0.22265625}
\definecolor{tomato4}{rgb}{0.54296875,0.2109375,0.1484375}
\definecolor{OrangeRed1}{rgb}{0.99609375,0.26953125,0}
\definecolor{OrangeRed2}{rgb}{0.9296875,0.25,0}
\definecolor{OrangeRed3}{rgb}{0.80078125,0.21484375,0}
\definecolor{OrangeRed4}{rgb}{0.54296875,0.14453125,0}
\definecolor{red1}{rgb}{0.99609375,0,0}
\definecolor{red2}{rgb}{0.9296875,0,0}
\definecolor{red3}{rgb}{0.80078125,0,0}
\definecolor{red4}{rgb}{0.54296875,0,0}
\definecolor{DeepPink1}{rgb}{0.99609375,0.078125,0.57421875}
\definecolor{DeepPink2}{rgb}{0.9296875,0.0703125,0.53515625}
\definecolor{DeepPink3}{rgb}{0.80078125,0.0625,0.4609375}
\definecolor{DeepPink4}{rgb}{0.54296875,0.0390625,0.3125}
\definecolor{HotPink1}{rgb}{0.99609375,0.4296875,0.703125}
\definecolor{HotPink2}{rgb}{0.9296875,0.4140625,0.65234375}
\definecolor{HotPink3}{rgb}{0.80078125,0.375,0.5625}
\definecolor{HotPink4}{rgb}{0.54296875,0.2265625,0.3828125}
\definecolor{pink1}{rgb}{0.99609375,0.70703125,0.76953125}
\definecolor{pink2}{rgb}{0.9296875,0.66015625,0.71875}
\definecolor{pink3}{rgb}{0.80078125,0.56640625,0.6171875}
\definecolor{pink4}{rgb}{0.54296875,0.38671875,0.421875}
\definecolor{LightPink1}{rgb}{0.99609375,0.6796875,0.72265625}
\definecolor{LightPink2}{rgb}{0.9296875,0.6328125,0.67578125}
\definecolor{LightPink3}{rgb}{0.80078125,0.546875,0.58203125}
\definecolor{LightPink4}{rgb}{0.54296875,0.37109375,0.39453125}
\definecolor{PaleVioletRed1}{rgb}{0.99609375,0.5078125,0.66796875}
\definecolor{PaleVioletRed2}{rgb}{0.9296875,0.47265625,0.62109375}
\definecolor{PaleVioletRed3}{rgb}{0.80078125,0.40625,0.53515625}
\definecolor{PaleVioletRed4}{rgb}{0.54296875,0.27734375,0.36328125}
\definecolor{maroon1}{rgb}{0.99609375,0.203125,0.69921875}
\definecolor{maroon2}{rgb}{0.9296875,0.1875,0.65234375}
\definecolor{maroon3}{rgb}{0.80078125,0.16015625,0.5625}
\definecolor{maroon4}{rgb}{0.54296875,0.109375,0.3828125}
\definecolor{VioletRed1}{rgb}{0.99609375,0.2421875,0.5859375}
\definecolor{VioletRed2}{rgb}{0.9296875,0.2265625,0.546875}
\definecolor{VioletRed3}{rgb}{0.80078125,0.1953125,0.46875}
\definecolor{VioletRed4}{rgb}{0.54296875,0.1328125,0.3203125}
\definecolor{magenta1}{rgb}{0.99609375,0,0.99609375}
\definecolor{magenta2}{rgb}{0.9296875,0,0.9296875}
\definecolor{magenta3}{rgb}{0.80078125,0,0.80078125}
\definecolor{magenta4}{rgb}{0.54296875,0,0.54296875}
\definecolor{orchid1}{rgb}{0.99609375,0.51171875,0.9765625}
\definecolor{orchid2}{rgb}{0.9296875,0.4765625,0.91015625}
\definecolor{orchid3}{rgb}{0.80078125,0.41015625,0.78515625}
\definecolor{orchid4}{rgb}{0.54296875,0.27734375,0.53515625}
\definecolor{plum1}{rgb}{0.99609375,0.73046875,0.99609375}
\definecolor{plum2}{rgb}{0.9296875,0.6796875,0.9296875}
\definecolor{plum3}{rgb}{0.80078125,0.5859375,0.80078125}
\definecolor{plum4}{rgb}{0.54296875,0.3984375,0.54296875}
\definecolor{MediumOrchid1}{rgb}{0.875,0.3984375,0.99609375}
\definecolor{MediumOrchid2}{rgb}{0.81640625,0.37109375,0.9296875}
\definecolor{MediumOrchid3}{rgb}{0.703125,0.3203125,0.80078125}
\definecolor{MediumOrchid4}{rgb}{0.4765625,0.21484375,0.54296875}
\definecolor{DarkOrchid1}{rgb}{0.74609375,0.2421875,0.99609375}
\definecolor{DarkOrchid2}{rgb}{0.6953125,0.2265625,0.9296875}
\definecolor{DarkOrchid3}{rgb}{0.6015625,0.1953125,0.80078125}
\definecolor{DarkOrchid4}{rgb}{0.40625,0.1328125,0.54296875}
\definecolor{purple1}{rgb}{0.60546875,0.1875,0.99609375}
\definecolor{purple2}{rgb}{0.56640625,0.171875,0.9296875}
\definecolor{purple3}{rgb}{0.48828125,0.1484375,0.80078125}
\definecolor{purple4}{rgb}{0.33203125,0.1015625,0.54296875}
\definecolor{MediumPurple1}{rgb}{0.66796875,0.5078125,0.99609375}
\definecolor{MediumPurple2}{rgb}{0.62109375,0.47265625,0.9296875}
\definecolor{MediumPurple3}{rgb}{0.53515625,0.40625,0.80078125}
\definecolor{MediumPurple4}{rgb}{0.36328125,0.27734375,0.54296875}
\definecolor{thistle1}{rgb}{0.99609375,0.87890625,0.99609375}
\definecolor{thistle2}{rgb}{0.9296875,0.8203125,0.9296875}
\definecolor{thistle3}{rgb}{0.80078125,0.70703125,0.80078125}
\definecolor{thistle4}{rgb}{0.54296875,0.48046875,0.54296875}
\definecolor{gray0}{rgb}{0,0,0}
\definecolor{grey0}{rgb}{0,0,0}
\definecolor{gray1}{rgb}{0.01171875,0.01171875,0.01171875}
\definecolor{grey1}{rgb}{0.01171875,0.01171875,0.01171875}
\definecolor{gray2}{rgb}{0.01953125,0.01953125,0.01953125}
\definecolor{grey2}{rgb}{0.01953125,0.01953125,0.01953125}
\definecolor{gray3}{rgb}{0.03125,0.03125,0.03125}
\definecolor{grey3}{rgb}{0.03125,0.03125,0.03125}
\definecolor{gray4}{rgb}{0.0390625,0.0390625,0.0390625}
\definecolor{grey4}{rgb}{0.0390625,0.0390625,0.0390625}
\definecolor{gray5}{rgb}{0.05078125,0.05078125,0.05078125}
\definecolor{grey5}{rgb}{0.05078125,0.05078125,0.05078125}
\definecolor{gray6}{rgb}{0.05859375,0.05859375,0.05859375}
\definecolor{grey6}{rgb}{0.05859375,0.05859375,0.05859375}
\definecolor{gray7}{rgb}{0.0703125,0.0703125,0.0703125}
\definecolor{grey7}{rgb}{0.0703125,0.0703125,0.0703125}
\definecolor{gray8}{rgb}{0.078125,0.078125,0.078125}
\definecolor{grey8}{rgb}{0.078125,0.078125,0.078125}
\definecolor{gray9}{rgb}{0.08984375,0.08984375,0.08984375}
\definecolor{grey9}{rgb}{0.08984375,0.08984375,0.08984375}
\definecolor{gray10}{rgb}{0.1015625,0.1015625,0.1015625}
\definecolor{grey10}{rgb}{0.1015625,0.1015625,0.1015625}
\definecolor{gray11}{rgb}{0.109375,0.109375,0.109375}
\definecolor{grey11}{rgb}{0.109375,0.109375,0.109375}
\definecolor{gray12}{rgb}{0.12109375,0.12109375,0.12109375}
\definecolor{grey12}{rgb}{0.12109375,0.12109375,0.12109375}
\definecolor{gray13}{rgb}{0.12890625,0.12890625,0.12890625}
\definecolor{grey13}{rgb}{0.12890625,0.12890625,0.12890625}
\definecolor{gray14}{rgb}{0.140625,0.140625,0.140625}
\definecolor{grey14}{rgb}{0.140625,0.140625,0.140625}
\definecolor{gray15}{rgb}{0.1484375,0.1484375,0.1484375}
\definecolor{grey15}{rgb}{0.1484375,0.1484375,0.1484375}
\definecolor{gray16}{rgb}{0.16015625,0.16015625,0.16015625}
\definecolor{grey16}{rgb}{0.16015625,0.16015625,0.16015625}
\definecolor{gray17}{rgb}{0.16796875,0.16796875,0.16796875}
\definecolor{grey17}{rgb}{0.16796875,0.16796875,0.16796875}
\definecolor{gray18}{rgb}{0.1796875,0.1796875,0.1796875}
\definecolor{grey18}{rgb}{0.1796875,0.1796875,0.1796875}
\definecolor{gray19}{rgb}{0.1875,0.1875,0.1875}
\definecolor{grey19}{rgb}{0.1875,0.1875,0.1875}
\definecolor{gray20}{rgb}{0.19921875,0.19921875,0.19921875}
\definecolor{grey20}{rgb}{0.19921875,0.19921875,0.19921875}
\definecolor{gray21}{rgb}{0.2109375,0.2109375,0.2109375}
\definecolor{grey21}{rgb}{0.2109375,0.2109375,0.2109375}
\definecolor{gray22}{rgb}{0.21875,0.21875,0.21875}
\definecolor{grey22}{rgb}{0.21875,0.21875,0.21875}
\definecolor{gray23}{rgb}{0.23046875,0.23046875,0.23046875}
\definecolor{grey23}{rgb}{0.23046875,0.23046875,0.23046875}
\definecolor{gray24}{rgb}{0.23828125,0.23828125,0.23828125}
\definecolor{grey24}{rgb}{0.23828125,0.23828125,0.23828125}
\definecolor{gray25}{rgb}{0.25,0.25,0.25}
\definecolor{grey25}{rgb}{0.25,0.25,0.25}
\definecolor{gray26}{rgb}{0.2578125,0.2578125,0.2578125}
\definecolor{grey26}{rgb}{0.2578125,0.2578125,0.2578125}
\definecolor{gray27}{rgb}{0.26953125,0.26953125,0.26953125}
\definecolor{grey27}{rgb}{0.26953125,0.26953125,0.26953125}
\definecolor{gray28}{rgb}{0.27734375,0.27734375,0.27734375}
\definecolor{grey28}{rgb}{0.27734375,0.27734375,0.27734375}
\definecolor{gray29}{rgb}{0.2890625,0.2890625,0.2890625}
\definecolor{grey29}{rgb}{0.2890625,0.2890625,0.2890625}
\definecolor{gray30}{rgb}{0.30078125,0.30078125,0.30078125}
\definecolor{grey30}{rgb}{0.30078125,0.30078125,0.30078125}
\definecolor{gray31}{rgb}{0.30859375,0.30859375,0.30859375}
\definecolor{grey31}{rgb}{0.30859375,0.30859375,0.30859375}
\definecolor{gray32}{rgb}{0.3203125,0.3203125,0.3203125}
\definecolor{grey32}{rgb}{0.3203125,0.3203125,0.3203125}
\definecolor{gray33}{rgb}{0.328125,0.328125,0.328125}
\definecolor{grey33}{rgb}{0.328125,0.328125,0.328125}
\definecolor{gray34}{rgb}{0.33984375,0.33984375,0.33984375}
\definecolor{grey34}{rgb}{0.33984375,0.33984375,0.33984375}
\definecolor{gray35}{rgb}{0.34765625,0.34765625,0.34765625}
\definecolor{grey35}{rgb}{0.34765625,0.34765625,0.34765625}
\definecolor{gray36}{rgb}{0.359375,0.359375,0.359375}
\definecolor{grey36}{rgb}{0.359375,0.359375,0.359375}
\definecolor{gray37}{rgb}{0.3671875,0.3671875,0.3671875}
\definecolor{grey37}{rgb}{0.3671875,0.3671875,0.3671875}
\definecolor{gray38}{rgb}{0.37890625,0.37890625,0.37890625}
\definecolor{grey38}{rgb}{0.37890625,0.37890625,0.37890625}
\definecolor{gray39}{rgb}{0.38671875,0.38671875,0.38671875}
\definecolor{grey39}{rgb}{0.38671875,0.38671875,0.38671875}
\definecolor{gray40}{rgb}{0.3984375,0.3984375,0.3984375}
\definecolor{grey40}{rgb}{0.3984375,0.3984375,0.3984375}
\definecolor{gray41}{rgb}{0.41015625,0.41015625,0.41015625}
\definecolor{grey41}{rgb}{0.41015625,0.41015625,0.41015625}
\definecolor{gray42}{rgb}{0.41796875,0.41796875,0.41796875}
\definecolor{grey42}{rgb}{0.41796875,0.41796875,0.41796875}
\definecolor{gray43}{rgb}{0.4296875,0.4296875,0.4296875}
\definecolor{grey43}{rgb}{0.4296875,0.4296875,0.4296875}
\definecolor{gray44}{rgb}{0.4375,0.4375,0.4375}
\definecolor{grey44}{rgb}{0.4375,0.4375,0.4375}
\definecolor{gray45}{rgb}{0.44921875,0.44921875,0.44921875}
\definecolor{grey45}{rgb}{0.44921875,0.44921875,0.44921875}
\definecolor{gray46}{rgb}{0.45703125,0.45703125,0.45703125}
\definecolor{grey46}{rgb}{0.45703125,0.45703125,0.45703125}
\definecolor{gray47}{rgb}{0.46875,0.46875,0.46875}
\definecolor{grey47}{rgb}{0.46875,0.46875,0.46875}
\definecolor{gray48}{rgb}{0.4765625,0.4765625,0.4765625}
\definecolor{grey48}{rgb}{0.4765625,0.4765625,0.4765625}
\definecolor{gray49}{rgb}{0.48828125,0.48828125,0.48828125}
\definecolor{grey49}{rgb}{0.48828125,0.48828125,0.48828125}
\definecolor{gray50}{rgb}{0.49609375,0.49609375,0.49609375}
\definecolor{grey50}{rgb}{0.49609375,0.49609375,0.49609375}
\definecolor{gray51}{rgb}{0.5078125,0.5078125,0.5078125}
\definecolor{grey51}{rgb}{0.5078125,0.5078125,0.5078125}
\definecolor{gray52}{rgb}{0.51953125,0.51953125,0.51953125}
\definecolor{grey52}{rgb}{0.51953125,0.51953125,0.51953125}
\definecolor{gray53}{rgb}{0.52734375,0.52734375,0.52734375}
\definecolor{grey53}{rgb}{0.52734375,0.52734375,0.52734375}
\definecolor{gray54}{rgb}{0.5390625,0.5390625,0.5390625}
\definecolor{grey54}{rgb}{0.5390625,0.5390625,0.5390625}
\definecolor{gray55}{rgb}{0.546875,0.546875,0.546875}
\definecolor{grey55}{rgb}{0.546875,0.546875,0.546875}
\definecolor{gray56}{rgb}{0.55859375,0.55859375,0.55859375}
\definecolor{grey56}{rgb}{0.55859375,0.55859375,0.55859375}
\definecolor{gray57}{rgb}{0.56640625,0.56640625,0.56640625}
\definecolor{grey57}{rgb}{0.56640625,0.56640625,0.56640625}
\definecolor{gray58}{rgb}{0.578125,0.578125,0.578125}
\definecolor{grey58}{rgb}{0.578125,0.578125,0.578125}
\definecolor{gray59}{rgb}{0.5859375,0.5859375,0.5859375}
\definecolor{grey59}{rgb}{0.5859375,0.5859375,0.5859375}
\definecolor{gray60}{rgb}{0.59765625,0.59765625,0.59765625}
\definecolor{grey60}{rgb}{0.59765625,0.59765625,0.59765625}
\definecolor{gray61}{rgb}{0.609375,0.609375,0.609375}
\definecolor{grey61}{rgb}{0.609375,0.609375,0.609375}
\definecolor{gray62}{rgb}{0.6171875,0.6171875,0.6171875}
\definecolor{grey62}{rgb}{0.6171875,0.6171875,0.6171875}
\definecolor{gray63}{rgb}{0.62890625,0.62890625,0.62890625}
\definecolor{grey63}{rgb}{0.62890625,0.62890625,0.62890625}
\definecolor{gray64}{rgb}{0.63671875,0.63671875,0.63671875}
\definecolor{grey64}{rgb}{0.63671875,0.63671875,0.63671875}
\definecolor{gray65}{rgb}{0.6484375,0.6484375,0.6484375}
\definecolor{grey65}{rgb}{0.6484375,0.6484375,0.6484375}
\definecolor{gray66}{rgb}{0.65625,0.65625,0.65625}
\definecolor{grey66}{rgb}{0.65625,0.65625,0.65625}
\definecolor{gray67}{rgb}{0.66796875,0.66796875,0.66796875}
\definecolor{grey67}{rgb}{0.66796875,0.66796875,0.66796875}
\definecolor{gray68}{rgb}{0.67578125,0.67578125,0.67578125}
\definecolor{grey68}{rgb}{0.67578125,0.67578125,0.67578125}
\definecolor{gray69}{rgb}{0.6875,0.6875,0.6875}
\definecolor{grey69}{rgb}{0.6875,0.6875,0.6875}
\definecolor{gray70}{rgb}{0.69921875,0.69921875,0.69921875}
\definecolor{grey70}{rgb}{0.69921875,0.69921875,0.69921875}
\definecolor{gray71}{rgb}{0.70703125,0.70703125,0.70703125}
\definecolor{grey71}{rgb}{0.70703125,0.70703125,0.70703125}
\definecolor{gray72}{rgb}{0.71875,0.71875,0.71875}
\definecolor{grey72}{rgb}{0.71875,0.71875,0.71875}
\definecolor{gray73}{rgb}{0.7265625,0.7265625,0.7265625}
\definecolor{grey73}{rgb}{0.7265625,0.7265625,0.7265625}
\definecolor{gray74}{rgb}{0.73828125,0.73828125,0.73828125}
\definecolor{grey74}{rgb}{0.73828125,0.73828125,0.73828125}
\definecolor{gray75}{rgb}{0.74609375,0.74609375,0.74609375}
\definecolor{grey75}{rgb}{0.74609375,0.74609375,0.74609375}
\definecolor{gray76}{rgb}{0.7578125,0.7578125,0.7578125}
\definecolor{grey76}{rgb}{0.7578125,0.7578125,0.7578125}
\definecolor{gray77}{rgb}{0.765625,0.765625,0.765625}
\definecolor{grey77}{rgb}{0.765625,0.765625,0.765625}
\definecolor{gray78}{rgb}{0.77734375,0.77734375,0.77734375}
\definecolor{grey78}{rgb}{0.77734375,0.77734375,0.77734375}
\definecolor{gray79}{rgb}{0.78515625,0.78515625,0.78515625}
\definecolor{grey79}{rgb}{0.78515625,0.78515625,0.78515625}
\definecolor{gray80}{rgb}{0.796875,0.796875,0.796875}
\definecolor{grey80}{rgb}{0.796875,0.796875,0.796875}
\definecolor{gray81}{rgb}{0.80859375,0.80859375,0.80859375}
\definecolor{grey81}{rgb}{0.80859375,0.80859375,0.80859375}
\definecolor{gray82}{rgb}{0.81640625,0.81640625,0.81640625}
\definecolor{grey82}{rgb}{0.81640625,0.81640625,0.81640625}
\definecolor{gray83}{rgb}{0.828125,0.828125,0.828125}
\definecolor{grey83}{rgb}{0.828125,0.828125,0.828125}
\definecolor{gray84}{rgb}{0.8359375,0.8359375,0.8359375}
\definecolor{grey84}{rgb}{0.8359375,0.8359375,0.8359375}
\definecolor{gray85}{rgb}{0.84765625,0.84765625,0.84765625}
\definecolor{grey85}{rgb}{0.84765625,0.84765625,0.84765625}
\definecolor{gray86}{rgb}{0.85546875,0.85546875,0.85546875}
\definecolor{grey86}{rgb}{0.85546875,0.85546875,0.85546875}
\definecolor{gray87}{rgb}{0.8671875,0.8671875,0.8671875}
\definecolor{grey87}{rgb}{0.8671875,0.8671875,0.8671875}
\definecolor{gray88}{rgb}{0.875,0.875,0.875}
\definecolor{grey88}{rgb}{0.875,0.875,0.875}
\definecolor{gray89}{rgb}{0.88671875,0.88671875,0.88671875}
\definecolor{grey89}{rgb}{0.88671875,0.88671875,0.88671875}
\definecolor{gray90}{rgb}{0.89453125,0.89453125,0.89453125}
\definecolor{grey90}{rgb}{0.89453125,0.89453125,0.89453125}
\definecolor{gray91}{rgb}{0.90625,0.90625,0.90625}
\definecolor{grey91}{rgb}{0.90625,0.90625,0.90625}
\definecolor{gray92}{rgb}{0.91796875,0.91796875,0.91796875}
\definecolor{grey92}{rgb}{0.91796875,0.91796875,0.91796875}
\definecolor{gray93}{rgb}{0.92578125,0.92578125,0.92578125}
\definecolor{grey93}{rgb}{0.92578125,0.92578125,0.92578125}
\definecolor{gray94}{rgb}{0.9375,0.9375,0.9375}
\definecolor{grey94}{rgb}{0.9375,0.9375,0.9375}
\definecolor{gray95}{rgb}{0.9453125,0.9453125,0.9453125}
\definecolor{grey95}{rgb}{0.9453125,0.9453125,0.9453125}
\definecolor{gray96}{rgb}{0.95703125,0.95703125,0.95703125}
\definecolor{grey96}{rgb}{0.95703125,0.95703125,0.95703125}
\definecolor{gray97}{rgb}{0.96484375,0.96484375,0.96484375}
\definecolor{grey97}{rgb}{0.96484375,0.96484375,0.96484375}
\definecolor{gray98}{rgb}{0.9765625,0.9765625,0.9765625}
\definecolor{grey98}{rgb}{0.9765625,0.9765625,0.9765625}
\definecolor{gray99}{rgb}{0.984375,0.984375,0.984375}
\definecolor{grey99}{rgb}{0.984375,0.984375,0.984375}
\definecolor{gray100}{rgb}{0.99609375,0.99609375,0.99609375}
\definecolor{grey100}{rgb}{0.99609375,0.99609375,0.99609375}
\definecolor{DarkGrey}{rgb}{0.66015625,0.66015625,0.66015625}
\definecolor{DarkGray}{rgb}{0.66015625,0.66015625,0.66015625}
\definecolor{DarkBlue}{rgb}{0,0,0.54296875}
\definecolor{DarkCyan}{rgb}{0,0.54296875,0.54296875}
\definecolor{DarkMagenta}{rgb}{0.54296875,0,0.54296875}
\definecolor{DarkRed}{rgb}{0.54296875,0,0}
\definecolor{LightGreen}{rgb}{0.5625,0.9296875,0.5625}

\usepackage{html}

\newcommand{\makecnts}{\tableofcontents\newpage}

\newcommand{\makeaddress}{
\noindent\textbf{Addresses:}
\begin{tabular}[t]{lcl}
E. Knill:&
Los Alamos National Laboratory&\htmladdnormallink{\texttt{knill@lanl.gov}}{mailto:knill@lanl.gov}\\
R. Laflamme:&
University of Waterloo and Perimeter Institute
&\htmladdnormallink{\texttt{laflamme@iqc.ca}}{mailto:laflamme@iqc.ca}
\\
A. Ashikhmin:&Bell Labs, Lucent&\htmladdnormallink{\texttt{aea@research.bell-labs.com}}{mailto:aea@research.bell-labs.com}
\\
H. Barnum: &Los Alamos National Laboratory&\htmladdnormallink{\texttt{barnum@lanl.gov}}{mailto:barnum@lanl.gov}
\\
L. Viola: &''&\htmladdnormallink{\texttt{lviola@lanl.gov}}{mailto:lviola@lanl.gov}
\\
W. H. Zurek: &''&\htmladdnormallink{\texttt{whz@lanl.gov}}{mailto:whz@lanl.gov}
\end{tabular}
}

\graphicspath{{./}{graphics/}}

\newcommand{\qvbar}{\mbox{$|\hspace*{-3pt}|\hspace*{-3pt}|$}}
\newcommand{\qrangle}{\mbox{$\rangle\hspace*{-4.3pt}\rangle\hspace*{-4.3pt}\rangle$}}
\newcommand{\qlangle}{\mbox{$\langle\hspace*{-4.3pt}\langle\hspace*{-4.3pt}\langle$}}

\newcommand{\physcolor}{brown}
\newcommand{\phys}[1]{{\color{\physcolor}#1}}
\newcommand{\physcolordescr}{in brown}

\newcommand{\errcolor}{red}
\newcommand{\err}[1]{{\color{\errcolor}#1}}
\newcommand{\errcolordescr}{in red}

\newcommand{\sysfnt}{\mathsf}

\newcommand{\ket}[1]{\qvbar{#1}\qrangle}
\newcommand{\bra}[1]{\qlangle{#1}\qvbar}
\newcommand{\braket}[2]{\qlangle{#1}\qvbar{#2}\qrangle}

\newcommand{\kets}[2]{\qvbar{#1}\qrangle_{{}_{\!\!{\sysfnt{#2}}}}}
\newcommand{\bras}[2]{{}^{\scriptscriptstyle\sysfnt{ #2}}\!\qlangle{#1}\qvbar}
\newcommand{\brakets}[3]{{}^{\scriptscriptstyle\sysfnt{ #3}}\!\qlangle{#1}\qvbar{#2}\qrangle_{{}_{\!\!{\sysfnt{#3}}}}}
\newcommand{\ketbras}[3]{\kets{#1}{#3}\!\!\bras{#2}{#3}}
\newcommand{\slb}[2]{{#1}^{({\sysfnt{#2}})}}

\newcommand{\qaop}[4]{\left(\begin{array}{cc}#1&#2\\ #3&#4\end{array}\right)}

\newcommand{\nputbox}[3]{\put(#1){\makebox(0,0)[#2]{#3}}}
\newcommand{\nputgr}[4]{\put(#1){\makebox(0,0)[#2]{\includegraphics[#3]{#4}}}}

\newlength{\elimdepthdim}
\newlength{\elimheightdim}
\newlength{\elimwidthdim}

\newlength{\strutdepthdim}
\newlength{\strutheightdim}
\newlength{\strutwidthdim}

\newcommand{\cB}{{\cal B}}

\newcommand{\cE}{{\cal E}}

\newcommand{\cH}{{\cal H}}

\newcommand{\tensor}{\otimes}
\newcommand{\trace}{\mbox{tr}}

\def\id{{\mathchoice {\rm 1\mskip-4mu l} {\rm 1\mskip-4mu l} {\rm
1\mskip-4.5mu l} {\rm 1\mskip-5mu l}}}

\newcommand{\mb}[1]{\mathbf{#1}}

\newcommand{\bitzero}{\mathfrak{0}}
\newcommand{\bitone}{\mathfrak{1}}

\newcommand{\idop}{\id}

\newenvironment{eqthm}[1]{\begin{minipage}[b]{6in}\textbf{#1.\ }}{\end{minipage}}

\newcounter{herefignum}
\newenvironment{herefig}{\begin{center}\refstepcounter{herefignum}}{\end{center}}
\newcommand{\herefigcap}[1]{\\\begin{minipage}{\textwidth}{FIG.~\theherefignum: #1}\end{minipage}}

\newcounter{hereboxnum}

\begin{document}

\title{Introduction to Quantum Error Correction}
\author{E. Knill, R. Laflamme, A. Ashikhmin, H. Barnum, L. Viola and W. H. Zurek}
\date{\today}

\maketitle

\begin{latexonly}
\makecnts
\end{latexonly}

\ignore{
In this introduction we motivate and explain the ``decoding'' and
``subsystems'' view of quantum error correction.  We explain how
quantum noise in QIP can be described and classified, and summarize
the requirements that need to be satisfied for fault tolerance. Considering
the capabilities of currently available quantum technology, the
requirements appear daunting. But the idea of ``subsystems'' shows
that these requirements can be met in many different, and often
unexpected ways.
}

When physically realized, quantum information processing (QIP) can be
used to solve problems in physics simulation, cryptanalysis and secure
communication for which there are no known efficient solutions based
on classical information processing. Numerous proposals exist for
building the devices required for QIP by using a variety of systems
that exhibit quantum properties. Examples include nuclear spins in
molecules, electron spins or charge in quantum dots, collective states
of superconductors, and photons~\cite{braunstein:qc2000b}. In all of
these cases, there are well established physical models that, under
ideal conditions, allow for exact realizations of quantum information
and its manipulation. However, real physical systems never behave
exactly like the ideal models.  The main problems are environmental
noise, which is due to incomplete isolation of the system from the
rest of the world, and control errors, which are caused by calibration
errors and random fluctuations in control parameters. Attempts to
reduce the effects of these errors are confronted by the conflicting
needs of being able to control and reliably measure the quantum
systems. These needs require strong interactions with control devices,
and systems sufficiently well isolated to maintain
coherence, which is the subtle relationship between the phases in a quantum
superposition.  The fact that quantum effects rarely persist on
macroscopic scales suggests that meeting these needs requires
considerable outside intervention.

Soon after P. Shor published the efficient quantum factoring algorithm
with its applications to breaking commonly used public-key
cryptosystems, A. Steane~\cite{steane:qc1995a} and
P. Shor~\cite{shor:qc1995b} gave the first constructions of quantum
error-correcting codes. These codes make it possible to store quantum
information so that one can reverse the effects of the most likely
errors. By demonstrating that quantum information can exist in
protected parts of the state space, they showed that, in principle, it
is possible to protect against environmental noise when storing or
transmitting information. Stimulated by these results and in order to
solve the problem of errors happening during computation with quantum
information, researchers initiated a series of investigations to
determine whether it was possible to quantum-compute in a
fault-tolerant manner. The outcome of these investigations was
positive and culminated in what are now known as ``accuracy threshold
theorems''~\cite{gottesman:qc1996a,calderbank:qc1996a,calderbank:qc1996b,shor:qc1996a,kitaev:qc1996a,knill:qc1996b,aharonov:qc1996a,aharonov:qc1999a,knill:qc1997a,knill:qc1998a,gottesman:qc1997a,preskill:qc1998a}.
According to these theorems, if the effects of all errors are
sufficiently small per quantum bit (qubit) and step of the
computation, then it is possible to process quantum information
arbitrarily accurately with reasonable resource overheads. The
requirement on errors is quantified by a maximum tolerable error rate
called the threshold.  The threshold value depends strongly on the
details of the assumed error model.  All threshold theorems require
that errors at different times and locations be independent and that
the basic computational operations can be applied in parallel.
Although the proven thresholds are well out of range of today's
devices there are signs that in practice, fault-tolerant quantum
computation may be realizable.

In retrospect, advances in quantum error correction and fault-tolerant
computation were made possible by the realization that accurate
computation does not require the state of the physical devices
supporting the computation to be perfect.  In classical information
processing, this observation is so obvious that it is often forgotten:
No two letters ``e'' on a written page are physically identical, and
the number of electrons used to store a bit in a computer's memory
varies substantially.  Nevertheless, we have no difficulty in
accurately identifying the desired letter or state. A crucial
conceptual difficulty with quantum information is that by its very
nature, it cannot be identified by being ``looked'' at. As a result,
the sense in which quantum information can be accurately stored in a
noisy system needs to be defined without reference to an observer.
There are two ways to accomplish this task.  The first
is to define stored information to be the information that
can, in principle, be extracted by a quantum decoding procedure.  The
second is to explicitly define ``subsystems'' (particle-like aspects
of the quantum device) that contain the desired information.  The
first approach is a natural generalization of the usual
interpretations of classical error-correction methods, whereas the
second is motivated by a way of characterizing quantum particles.

In this introduction we motivate and explain the ``decoding'' and
``subsystems'' view of quantum error correction.  We explain how
quantum noise in QIP can be described and classified, and summarize
the requirements that need to be satisfied for fault tolerance. Considering
the capabilities of currently available quantum technology, the
requirements appear daunting. But the idea of ``subsystems'' shows
that these requirements can be met in many different, and often
unexpected ways.

Our introduction is structured as follows: The basic concepts are
introduced by example, first for classical and then for quantum codes. We
then show how the concepts are defined in general. Following a
discussion of error models and analysis
(Sect.~\ref{sec:error_models}), we state and explain the necessary and
sufficient conditions for detectability of errors and correctability
of error sets (Sect.~\ref{sec:fromqedtoec}). This is followed by a
brief introduction to two of the most important methods for
constructing error-correcting codes and subsystems
(Sect.~\ref{sec:constructing}). For a basic overview, it suffices to
read the beginnings of these more technical sections. The principles
of fault-tolerant quantum computation are outlined in the last
section.

\section{Concepts and Examples}
\label{sec:concepts_and_examples}

Communication is the prototypical application of error-correction methods.
To communicate, a sender needs to convey information
to a receiver over a noisy ``communication channel''.  Such a channel
can be thought of as a means of transmitting an information-carrying
physical system from one place to another.  During transmission, the
physical system is subject to disturbances  that can affect the
information carried.  To use a communication channel, the sender needs
to ``encode'' the information to be transmitted in the physical
system. After transmission, the receiver ``decodes'' the information.
The procedure is shown in Fig.~\ref{fig:cchannel}.

\begin{herefig}
\begin{picture}(7,2.5)(-3.5,-2.5)
\nputbox{-3.1,-1.5}{t}{
\begin{minipage}[t]{.6in}\large
\setlength{\baselineskip}{0pt}
$\bitone$ $\bitzero$ $\bitone$ $\bitzero$ $\bitzero$ $\bitone$
$\bitzero$ $\bitone$ $\bitone$ $\bitone$ $\bitone$ $\bitzero$
$\bitone$ $\bitone$ $\bitzero$ $\bitzero$ $\bitone$ $\bitzero$
\end{minipage}
}
\nputbox{+3.2,-.1}{t}{
\begin{minipage}[t]{.5in}
\setlength{\baselineskip}{0pt}
$\bitone$ $\bitzero$
$\bitone$ $\bitzero$ $\bitzero$
$\bitone$ $\bitzero$ $\bitone$
$\bitone$ 
$\bitone$ 
$\bitone$ $\bitzero$ $\bitone$
$\bitone$ $\bitzero$ $\bitzero$
$\bitone$ $\bitzero$
\end{minipage}
}
\nputbox{-3.3,-.85}{l}{\large Encode}
\nputbox{-0.5,-.4}{t}{\large Transmit}
\nputbox{+3.5,-1.35}{r}{\large Decode}
\nputgr{0,0}{t}{width=7in}{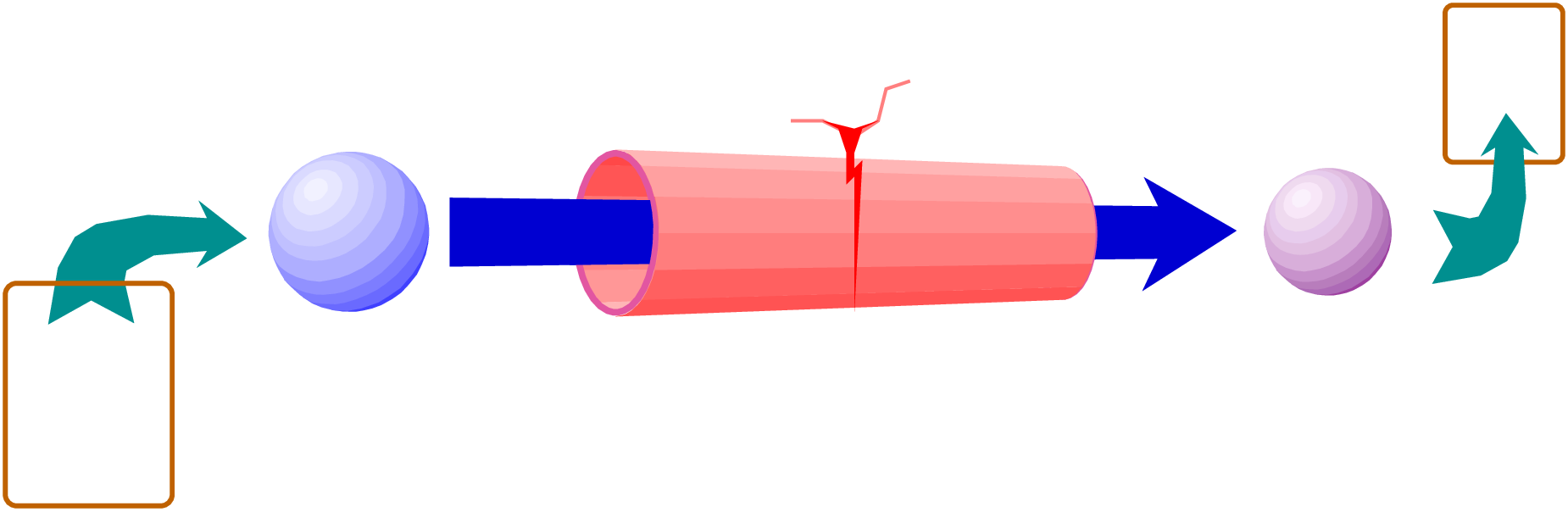}
\end{picture}
\label{fig:cchannel}
\herefigcap{A typical application of error-correction methods:
The illustration shows the three main steps required for communication.
Information is first encoded in a physical system, then transmitted
over the noisy communication channel and finally decoded. The
combination of encoding and decoding is chosen so that
errors have no effect on the transmitted information.
}
\end{herefig}

The protection of stored information
is an other important application of error-correction methods.
In this case, the user encodes the
information in a storage system and retrieves it at a later
time. Provided that there is no communication from the
receiver to the sender, any error-correction method applicable to
communication is also applicable to storage and vice versa.  In
Sect.~\ref{sec:fault_tolerance} we discuss the problem of
fault-tolerant computation, which requires enhancing error-correction
methods in order to enable applying operations to encoded information
without losing protection against errors.

To illustrate the different features of error-correction methods we
consider three examples.  We begin by describing them for classical
information, but in each case, there is a quantum analogue that will
be introduced later.

\subsection{Trivial Two-Bit Example}

Consider a physical system consisting of two bits with state space
$\{\phys{\bitzero\bitzero},\phys{\bitzero\bitone},\phys{\bitone\bitzero},
\phys{\bitone\bitone}\}$.
We use the convention that state symbols for physical
systems subject to errors are \physcolordescr. States changed by
errors are shown 
\errcolordescr\footnote{These graphical conventions are not crucial
for understanding what the symbols mean and are for emphasis only.}.
In this example, the system is subject to errors that flip (apply the
$\mb{not}$ operator to) the first bit with probability $.5$.  We wish
to safely store one bit of information.  To this end,
we store the information in the second physical bit, because this bit is
unaffected by the errors (Fig.~\ref{fig:trivialccode}).

\begin{herefig}
\begin{picture}(6.5,6.3)(-3.25,-6.1)\Large
\nputbox{-3.25,-.0}{tl}{\large Physical system and error model:}
\nputbox{-2.0,-1}{c}{$\phys{a\;b}$}
\nputbox{+2,-1.5}{c}{$\phys{\mb{not}(a)\;b}$}
\nputbox{+2,-.5}{c}{$\phys{a\;b}$}
\nputbox{+.7,-.3}{t}{\large prob.=$.5$}
\nputbox{+.7,-1.8}{b}{\large prob.=$.5$}
\nputgr{0,0}{t}{}{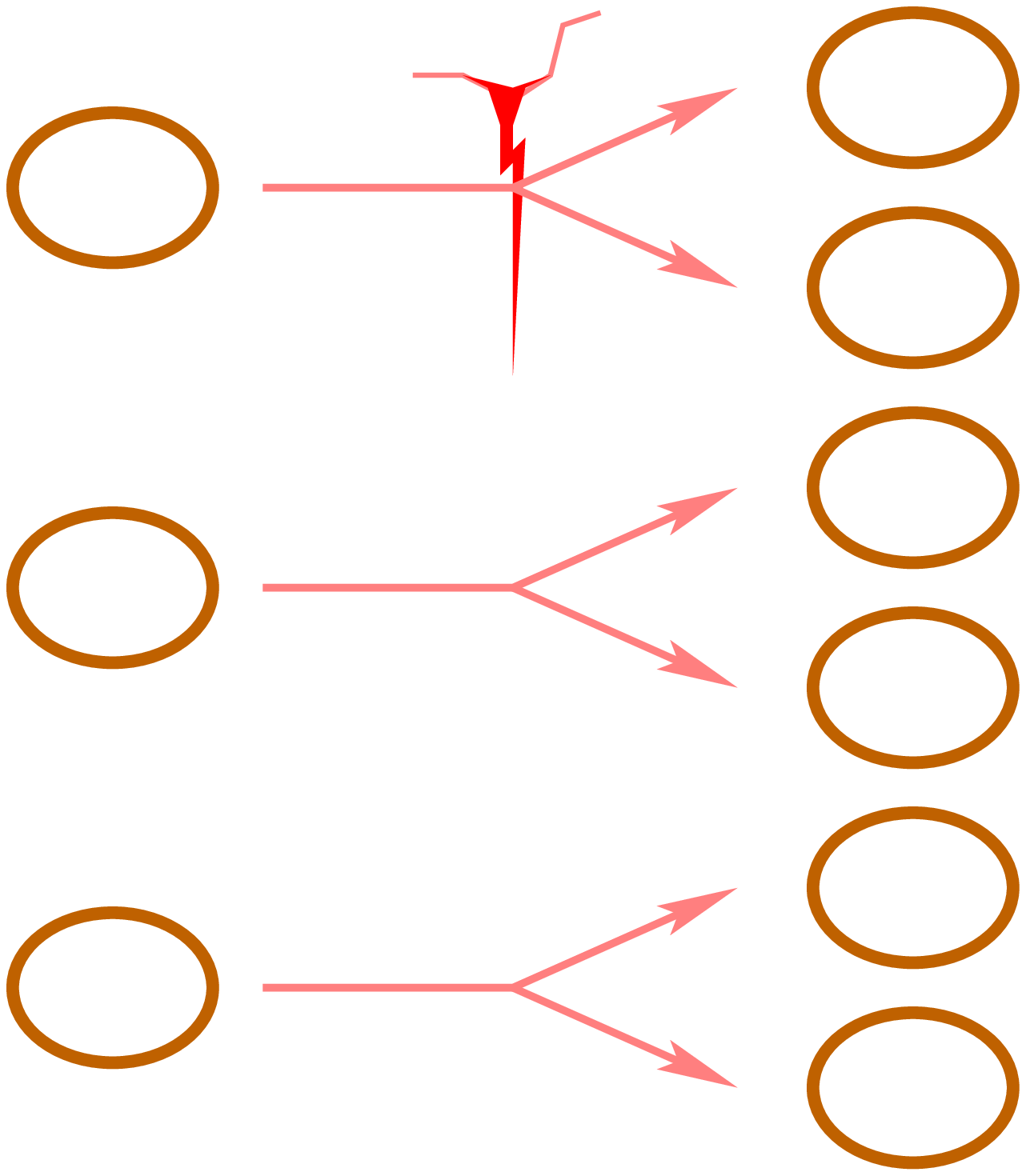}
\nputbox{-3.25,-2.1}{tl}{\large Usage examples.}
\nputbox{-3,-2.4}{tl}{\large Store $\bitzero$ in the second bit:}
\nputbox{-2,-3}{c}{$\phys{\bitzero\;\bitzero}$}
\nputbox{+2,-2.5}{c}{$\phys{\bitzero\;\bitzero}$}
\nputbox{+2,-3.5}{c}{$\phys{\err{\bitone}\;\bitzero}$}
\nputbox{-3,-4.4}{tl}{\large Store $\bitone$ in the second bit:}
\nputbox{-2,-5}{c}{$\phys{\bitzero\;\bitone}$}
\nputbox{+2,-4.5}{c}{$\phys{\bitzero\;\bitone}$}
\nputbox{+2,-5.5}{c}{$\phys{\err{\bitone}\;\bitone}$}
\end{picture}
\label{fig:trivialccode}
\herefigcap{A simple error model.
Errors affect only the first bit of a physical two bit system. All
\emph{joint} states of the two bits are affected by errors. For
example, the joint state $\phys{\bitzero\bitzero}$ is changed by the
error to $\phys{\err{\bitone}\bitzero}$.  Nevertheless the value of
the information represented in the second physical bit is unchanged.}
\end{herefig}

As suggested by the usage example in Fig.~\ref{fig:cchannel}, one can
``encode'' one bit of information in the physical system by the map
that takes $\bitzero\rightarrow\phys{\bitzero\bitzero}$ and
$\bitone\rightarrow\phys{\bitzero\bitone}$.  This means that the
states $\bitzero$ and $\bitone$ of an ideal bit are represented by the
states $\phys{\bitzero\bitzero}$ and $\phys{\bitzero\bitone}$ of the
noisy physical system, respectively.

To ``decode'' the information one can
extract the second bit by
the following map:
\begin{equation}
\begin{array}{rcl}
\phys{\bitzero\bitzero} &\rightarrow \bitzero \\
{} \phys{\bitone\bitzero} &\rightarrow \bitzero \\
{} \phys{\bitzero\bitone} &\rightarrow \bitone \\
{} \phys{\bitone\bitone} &\rightarrow \bitone 
\end{array}
\end{equation}
This procedure ensures that the encoded bit is recovered by the
decoding regardless of the error.  There are other combinations of
encoding and decoding that work.  For example, in the encoding, we
could swap the meaning of $\bitzero$ and $\bitone$ by using the map
$\bitzero\rightarrow\phys{\bitzero\bitone}$ and
$\bitone\rightarrow\phys{\bitzero\bitzero}$.  The new decoding
procedure adds a bit flip to the one shown above. The only difference
between this combination of encoding/decoding and the previous one
lies in the way in which the information is represented inside the range
of the encoding. The range consists of the two states
$\phys{\bitzero\bitzero}$ and $\phys{\bitzero\bitone}$ and is called the
``code''. The states in the code are called ``code words''.

Although trivial, the example just given is typical of ways for
dealing with errors.  That is, there is always a way of viewing the
physical system as a pair of abstract systems: The first member of the
pair experiences the errors and the second carries the information to
be protected. The two abstract systems are called ``subsystems'' of the
physical system and are usually not identifiable with any of the system's
physical components. The first is the ``syndrome'' subsystem and the
second is the ``information-carrying'' subsystem.  Encoding consists
of initializing the first system and storing the information in the
second.  Decoding is accomplished by extraction of the second system.
In the example, the two subsystems are readily identified as the two
physical bits that make up the physical system.  The first is the
syndrome subsystem and is initialized to $\phys{\bitzero}$ by the
encoding. The second carries the encoded information.

\subsection{The Repetition Code}

The next example is a special case of the main problem of classical
error-correction and occurs in typical communication settings and in
computer memories. Let the physical system consist of three bits. The
effect of the errors is to independently flip each bit with
probability $p$, which we take to be $p=.25$.  The repetition code results
from triplicating the information to be protected.  An encoding
is given by the map
$\bitzero\rightarrow\phys{\bitzero\bitzero\bitzero},
\bitone\rightarrow\phys{\bitone\bitone\bitone}$.
The repetition code is the set
$\{\phys{\bitzero\bitzero\bitzero},\phys{\bitone\bitone\bitone}\}$,
which is the range of the encoding.  The information can be
decoded with majority logic: If the majority of the three bits is
$\phys{\bitzero}$, output $\bitzero$, otherwise output $\bitone$.

How well does this encoding/decoding combination work for protecting one bit
of information against the errors?  The decoding fails to extract the
bit of information correctly if two or three of the bits were flipped
by the error.  We can calculate the probability of incorrect decoding
as follows: The probability of a given pair of bits having flipped is
$.25^2*.75$. There are three different pairs. The probability of three
bits having flipped is $.25^3$.  Thus the probability of error in the
encoded bit is $3\cdot.25^2*.75+.25^3=0.15625$.  This is an
improvement over $.25$, which is the probability that the information
represented in one of the three physical bits is corrupted by error.

To see that one can interpret this example by viewing the physical
system as a pair of subsystems, it suffices to identify the physical
system's states with the states of a suitable pair.
The following shows such a ``subsystem identification'':
\begin{equation}
\begin{array}{rcl}
\phys{\bitzero\bitzero\bitzero}&\leftrightarrow&\phys{\bitzero\bitzero}\cdot\bitzero
\\
\phys{\bitzero\bitzero\bitone}&\leftrightarrow&\phys{\bitone\bitone}\cdot\bitzero
\\
\phys{\bitzero\bitone\bitzero}&\leftrightarrow&\phys{\bitzero\bitone}\cdot\bitzero
\\
\phys{\bitone\bitzero\bitzero}&\leftrightarrow&\phys{\bitone\bitzero}\cdot\bitzero
\\
\phys{\bitzero\bitone\bitone}&\leftrightarrow&\phys{\bitone\bitzero}\cdot\bitone
\\
\phys{\bitone\bitzero\bitone}&\leftrightarrow&\phys{\bitzero\bitone}\cdot\bitone
\\
\phys{\bitone\bitone\bitzero}&\leftrightarrow&\phys{\bitone\bitone}\cdot\bitone
\\
\phys{\bitone\bitone\bitone}&\leftrightarrow&\phys{\bitzero\bitzero}\cdot\bitone
\end{array}
\end{equation}
The left side consists of the 8 states of the physical system,
which are the possible states for the three physical bits making up
the system. The right side shows the corresponding states for the
subsystem pair. The syndrome subsystem is a two bit
subsystem, whose states are shown first. The syndrome subsystem's states
are called ``syndromes''.  After the ``$\cdot$'' symbol are the states of the
information-carrying one-bit subsystem.

In the subsystem identification above, the repetition code consists of
the two states for which the syndrome is $\phys{\bitzero\bitzero}$.
That is, the code states $\phys{\bitzero\bitzero\bitzero}$ and
$\phys{\bitone\bitone\bitone}$ correspond to the states
$\phys{\bitzero\bitzero}\cdot\bitzero$ and
$\phys{\bitzero\bitzero}\cdot\bitone$ of the subsystem pair. For a
state in this code, single-bit flips do not change the
information-carrying bit, only the syndrome.  For example, a bit flip
of the second bit changes $\phys{\bitzero\bitzero\bitzero}$ to
$\phys{\bitzero\err{\bitone}\bitzero}$ which is identified with
$\phys{\bitzero\bitone}\cdot\bitzero$. The syndrome has changed from
$\phys{\bitzero\bitzero}$ to $\phys{\bitzero\bitone}$. Similarly, this
error changes $\phys{\bitone\bitone\bitone}$ to
$\phys{\bitone\err{\bitzero}\bitone}\leftrightarrow
\phys{\bitzero\bitone}\cdot\bitone$. The following diagram
shows these effects:
\begin{equation}
\begin{array}{ccc@{\hspace*{.6in}}rcr}
\phys{\bitzero\bitzero\bitzero} &\leftrightarrow & \phys{\bitzero\bitzero}\cdot\bitzero &
\phys{\bitone\bitone\bitone} &\leftrightarrow & \phys{\bitzero\bitzero}\cdot\bitone
  \\
\err{\downarrow} & &  &\err{\downarrow} & & 
  \\
\phys{\bitzero}\err{\bitone}\phys{\bitzero} &\leftrightarrow & \phys{\bitzero\bitone}\cdot\bitzero &
\phys{\bitone}\err{\bitzero}\phys{\bitone} &\leftrightarrow & \phys{\bitzero\bitone}\cdot\bitone
\end{array}
\end{equation}
Note that the syndrome change is
the same. 
In general, with this subsystem identification, we can infer from the
syndrome which single bit was flipped on an encoded state.

Errors usually act cumulatively over time. For the repetition code
this is a problem in the sense that it takes only a few actions of the
above error model for the two- and three-bit errors to overwhelm the
encoded information. One way to delay the loss of information is to
decode and re-encode sufficiently frequently.  Instead of explicitly
decoding and re-encoding, the subsystem identification can be used
directly for the same effect, namely, that of resetting the syndrome
subsystem's state to $\phys{\bitzero\bitzero}$.  For example, if the
state is $\phys{\bitone\bitzero}\cdot\bitone$, it needs to be
reset to $\phys{\bitzero\bitzero}\cdot\bitone$.  Therefore, using
the subsystem identification, resetting the syndrome subsystem
requires changing the state $\phys{\bitzero\bitone\bitone}$ to
$\phys{\bitone\bitone\bitone}$. It can be checked that, in every
case, what is required is to set all bits of the physical system to
the majority of the bits.  After the the syndrome subsystem has been reset,
the information is again protected against the next one-bit error.

\subsection{A Code for a Cyclic System}

We next consider a physical system that does not consist
of bits. This system has seven states symbolized by
$\phys{0},\phys{1},\phys{2},\phys{3},\phys{4},\phys{5}$ and $\phys{6}$.
Let $s_1$ be the right-circular shift operator defined by
$s_1(\phys{l}) = \phys{l+1}$ for
$\phys{0}\leq \phys{l}\leq \phys{5}$ and
$s_1(\phys{6})=\phys{0}$. 
Define $s_0=\idop$ (the identity operator),
\begin{equation}
s_k=\underbrace{s_1\ldots s_1}_{\mbox{$k$ times}},
\end{equation}
and $s_{-k} = s_k^{-1}$ (left-circular shift by $k$). 
The model can be visualized as a pointer on a dial with
seven positions as shown in Fig.~\ref{fig:cyclic}.
\begin{herefig}
\begin{picture}(6.5,4.3)(-3.25,-4.3)
\nputgr{0,0}{t}{}{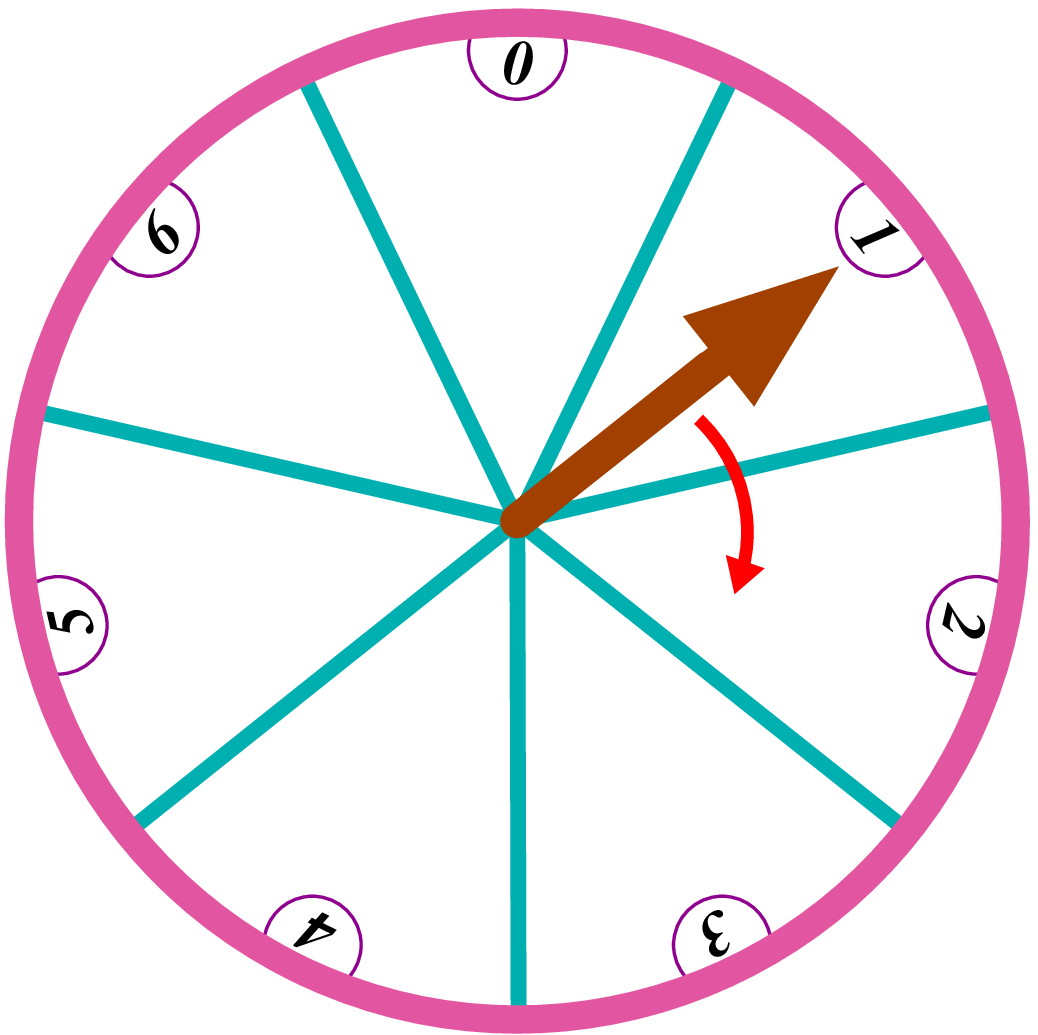}
\nputbox{1.05,-2}{c}{\Large $s_1$}
\end{picture}
\label{fig:cyclic}
\herefigcap{
A seven state cyclic system. The position of the pointer on
the seven-position dial determines the state of the system.
With the pointer in the position shown, the state is $\phys{1}$.
Errors have the effect of rotating the pointer clockwise (to the ``right'')
or counter-clockwise (to the ``left''). The effect of $s_1$ is to rotate
the pointer clockwise as shown by the red arrow.
}
\end{herefig}
Suppose that
the errors consist of applying $s_k$ with probability $q e^{-k^2}$,
where $q=0.5641$ is chosen so that the probabilities sum to $1$, that
is $\sum_{k=-\infty}^{\infty} q e^{-k^2}=1$.  Thus $s_0$ has
probability $0.5641$, and each of $s_{-1}$ and $s_{1}$ has probability
$0.2075$. These are the main errors that we need to protect
against. Continuous versions of this error model in the context of
communication channels are known as ``Gaussian channels''.

\ignore{
z = sum(exp(-([-20:1:20].^2)));
ps = exp(-([-2:1:2].^2))/z;
ps =[0.01033242382528,0.20753228024875,0.56413122621884,0.20753228024875, ...
     0.01033242382528];
sum(ps(2:4)) 
}

One bit can be encoded in this physical system by the map
$\bitzero\rightarrow\phys{1},
\bitone\rightarrow\phys{4}$. 
To decode with protection against $s_{0}$, $s_{-1}$ and $s_{1}$,
use the mapping:
\begin{equation}
\begin{array}{rcl}
\phys{0}&\rightarrow&\bitzero \\
\phys{1}&\rightarrow&\bitzero \\
\phys{2}&\rightarrow&\bitzero \\
\phys{3}&\rightarrow&\bitone \\
\phys{4}&\rightarrow&\bitone \\
\phys{5}&\rightarrow&\bitone \\
\phys{6} &\rightarrow& \mbox{fail}
\end{array}
\end{equation}
If state $\phys{6}$ is encountered, we know that an error
involving a shift of at least $2$ (left or right)  occurred, but there is no
reasonable way of decoding it to the state of a bit.  This means that
the error is detected, but we cannot correct it. Error detection can
be used by the receiver of information to ask for it to be sent again.
The probability of correctly decoding with this code is at least
$0.9792$, which is the probability that the error caused a shift of at
most one.

As before, a pair of syndrome and information-carrying subsystems can
be identified as being used by the encoding and decoding procedures. It
suffices to correctly identify the syndrome states, which we name
$\phys{{-}\bitone}$, $\phys{\bitzero}$ and $\phys{\bitone}$, because
they indicate which of the likeliest shifts happened.
The resulting subsystem identification is
\begin{equation}
\begin{array}{rcr}
\phys{0}&\leftrightarrow&\phys{{-}\bitone}\cdot\bitzero \\
\phys{1}&\leftrightarrow&\phys{\bitzero}\cdot\bitzero \\
\phys{2}&\leftrightarrow&\phys{\bitone}\cdot\bitzero \\
\phys{3}&\leftrightarrow&\phys{{-}\bitone}\cdot\bitone \\
\phys{4}&\leftrightarrow&\phys{\bitzero}\cdot\bitone \\
\phys{5}&\leftrightarrow&\phys{\bitone}\cdot\bitone
\end{array}
\label{eq:cyclic_id}
\end{equation}
A new feature of this subsystem identification is that it is
incomplete: Only a subset of the state space is identified. In this
case, the complement can be used for error detection. 

Like the repetition code, this code can be used in a setting
where the errors happen repeatedly. Again it suffices to reset the
syndrome subsystem, in this case to $\phys{\bitzero}$, to keep
the encoded information protected.  After the syndrome subsystem
has been reset, a subsequent $s_{1}$ or $s_{-1}$ error affects only
the syndrome.

\section{Principles of Error Correction}
\label{sec:prec}

When considering the problem of limiting the effects of errors in information
processing, the first task is to establish the properties of the
physical systems that are available for representing and computing
with information. Thus it is necessary to learn the following:
\begin{itemize}
\item[1.] The physical system to be used, in particular the structure
of its state space.  
\item[2.] The available means for controlling this system.  
\item[3.] The type of information to be processed.
\item[4.] The nature of the errors, that is, the error model. 
\end{itemize}
With this information, the approaches used  to correct errors in the
three examples provided in the previous section involve the following:
\begin{itemize}
\item[1.] Determine a code, which is a subspace of the
physical system that can represent the information to be processed.
\item[2.a] Identify a decoding procedure that can restore
the information represented in the code after any one of the most likely
errors occurred.
\item[2.b] Or,
determine a pair of syndrome and information-carrying subsystems such
that the code corresponds to a ``base'' state of the syndrome subsystem
and the primary errors act only on the syndrome.
\item[3.] Analyze the error behavior of the code and subsystem.
\end{itemize}

The tasks of determining a code and of identifying decoding procedures
or subsystems are closely related. As a result, the following
questions are at the foundation of the theory of error-correction:
What properties must a code satisfy so that it can be used to
protect well against a given error model? How does one obtain the
decoding or subsystem identification that achieves this protection?
In many cases, the answers can be based on choosing a fixed set of
error operators that represents well the most likely errors and then
determining whether these errors can be protected against without any
loss of information. Once an error set is fixed, determining whether
it is ``correctable'' can be cast in terms of the idea of ``detectable''
errors.  This idea works equally well for both classical and quantum
information.  We introduce it using classical information concepts.

\subsection{Error Detection}

Error detection was used in the cyclic system example to reject a
state that could not be properly decoded. In the communication
setting, error control methods based on error detection alone work as
follows: The encoded information is transmitted.  The receiver checks
whether the state is still in the code, that is, whether it could have
been obtained by encoding. If not, the result is rejected.  The sender
can be informed of the failure so that the information can be
retransmitted.  Given a set of error operators that need to be
protected against, the scheme is successful if for each error
operator, either the information is unchanged, or the error is
detected.  Thus we can say that an operator $\err{E}$ is
``detectable'' by a code if for each state $x$ in the code, either
$\err{E}x=x$ or $\err{E}x$ is not in the code. See
Fig.~\ref{fig:cdetect}

\pagebreak
\begin{herefig}
\begin{picture}(7,2.25)(-3.5,-2.25)
  \nputgr{0,0}{t}{width=7in}{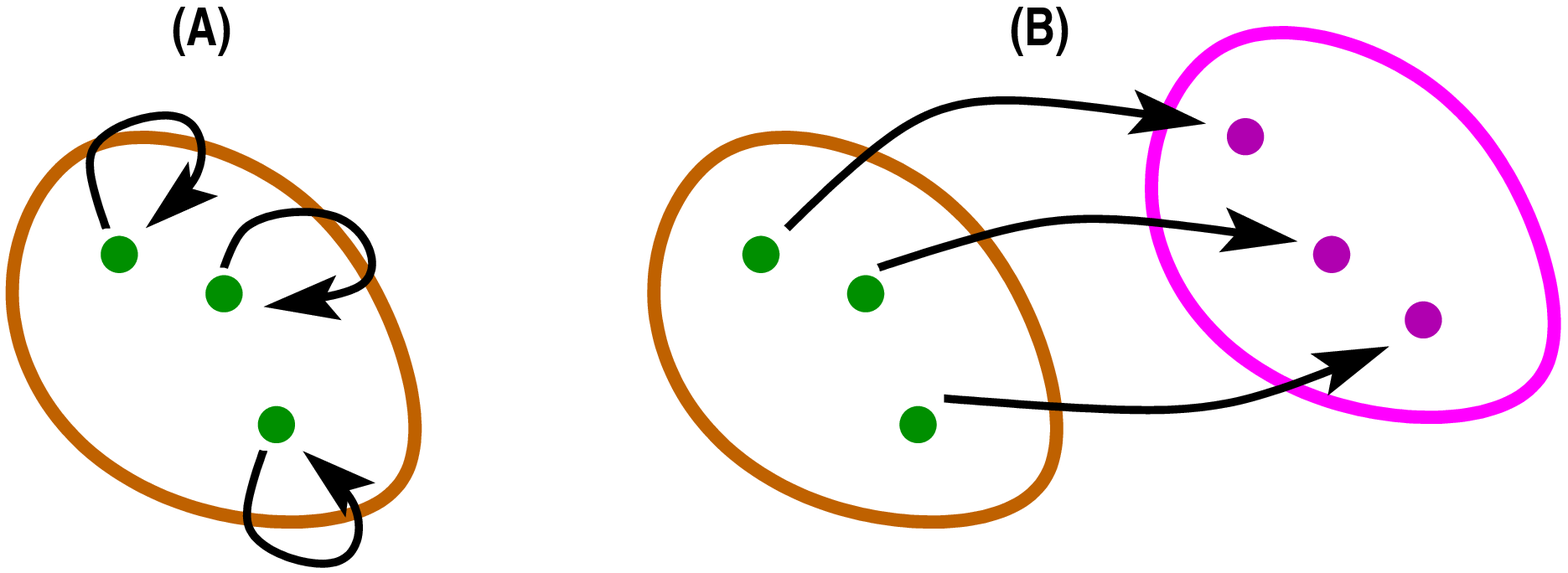}
\end{picture}
\label{fig:cdetect}
\herefigcap{
Pictorial representation of typical detectable and undetectable errors
for a code. Three examples are shown.  In each, the code is
represented by a brown oval containing three code words (green points). The
effect of the error operator is shown as arrows.  In \textsf{(A)}, the
error does not change the code words and is therefore considered
detectable.  In \textsf{(B)}, the error maps the code words outside the
code, so that it is detected.  In \textsf{(C)}, one code word is
mapped to another, as shown by the red arrow. Finding that a received
word is still in the code does not guarantee that it was the
originally encoded word. The error is therefore not detectable.  }
\end{herefig}

What errors are detectable by the codes in the examples?  The code in
the first example consists of $\phys{\bitzero\bitzero}$ and
$\phys{\bitzero\bitone}$.  Every operator that affects only the first
bit is therefore detectable. In particular, all operators in the error
model are detectable.  In the second example, the code consists of the
states $\phys{\bitzero\bitzero\bitzero}$ and
$\phys{\bitone\bitone\bitone}$.  The identity operator has no effect
and is therefore detectable.  Any flips of exactly one or two bits are
detectable because the states in the code are changed to states outside
the code.  The error that flips all bits is not detectable because it
preserves the code but changes the states in the code.  With the code
for the cyclic system, shifts by $-2,-1,0,1,2$ are detectable but not
shifts by $3$.

To conclude the section, we state a characterization of detectability,
which has a natural generalization to the case of quantum information.
\begin{equation}
\begin{eqthm}{Theorem}
$\err{E}$ is detectable by a code if and only if for all
$x\not=y$ in the code, $\err{E}x\not=y$.
\end{eqthm}
\label{thm:classdetect}
\end{equation}

\subsection{From Error Detection to Error Correction}

Given a code $C$ and a set of error operators $\err{\cE}=\{\idop =
\err{E}_0,\err{E}_1,\err{E}_2,\ldots\}$ is it possible to determine 
whether a decoding procedure or subsystem exists such that $\err{\cE}$
is ``correctable'' (by $C$), that is, such that the errors in $\err{\cE}$ do
not affect the encoded information? 
As explained below, the answer is
yes and the solution is to check the condition in the following theorem:
\begin{equation}
\begin{eqthm}{Theorem}
$\err{\cE}$ is correctable by $C$ if and only if
for all $x\not=y$ in the code
and all $i,j$, it is true that $\err{E}_ix\not=\err{E}_jy$.
\end{eqthm}
\label{thm:cecc}
\end{equation}
Observe that the notion of correctability depends on all the errors in
the set under consideration and, unlike detectability, cannot be
applied to individual errors.

To see that the condition for correctability in Thm.~\ref{thm:cecc} is
necessary, suppose that for some $x\not=y$ in the code and some $i$
and $j$, we have $z=\err{E}_ix=\err{E}_jy$.  If the state $z$ is
obtained after an unkown error in $\err{\cE}$, then it is not possible
to determine whether the original code word was $x$ or $y$, because we
cannot tell whether $\err{E}_i$ or $\err{E}_j$ occurred.

To see that the condition for correctability in Thm.~\ref{thm:cecc} is
sufficient, we assume it and construct a decoding method $z\rightarrow
\textrm{dec}(z)$. Suppose that after an unknown error occurred, the
state $z$ is obtained. There can be one and only one $x$ in the code
for which some $\err{E}_{i(z)}\in\err{\cE}$ satisfies the condition
that $\err{E}_{i(z)}x=z$.  Thus $x$ must be the original code word and
we can decode $z$ by defining $x=\textrm{dec}(z)$.  Note that it is
possible for two errors to have the same effect on some code words.  A
subsystem identification for this decoding is given by
$z\leftrightarrow i(z)\cdot
\textrm{dec}(z)$, where the syndrome subsystem's state space consists of
error operator indices $i(z)$, and the information-carrying system's
consists of the code words $\textrm{dec}(z)$ returned by the decoding.
The subsystem identification thus constructed is not necessarily onto
the state space of the subsystem pair. That is, for different code
words $x$, the set of $i(z)$ such that $\textrm{dec}(z)=x$ can vary
and need not be all of the error indices. As we will show, the
subsystem identification is onto the state space of the subsystem pair
in the case of quantum information.  It is instructive to check that,
when applied to the examples, this subsystem construction does give a
version of the subsystem identifications provided earlier.

It is possible to relate the condition for correctability of an error
set to detectability.  For simplicity, assume that each $\err{E}_i$ is
invertible. (This assumption is satisfied by our examples, but not by
error operators such as ``reset bit one to $\bitzero$''.)  In this case,
the correctability condition is equivalent to the statement that all
products $\err{E}_j^{-1}\err{E}_i$ are detectable. To see the
equivalence, first suppose that some $\err{E}_j^{-1}\err{E}_i$ is not
detectable. Then there are $x\not=y$ in the code such that
$\err{E}_j^{-1}\err{E}_ix = y$. Consequently $\err{E}_ix=\err{E}_jy$
and the error set is not correctable. This argument can be reversed to
complete the proof of equivalence.

If the assumption that the errors are invertible does not hold, the
relationship between detectability and correctability becomes more
complicated, requiring a generalization of the inverse operation. This
generalization is simpler in the quantum setting.

\section{Quantum Error Correction}

The principles of error correction outlined in Sec.~\ref{sec:prec}
apply to the quantum setting as readily as to the classical
setting.  The main difference is that the physical system to be used
for representing and processing information behaves quantum
mechanically and the type of information is quantum. The question of
how classical information can be protected in quantum systems is
also interesting but will not be discussed here. We illustrate the
principles of quantum error correction by considering quantum versions
of the three examples of Sect.~\ref{sec:concepts_and_examples} and
then add a uniquely quantum example with potentially practical
applications in, for example, quantum dot technologies. For an
explanation of the basic quantum information concepts and conventions,
see~\cite{knill:qc2001c}.

\subsection{Trivial Two-Qubit Example}
\label{sec:qtriv}

A quantum version of the two bit example from the previous section
consists of two physical qubits, where the errors
randomly apply the identity or one of the Pauli
operators to the first qubit. The Pauli operators are defined
by
\begin{equation}
\begin{array}{rcl}
\idop &=& \qaop{1}{0}{0}{1},\\
\sigma_x &=& \qaop{0}{1}{1}{0},\\
\sigma_y &=& \qaop{0}{-i}{i}{0},\\
\sigma_z &=& \qaop{1}{0}{0}{-1}.
\end{array}
\label{eq:paulidef}
\end{equation}
Explicitly, the errors have the effect
\begin{equation}
\kets{\phys{\psi}}{12} \rightarrow
  \left\{
     \begin{array}{rl}
        \idop\kets{\phys{\psi}}{12} & \mbox{Prob. $.25$}\\
        \slb{\sigma_x}{1}\kets{\phys{\psi}}{12} & \mbox{Prob. $.25$}\\
        \slb{\sigma_y}{1}\kets{\phys{\psi}}{12} & \mbox{Prob. $.25$}\\
        \slb{\sigma_z}{1}\kets{\phys{\psi}}{12} & \mbox{Prob. $.25$}
     \end{array}
  \right.,
\label{eq:cdep}
\end{equation}
where the superscripts in parentheses specify the qubit
that an operator acts on.  This error model is called ``completely
depolarizing'' on qubit $\sysfnt{1}$.  Obviously, a one-qubit
state can be stored in the second physical qubit without being affected
by the errors.  An encoding operation that implements this observation
is
\begin{equation}
\ket{\psi}\rightarrow\kets{\phys{\bitzero}}{1}\kets{\phys{\psi}}{2},
\end{equation}
which realizes an ideal qubit as a two-dimensional subspace of the
physical qubits. This subspace is the ``quantum code'' for this
encoding.  To decode one can discard physical qubit $\sysfnt{1}$ and
return qubit $\sysfnt{2}$, which is considered a natural
subsystem of the physical system. In this case, the identification of
syndrome and information-carrying subsystems is the obvious one
associated with the two physical qubits.

\subsection{Quantum Repetition Code}
\label{sec:qrc}

The repetition code can be used to protect quantum information in
the presence of a restricted error model.  Let the physical system
consist of three qubits. Errors act by independently applying, to each
qubit, the flip operator $\sigma_x$ with probability $.25$.
The classical code can be made into a quantum code by the
superposition principle. Encoding one qubit is accomplished by
\begin{equation}
\alpha\ket{\bitzero}+\beta\ket{\bitone}
 \rightarrow \alpha\ket{\phys{\bitzero\bitzero\bitzero}} +
             \beta\ket{\phys{\bitone\bitone\bitone}}.
\end{equation}
The associated quantum code is the range of the encoding, that is, the
two-dimensional subspace spanned by the encoded states
$\ket{\phys{\bitzero\bitzero\bitzero}}$ and
$\ket{\phys{\bitone\bitone\bitone}}$.

As in the classical case, decoding is accomplished by majority logic.
However, it must be implemented carefully to avoid destroying quantum
coherence in the stored information. One way to do that is to use only
unitary operations to transfer the stored information to the output
qubit.  Fig.~\ref{fig:majdec} shows a quantum network that accompishes
this task.

\begin{herefig}
\begin{picture}(7.2,4)(-3.6,-4)
\nputgr{0,0}{t}{}{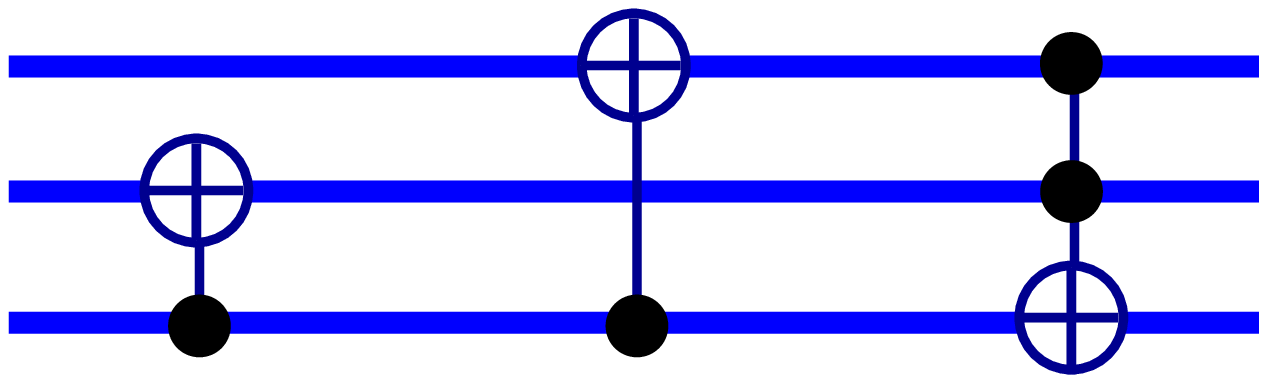}
\nputbox{-2.5,-.2}{bl}{\large$\sysfnt{1}$}
\nputbox{-2.5,-.7}{bl}{\large$\sysfnt{2}$}
\nputbox{-2.5,-1.2}{bl}{\large$\sysfnt{3}$}
\nputbox{+2.6,-1.28}{l}{\scalebox{1.3}{$\ket{\psi_{\textrm{out}}}$}}
\nputbox{0.08,-1.7}{t}{
\begin{tabular}{|@{\hspace*{.4in}}c@{\hspace*{.4in}}|@{\hspace*{.7in}}c@{\hspace*{.7in}}|@{\hspace*{.7in}}c@{\hspace*{.7in}}|@{\hspace*{.4in}}c@{\hspace*{.4in}}|}
\hline
$\ket{\phys{\bitzero\bitzero\bitzero}}$&$\ket{\phys{\bitzero\bitzero\bitzero}}$&$\ket{\phys{\bitzero\bitzero\bitzero}}$&$\ket{\phys{\bitzero\bitzero}}\ket{\bitzero}$\\
$\ket{\phys{\bitzero\bitzero\bitone}}$&$\ket{\phys{\bitzero\bitone\bitone}}$&$\ket{\phys{\bitone\bitone\bitone}}$&$\ket{\phys{\bitone\bitone}}\ket{\bitzero}$\\
$\ket{\phys{\bitzero\bitone\bitzero}}$&$\ket{\phys{\bitzero\bitone\bitzero}}$&$\ket{\phys{\bitzero\bitone\bitzero}}$&$\ket{\phys{\bitzero\bitone}}\ket{\bitzero}$\\
$\ket{\phys{\bitone\bitzero\bitzero}}$&$\ket{\phys{\bitone\bitzero\bitzero}}$&$\ket{\phys{\bitone\bitzero\bitzero}}$&$\ket{\phys{\bitone\bitzero}}\ket{\bitzero}$\\
\hline
$\ket{\phys{\bitone\bitone\bitone}}$&$\ket{\phys{\bitone\bitzero\bitone}}$&$\ket{\phys{\bitzero\bitzero\bitone}}$&$\ket{\phys{\bitzero\bitzero}}\ket{\bitone}$
\\
$\ket{\phys{\bitone\bitone\bitzero}}$&$\ket{\phys{\bitone\bitone\bitzero}}$&$\ket{\phys{\bitone\bitone\bitzero}}$&$\ket{\phys{\bitone\bitone}}\ket{\bitone}$\\
$\ket{\phys{\bitone\bitzero\bitone}}$&$\ket{\phys{\bitone\bitone\bitone}}$&$\ket{\phys{\bitzero\bitone\bitone}}$&$\ket{\phys{\bitzero\bitone}}\ket{\bitone}$\\
$\ket{\phys{\bitzero\bitone\bitone}}$&$\ket{\phys{\bitzero\bitzero\bitone}}$&$\ket{\phys{\bitone\bitzero\bitone}}$&$\ket{\phys{\bitone\bitzero}}\ket{\bitone}$\\
\hline
\end{tabular}
}
\end{picture}
\label{fig:majdec}
\herefigcap{Quantum network for 
majority logic decoding into the output qubit $\sysfnt{3}$.  The
effect of the quantum network on the basis states is shown. The top
half shows the states with majority $\phys{\bitzero}$. The decoded
qubit is separated in the last step. The conventions for illustrating
quantum networks are explained in~\cite{knill:qc2001c}.}
\end{herefig}

As shown, the decoding network establishes an identification between
the three physical qubits and a pair of subsystems consisting of two
qubits representing the syndrome subsystem and one qubit for the
information-carrying subsystem.  On the left side of the
correspondence, the information-carrying subsystem is not identifiable
with any one (or two) of the physical qubits. Nevertheless it exists
there through the identification.

To obtain a network for encoding, we reverse the decoding
network and initialize qubits $\sysfnt{2},\sysfnt{3}$ in the state
$\ket{\phys{\bitzero\bitzero}}$. Because of the initialization, the Toffoli
gate becomes unnecessary. The complete system with a typical
error is shown in Fig.~\ref{fig:repsys}.

\begin{herefig}
\begin{picture}(6.25,3.3)(-3.125,-3)
\nputgr{-.3,0}{t}{}{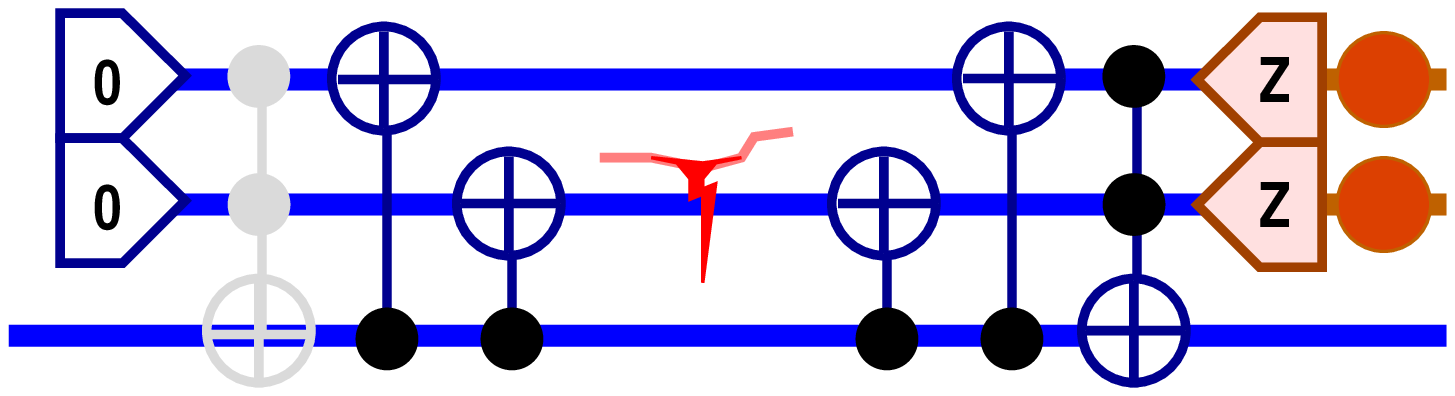}
\nputbox{-2,.05}{b}{$\overbrace{\rule{2in}{0in}}^{\mbox{\large Encode}}$}
\nputbox{+.8,.05}{b}{$\overbrace{\rule{1.5in}{0in}}^{\mbox{\large Decode}}$}
\nputbox{-3.2,-1.35}{r}{\scalebox{1.3}{$\ket{\psi}$}}
\nputbox{-2.5,-1.5}{t}{$
\alpha\ket{\bitzero}+\beta\ket{\bitone}\rightarrow \left\{\begin{array}{l}
  \alpha\ket{\phys{\bitzero\bitzero}}\ket{\bitzero}\\
  +\\
  \beta\ket{\phys{\bitzero\bitzero}}\ket{\bitone}
\end{array}\right.$}
\nputbox{-1.15,-1.5}{tl}{$
\begin{array}{l}
  \alpha\ket{\phys{\bitzero\bitzero\bitzero}}\\
  +\\
  \beta\ket{\phys{\bitone\bitone\bitone}}
\end{array}$}
\nputbox{+.25,-1.5}{tr}{$
\begin{array}{l}
  \alpha\ket{\phys{\bitzero\err{\bitone}\bitzero}}\\
  +\\
  \beta\ket{\phys{\bitone\err{\bitzero}\bitone}}
\end{array}$}
\nputbox{+1.2,-1.5}{tl}{$
\left.\begin{array}{l}
  \alpha\ket{\phys{\bitzero\err{\bitone}}}\ket{\bitzero}\\
  +\\
  \beta\ket{\phys{\bitzero\err{\bitone}}}\ket{\bitone}
\end{array}\right\}=\ket{\phys{\bitzero\err{\bitone}}}(\alpha\ket{\bitzero}+\beta\ket{\bitone})
$}
\end{picture}
\label{fig:repsys}
\herefigcap{Encoding and decoding networks for the quantum repetition
code with a typical error. The error that occurred can be determined
from the state of the syndrome subsystem, which consists of the top two
qubits.  The encoding is shown as the reverse of the decoding,
starting with an initialized syndrome subsystem.  When the decoding is
reversed to obtain the encoding, there is an initial Toffoli gate
(shown in gray). Because of the initialization, this gate has no
effect and is therefore omitted in an implementation.  }
\end{herefig}

As in the case of the classical repetition code, we can protect
against cumulative errors without explicitly decoding and then
re-encoding, which would cause a temporary loss of protection. Instead,
one can find a means for directly resetting the syndrome subsystem to
$\ket{\phys{\bitzero\bitzero}}$ (thus returning the information to the
code) before the errors happen again. After resetting in this way,
the errors in the correctable set have no effect on the encoded
information because they act only on the syndrome subsystem.

Part of the task of designing error-correcting systems is to determine
how well the system performs. An important performance measure is the
probability of error. In quantum systems, the probability of error is
intuitively interpreted as the maximum probability with which we can
see a result different from the expected one in any measurement.
Specifically, to determine the error, one compares the output
$\ket{\psi_o}$ of the system to the input $\ket{\psi}$. An upper bound
is obtained if the output is written as a combination of the input state
and an ``error'' state. For quantum information, combinations are
linear combinations (that is, superpositions).  Thus
$\ket{\psi_o}=\gamma\ket{\psi}+\ket{\err{e}}$ (see
Fig.~\ref{fig:error}).The probability of error is bounded by
$\err{\epsilon}=|\ket{\err{e}}|^2$ (which we call an ``error estimate''). In
general, there are many different ways of writing the output as a
combination of an acceptable state and an error term. One attempts to
choose the combination that minimizes the error estimate.  This choice
yields the number $\err{\epsilon}$, for which $1-\err{\epsilon}$ is
called the ``fidelity''. A fidelity of $1$ means that the output is
the same (up to a phase factor) as the input.

\begin{herefig}
\begin{picture}(7,2)(-3.5,-2)
\nputgr{0,0}{t}{height=2in}{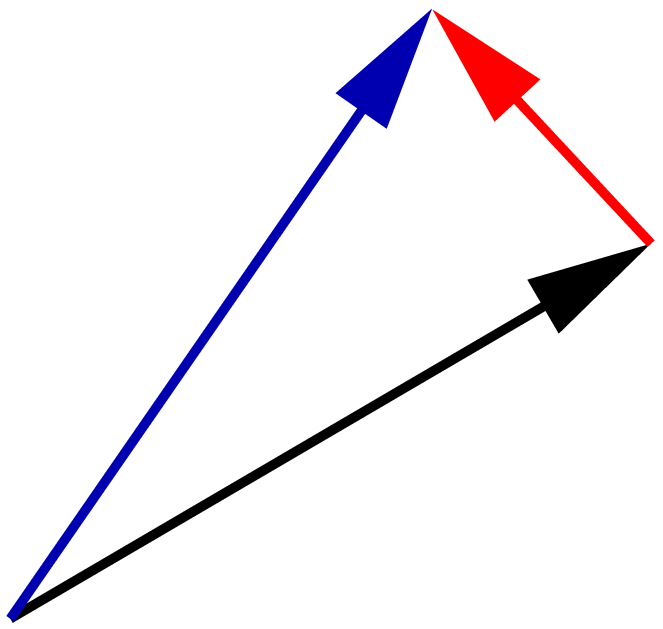}
\nputbox{.8,-.4}{l}{\scalebox{1.2}{$\ket{\err{e}}$}}
\nputbox{-.3,-.7}{r}{\scalebox{1.2}{$\ket{\psi_o}$}}
\nputbox{.5,-1.5}{r}{\scalebox{1.2}{$\gamma\ket{\psi}$}}
\end{picture}
\label{fig:error}
\herefigcap{Representation of an error estimate.
Any decomposition of the output state $\ket{\psi_o}$ into a ``good''
state $\gamma\ket{\psi}$ and an (unnormalized) error term
$\ket{\err{e}}$ gives an estimate $\err{\epsilon}=|\ket{\err{e}}|^2$.
For pure states, the optimum estimate is obtained when the error term
is orthogonal to the input state.  To obtain an error estimate for
mixtures, one can use any representation of the state as a
probabilistic combination of pure states and calculate the
probabilistic sum of the pure state errors.  }
\end{herefig}

To illustrate error analysis, we calculate the error for the
repetition code example for the two initial states $\ket{\bitzero}$ and
${1\over\sqrt{2}}\left(\ket{\bitzero}+\ket{\bitone}\right)$.
\begin{eqnarray}
\ket{\bitzero}&\stackrel{\mbox{\small encode}}{\longrightarrow}&\ket{\phys{\bitzero\bitzero\bitzero}}\nonumber\\
  &\err{\longrightarrow}&
  \left\{
     \begin{array}{rl}
        .75^3:&\ket{\phys{\bitzero\bitzero\bitzero}},\\
        .25*.75^2:&\ket{\phys{\err{\bitone}\bitzero\bitzero}},\\
        .25*.75^2:&\ket{\phys{\bitzero\err{\bitone}\bitzero}},\\
        .25*.75^2:&\ket{\phys{\bitzero\bitzero\err{\bitone}}},\\
        .25^2*.75:&\ket{\phys{\err{\bitone\bitone}\bitzero}},\\
        .25^2*.75:&\ket{\phys{\err{\bitone}\bitzero\err{\bitone}}},\\
        .25^2*.75:&\ket{\phys{\bitzero\err{\bitone\bitone}}},\\
        .25^3:&\ket{\phys{\err{\bitone\bitone\bitone}}}
     \end{array}
  \right.\nonumber\\
  &\stackrel{\mbox{\small decode}}{\longrightarrow}&
  \left\{
     \begin{array}{rl}
        .4219:&\ket{\phys{\bitzero\bitzero}}\cdot\ket{\bitzero},\\
        .1406:&\ket{\phys{\bitone\bitzero}}\cdot\ket{\bitzero},\\
        .1406:&\ket{\phys{\bitzero\bitone}}\cdot\ket{\bitzero},\\
        .1406:&\ket{\phys{\bitone\bitone}}\cdot\ket{\bitzero},\\
        .0469:&\ket{\phys{\bitone\bitone}}\cdot\ket{\bitone},\\
        .0469:&\ket{\phys{\bitzero\bitone}}\cdot\ket{\bitone},\\
        .0469:&\ket{\phys{\bitone\bitzero}}\cdot\ket{\bitone},\\
        .0156:&\ket{\phys{\bitzero\bitzero}}\cdot\ket{\bitone}
     \end{array}
  \right.
\end{eqnarray}
The final state is a mixture consisting of four correctly
decoded components and four incorrectly decoded ones.
The probability of each state in the mixture is shown
before the colon.
The incorrectly decoded information is orthogonal to the encoded
information, and its probability is
$0.1563$, an improvement over the one-qubit error-probability of $0.25$.
The second state behaves quite differently:
\begin{eqnarray}
{1\over\sqrt{2}}\left(\ket{\bitzero}+\ket{\bitone}\right)
  &\stackrel{\mbox{\small encode}}{\longrightarrow}&
  {1\over\sqrt{2}}\left(\ket{\phys{\bitzero\bitzero\bitzero}}+\ket{\phys{\bitone\bitone\bitone}}\right)\nonumber \\
  &\err{\longrightarrow}&
  \left\{
     \begin{array}{rl}
        \vdots&\\
        .25^2*.75:&{1\over\sqrt{2}}\left(
            \ket{\phys{\err{\bitone\bitone}\bitzero}}+
            \ket{\phys{\err{\bitzero\bitzero}\bitone}}\right)
            ,\\
        \vdots&
     \end{array}
  \right.\nonumber \\
  &\stackrel{\mbox{\small decode}}{\longrightarrow}&
  \left\{
     \begin{array}{rl}
        \vdots&\\
        .0469:&{1\over\sqrt{2}}
          \ket{\phys{\bitone\bitone}}\cdot
             \left(\ket{\bitone}+\ket{\bitzero}\right),\\
        \vdots&
     \end{array}
  \right.
\end{eqnarray}
Not all error events have been shown, but in each case it can be seen
that the state is decoded correctly, so the error is $0$. This shows
that the error probability can depend significantly on the initial
state.  To remove this dependence and give a state independent error
quantity, one can use the ``worst-case'', the ``average'' or the
``entanglement'' error.  See Sect.~\ref{sec:noise_analysis}.

\subsection{Quantum Code for a Cyclic System}

The shift operators introduced earlier act as permutations of the
seven states of the cyclic system. They can therefore be extended to
unitary operators on a seven-state cyclic quantum system with logical
basis
$\ket{\phys{0}},\ket{\phys{1}},\ket{\phys{2}},\ket{\phys{3}},\ket{\phys{4}},\ket{\phys{5}},\ket{\phys{6}}$. The
error model introduced earlier makes sense here without modification,
as does the encoding. The subsystem identification now
takes the six-dimensional subspace spanned by
$\ket{\phys{0}},\ldots\ket{\phys{5}}$ to a pair consisting of a
three-state system with basis
$\ket{\phys{-1}},\ket{\phys{0}},\ket{\phys{1}}$ and a qubit.  The
identification of Eq.~\ref{eq:cyclic_id} extends linearly to a unitary
subsystem identification. The procedure for decoding is modified as
follows: First,
a measurement is performed to determine whether the state is in the
six-dimensional subspace or not. If it is, the identification is used
to extract the qubit. Here is an outline of what happens 
when the state
${1\over\sqrt{2}}\left(\ket{\bitzero}+\ket{\bitone}\right)$ is encoded:
\begin{eqnarray}
{1\over\sqrt{2}}\left(\ket{\bitzero}+\ket{\bitone}\right)
  &\stackrel{\mbox{\small encode}}{\longrightarrow}&
  {1\over\sqrt{2}}\left(\ket{\phys{1}}+\ket{\phys{4}}\right)\nonumber\\
  &\err{\longrightarrow}&
  \left\{
     \begin{array}{rl}
        \vdots&\\
        .05641 e^{-4}:&{1\over\sqrt{2}}\left(
            \ket{\phys{\err{3}}}+
            \ket{\phys{\err{6}}}\right)
            ,\\
        \vdots&
     \end{array}
  \right.\nonumber\\
  &\stackrel{\mbox{\small detect}}{\longrightarrow}&
  \left\{
     \begin{array}{rl}
        \vdots&\\
        .001:\left\{
             \begin{array}{rl}
               .5:& \mbox{fail}\\
               .5:& \ket{\phys{\err{3}}}
             \end{array}     
             \right.
        \vdots&
     \end{array}
  \right.\nonumber\\
  &\stackrel{\mbox{\small decode}}{\longrightarrow}&
   \left\{
     \begin{array}{rl}
        \vdots&\\
        .0005:& \mbox{fail}\\
        .0005:& \ket{\phys{-1}}\cdot\ket{\bitone}\\
        \vdots&
     \end{array}
   \right.\nonumber\\
  &=&
   \left\{
     \begin{array}{rl}
        \vdots&\\
        .0005:& \mbox{fail}\\
        .0005:& \ket{\phys{-1}}\cdot \left(
                 {1\over{2}}\left(\ket{\bitzero}+\ket{\bitone}\right) +
                 \err{{1\over{2}}\left(-\ket{\bitzero}+\ket{\bitone}\right)}
                 \right)\\
        \vdots&
     \end{array}
   \right.
\end{eqnarray}
A ``good'' state was separated from the output in the case that is
shown. The leftover error term has probability amplitude
$.0005*((1/2)^2+(1/2)^2)=.00025$, which contributes to the total error
(not shown).

\subsection{Three Quantum Spin-$1/2$ Particles}
\label{sec:threeq1/2}

Quantum physics provides a rich source of systems with many
opportunities for representing and protecting quantum
information. Sometimes it is possible to encode information in such a
way that it is protected from the errors indefinitely, without
intervention. An example is the trivial two-qubit system discussed
before.  Whenever error protection without intervention is possible,
there is an information-carrying subsystem such that errors act only
on the associated syndrome subsystem regardless of the current
state. An information-carrying subsystem with this property is called
``noiseless''.  A physically motivated example of a one-qubit
noiseless subsystem can be found in three spin-$1\over 2$ particles
with errors due to random fluctuations in an external field.

A spin-$1\over 2$ particle's state space is spanned by two states
$\ket{\phys{\uparrow}}$ and $\ket{\phys{\downarrow}}$. Intuitively,
these states correspond to the spin pointing ``up'' ($\ket{\phys{\uparrow}}$)
or ``down'' ($\ket{\phys{\downarrow}}$) in some chosen
reference frame. The state space is therefore the same as that of a
qubit and we can make the identifications
$\ket{\phys{\uparrow}}\leftrightarrow\ket{\bitzero}$ and
$\ket{\phys{\downarrow}}\leftrightarrow\ket{\bitone}$.  An external
field causes the spin to ``rotate'' according to an evolution of the form
\begin{equation}
\ket{\psi_t} = e^{-i(u_x\sigma_x+u_y\sigma_y+u_z\sigma_z)t/2}\ket{\psi}.
\end{equation}
The vector $\vec u = (u_x,u_y,u_z)$ characterizes the direction of the
field and the strength of the spin's interaction with the field. 
This situation arises, for example, in nuclear magnetic
resonance with spin-$1\over 2$ nuclei, where the fields are magnetic
fields (see~\cite{knill:qc2001f}).

Now consider the physical system composed of three spin-$1\over 2$
particles with errors acting as identical rotations of the three
particles. Such errors occur if they are due to a uniform external
field that fluctuates randomly in direction and strength. 
The evolution caused by a uniform field is given by
\begin{eqnarray}
\kets{\psi_t}{123} &=& 
e^{-i(u_x\slb{\sigma_x}{1}+u_y\slb{\sigma_y}{1}+u_z\slb{\sigma_z}{1})t/2}
e^{-i(u_x\slb{\sigma_x}{2}+u_y\slb{\sigma_y}{2}+u_z\slb{\sigma_z}{2})t/2}
e^{-i(u_x\slb{\sigma_x}{3}+u_y\slb{\sigma_y}{3}+u_z\slb{\sigma_z}{3})t/2}
\kets{\psi}{123}
\nonumber\\
&=& 
e^{-i(u_x(\slb{\sigma_x}{1}+\slb{\sigma_x}{2}+\slb{\sigma_x}{3})
       + u_y(\slb{\sigma_y}{1}+\slb{\sigma_y}{2}+\slb{\sigma_y}{3})
       + u_z(\slb{\sigma_z}{1}+\slb{\sigma_z}{2}+\slb{\sigma_z}{3}))t/2}
\kets{\psi}{123}
\nonumber\\
 &=&
 e^{-i(u_x J_x+u_y J_y + u_z J_z)t}\kets{\psi}{123},
\end{eqnarray}
with
$J_u=\left(\slb{\sigma_u}{1}+\slb{\sigma_u}{2}+\slb{\sigma_u}{3}\right)/2$
for $u=x,y$ and $z$.  We can exhibit the error operators arising from a
uniform field in a compact form by defining $\vec J=(J_x,J_y,J_z)$ and
$\vec v = (u_x,u_y,u_z)t$.  Then the error operators are given by
$\err{E}(\vec v)= e^{-i\vec v\cdot\vec J}$, where the dot product in
the exponent is calculated like the standard vector dot product.

For a one-qubit noiseless subsystem, the
key property of the error model is that the errors are
symmetric under any permutation of the three particles. A
permutation of the particles acts on the particles' state space by
permuting the labels in the logical states.  For example, the
permutation $\pi$ that swaps the first two particles acts on logical
states as
\begin{equation}
\pi\kets{a}{1}\kets{b}{2}\kets{c}{3}=\kets{a}{2}\kets{b}{1}\kets{c}{3}=
\kets{b}{1}\kets{a}{2}\kets{c}{3}.
\end{equation}
To say that the errors are symmetric under particle permutations means
that each error $\err{E}$ satisfies $\pi^{-1}\err{E}\pi = \err{E}$, or
equivalently $\err{E}\pi=\pi\err{E}$ ($\err{E}$ ``commutes'' with
$\pi$).  To see that this condition is satisfied, write
\begin{eqnarray}
\pi^{-1}\err{E}(\vec v)\pi &=&
  \pi^{-1}e^{-i\vec v\cdot\vec J}\pi\nonumber\\
  &=&
  e^{-i\pi^{-1}(\vec v\cdot\vec J)\pi}\nonumber\\
  &=& e^{-i\vec v\cdot (\pi^{-1}\vec J\pi)}.
\end{eqnarray}
If $\pi$ permutes particle $\sysfnt{a}$ to $\sysfnt{b}$, then
$\pi^{-1}\slb{\sigma_u}{a}\pi = \slb{\sigma_u}{b}$.  It follows that
$\pi^{-1}\vec J\pi=\vec J$.  This expression shows that the errors commute with
the particle permutations and therefore cannot distinguish between the
particles. An error model satisfying this property is called
a ``collective'' error model.

If a noiseless subsystem exists, then it suffices to learn the
symmetries of the error model to construct the subsystem.  This
procedure is explained in Sect.~\ref{sec:symm}.  For the three
spin-${1\over 2}$ system, the procedure results in a one-qubit
noiseless subsystem protected from all collective errors.  We first
exhibit the subsystem identification and then discuss its properties
to explain why it is noiseless. As in the case of the seven-state
cyclic system, the identification involves a proper subspace of the
physical system's state space. The subsystem identification
involves a four-dimensional subspace and is defined by the following
correspondence:
\begin{equation}
\begin{array}{rcrcrcl}
{1\over \sqrt{3}}\Big(\kets{\phys{\downarrow}}{1}\kets{\phys{\uparrow}}{2}\kets{\phys{\uparrow}}{3}
  &+&
 e^{-i2\pi/3}
 \kets{\phys{\uparrow}}{1}\kets{\phys{\downarrow}}{2}\kets{\phys{\uparrow}}{3}
  &+&
 e^{i2\pi/3}
 \kets{\phys{\uparrow}}{1}\kets{\phys{\uparrow}}{2}\kets{\phys{\downarrow}}{3}\Big)
&\leftrightarrow& \ket{\phys{\uparrow}}\cdot\ket{\bitzero}
\\
{1\over \sqrt{3}}\Big(\kets{\phys{\downarrow}}{1}\kets{\phys{\uparrow}}{2}\kets{\phys{\uparrow}}{3}
  &+&
 e^{i2\pi/3}
 \kets{\phys{\uparrow}}{1}\kets{\phys{\downarrow}}{2}\kets{\phys{\uparrow}}{3}
  &+&
 e^{-i2\pi/3}
 \kets{\phys{\uparrow}}{1}\kets{\phys{\uparrow}}{2}\kets{\phys{\downarrow}}{3}\Big)
&\leftrightarrow& \ket{\phys{\uparrow}}\cdot\ket{\bitone}
\\
-{1\over \sqrt{3}}\Big(\kets{\phys{\uparrow}}{1}\kets{\phys{\downarrow}}{2}\kets{\phys{\downarrow}}{3}
  &+&
 e^{-i2\pi/3}
 \kets{\phys{\downarrow}}{1}\kets{\phys{\uparrow}}{2}\kets{\phys{\downarrow}}{3}
  &+&
 e^{i2\pi/3}
 \kets{\phys{\downarrow}}{1}\kets{\phys{\downarrow}}{2}\kets{\phys{\uparrow}}{3}\Big)
&\leftrightarrow& \ket{\phys{\downarrow}}\cdot\ket{\bitzero}
\\
-{1\over \sqrt{3}}\Big(\kets{\phys{\uparrow}}{1}\kets{\phys{\downarrow}}{2}\kets{\phys{\downarrow}}{3}
  &+&
 e^{i2\pi/3}
 \kets{\phys{\downarrow}}{1}\kets{\phys{\uparrow}}{2}\kets{\phys{\downarrow}}{3}
  &+&
 e^{-i2\pi/3}
 \kets{\phys{\downarrow}}{1}\kets{\phys{\downarrow}}{2}\kets{\phys{\uparrow}}{3}\Big)
&\leftrightarrow& \ket{\phys{\downarrow}}\cdot\ket{\bitone}
\end{array}
\label{eq:3spinid}
\end{equation}
The state labels for the syndrome subsystem (before the dot in the
expressions on the right side) identify it as a spin-${1\over 2}$
subsystem. In particular, it responds to the errors caused by uniform
fields in the same way as the physical spin-${1\over 2}$ particles.
This behavior is caused by $2J_u$ acting as the $u$-Pauli operator on
the syndrome subsystem. To confirm this property, we apply $2J_u$ to
the logical states of Eq.~\ref{eq:3spinid} for $u=z,x$.  The property
for $u=y$ then follows because $i\sigma_y=\sigma_z\sigma_x$.  Consider
$2J_z$. Each of the four states shown in Eq.~\ref{eq:3spinid} is an
eigenstate of $2J_z$.  For example, the physical state for
$\ket{\phys{\uparrow}}\cdot\ket{\bitzero}$ is a superposition of
states with two spins up ($\phys{\uparrow}$) and one spin down
($\phys{\downarrow}$). The eigenvalue of such a state with respect to
$2J_z$ is the difference $\Delta$ between the number of spins that are
up and down.  Thus,
$2J_z\ket{\phys{\uparrow}}\cdot\ket{\bitzero}=\ket{\phys{\uparrow}}\cdot\ket{\bitzero}$. The
difference is also $\Delta=1$ for
$\ket{\phys{\uparrow}}\cdot\ket{\bitone}$ and $\Delta=-1$ for
$\ket{\phys{\downarrow}}\cdot\ket{\bitzero}$ and
$\ket{\phys{\downarrow}}\cdot\ket{\bitone}$. Therefore, $2J_z$
acts as the $z$-Pauli operator on the syndrome subsystem.  To confirm
this behavior for $2J_x$, we compute
$2J_x\ket{\phys{\uparrow}}\cdot\ket{\bitzero}$.
\begin{equation}
\begin{array}[b]{rcl}
2J_x\ket{\phys{\uparrow}}\cdot\ket{\bitzero}
  &=& 
2J_x{1\over\sqrt{3}}\Big(\kets{\phys{\downarrow}}{1}\kets{\phys{\uparrow}}{2}\kets{\phys{\uparrow}}{3}
  +
 e^{-i2\pi/3}
 \kets{\phys{\uparrow}}{1}\kets{\phys{\downarrow}}{2}\kets{\phys{\uparrow}}{3}
  +
 e^{i2\pi/3}
 \kets{\phys{\uparrow}}{1}\kets{\phys{\uparrow}}{2}\kets{\phys{\downarrow}}{3}\Big) \\
  &=&
  \begin{array}[t]{c@{}r@{}r@{}r@{}r}
  &&{1\over\sqrt{3}}&\Big(\slb{\sigma_x}{1}+\slb{\sigma_x}{2}+\slb{\sigma_x}{3}\Big)&
  \kets{\phys{\downarrow}}{1}\kets{\phys{\uparrow}}{2}\kets{\phys{\uparrow}}{3}
       \\
  +&e^{-i2\pi/3}&{1\over\sqrt{3}}&\Big(\slb{\sigma_x}{1}+\slb{\sigma_x}{2}+\slb{\sigma_x}{3}\Big)
  &\kets{\phys{\uparrow}}{1}\kets{\phys{\downarrow}}{2}\kets{\phys{\uparrow}}{3}
       \\
  +&e^{i2\pi/3}&{1\over\sqrt{3}}&\Big(\slb{\sigma_x}{1}+\slb{\sigma_x}{2}+\slb{\sigma_x}{3}\Big)&
  \kets{\phys{\uparrow}}{1}\kets{\phys{\uparrow}}{2}\kets{\phys{\downarrow}}{3}
  \end{array} \\
  &=&
  \begin{array}[t]{c@{}r@{}r@{}rcrcr}
  &&{1\over\sqrt{3}}&\Big(
  \kets{\phys{\uparrow}}{1}\kets{\phys{\uparrow}}{2}\kets{\phys{\uparrow}}{3}
  &+&
  \kets{\phys{\downarrow}}{1}\kets{\phys{\downarrow}}{2}\kets{\phys{\uparrow}}{3}
  &+&
  \kets{\phys{\downarrow}}{1}\kets{\phys{\uparrow}}{2}\kets{\phys{\downarrow}}{3}\Big)
       \\
  +&e^{-i2\pi/3}&{1\over\sqrt{3}}&\Big(
  \kets{\phys{\downarrow}}{1}\kets{\phys{\downarrow}}{2}\kets{\phys{\uparrow}}{3}
  &+&
  \kets{\phys{\uparrow}}{1}\kets{\phys{\uparrow}}{2}\kets{\phys{\uparrow}}{3}
  &+&
  \kets{\phys{\uparrow}}{1}\kets{\phys{\downarrow}}{2}\kets{\phys{\downarrow}}{3}
  \Big)
       \\
  +&e^{i2\pi/3}&{1\over\sqrt{3}}&\Big(
  \kets{\phys{\downarrow}}{1}\kets{\phys{\uparrow}}{2}\kets{\phys{\downarrow}}{3}
  &+&
  \kets{\phys{\uparrow}}{1}\kets{\phys{\downarrow}}{2}\kets{\phys{\downarrow}}{3}
  &+&
  \kets{\phys{\uparrow}}{1}\kets{\phys{\uparrow}}{2}\kets{\phys{\uparrow}}{3}
  \Big)
  \end{array} 
  \\
  &=&
  \begin{array}[t]{c@{}r@{}rcrcr@{}r}
  &{1\over \sqrt{3}}&\Big(1&+&e^{-i2\pi/3}&+&e^{i2\pi/3}\Big)&\kets{\phys{\uparrow}}{1}\kets{\phys{\uparrow}}{2}\kets{\phys{\uparrow}}{3} \\
  +&{1\over \sqrt{3}}&\Big(&&e^{-i2\pi/3}&+&e^{i2\pi/3}\Big)&\kets{\phys{\uparrow}}{1}\kets{\phys{\downarrow}}{2}\kets{\phys{\downarrow}}{3} \\
  +&{1\over \sqrt{3}}&\Big(1&+&&&e^{i2\pi/3}\Big)&\kets{\phys{\downarrow}}{1}\kets{\phys{\uparrow}}{2}\kets{\phys{\downarrow}}{3} \\
  +&{1\over \sqrt{3}}&\Big(1&+&e^{-i2\pi/3}&&\Big)&\kets{\phys{\downarrow}}{1}\kets{\phys{\downarrow}}{2}\kets{\phys{\uparrow}}{3} \\
  \end{array}
  \\
  &=&
-{1\over \sqrt{3}}\Big(\kets{\phys{\uparrow}}{1}\kets{\phys{\downarrow}}{2}\kets{\phys{\downarrow}}{3}
  +
 e^{-i2\pi/3}
 \kets{\phys{\downarrow}}{1}\kets{\phys{\uparrow}}{2}\kets{\phys{\downarrow}}{3}
  +
 e^{i2\pi/3}
 \kets{\phys{\downarrow}}{1}\kets{\phys{\downarrow}}{2}\kets{\phys{\uparrow}}{3}\Big)\\
  &=& \ket{\phys{\downarrow}}\cdot\ket{\bitzero}.
\end{array}
\end{equation}
Similarly, one can check that, for the other logical states, the effect
of $2J_x$ is to flip the orientation of the syndrome
spin. 
The fact that the subsystem identified in Eq.~\ref{eq:3spinid} is
noiseless now follows from the fact that the errors $\err{E}(\vec v)$
are exponentials of sums of the syndrome spin operators
$J_u$. The errors therefore act as the identity on the
information-carrying subsystem.

The noiseless qubit supported by three spin-$1\over 2$ particles with
collective errors is another example in which the subsystem
identification does not involve the whole state space of the system.
In this case, the errors of the error model cannot remove amplitude
from the subspace.  As a result, if we detect an error, that is, if we
find that the system's state is in the orthogonal complement of the
subspace of the subsystem identification, we can deduce that either
the error model is inadequate, or we introduced errors in the
manipulations required for transferring information to the noiseless
qubit.

The noiseless subsystem of three spin-${1\over 2}$ particles can be
physically motivated by an analysis of quantum spin numbers.
The physical motivation is outlined in Fig.~\ref{fig:3spin1/2}.

\begin{herefig}
\begin{picture}(7,4.5)(-3.5,-4.3)
\nputgr{0,0}{t}{width=7in}{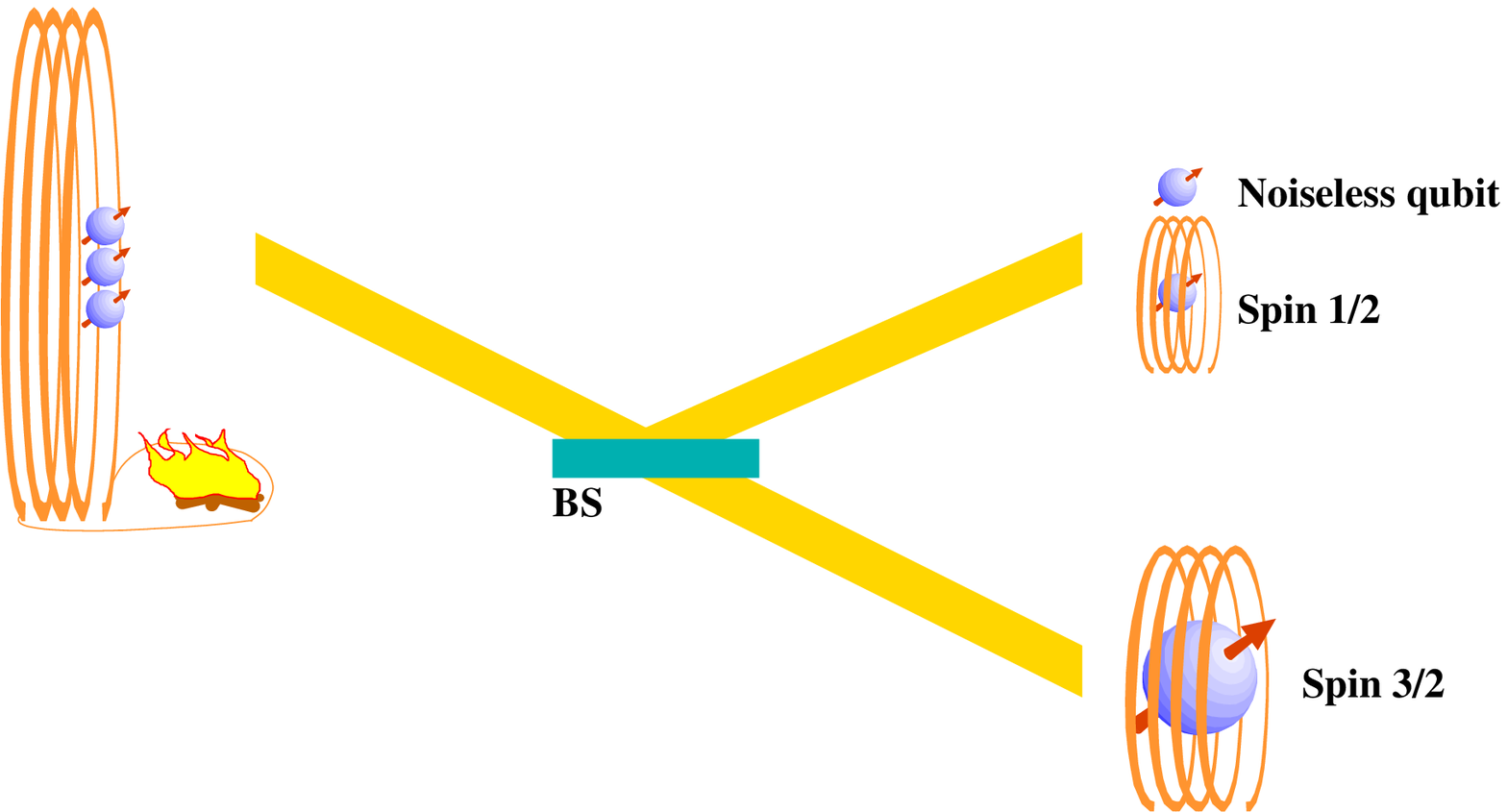}
\end{picture}
\label{fig:3spin1/2}
\herefigcap{
One noiseless qubit from three spin-$1\over 2$ particles.  The
left side shows the three particles, with errors caused by
fluctuations in a uniform magnetic field depicted by a noisy coil.
The spin along direction $u$ ($u=x,y,z$) can be measured and its
expectation is given by $\bra{\psi}J_u\ket{\psi}$, where $\ket{\psi}$
is the quantum state of the particles and $J_u$ is the total spin
observable along the $u$-axis given by the half-sum of the $u$-Pauli
matrices of the particles as defined in the text. The squared
magnitude of the total spin is given by the expectation of the
observable $J^2=\vec J \cdot \vec J = J_x^2+ J_y^2+J_z^2$.  The
observable $J^2$ commutes with the $J_u$ and therefore also with the
errors $\err{E}(\vec v)=e^{-i\vec v \cdot \vec J}$ caused by uniform
field fluctuations.  This can be verified directly, or one can note
that $\err{E}(\vec v)$ acts on $\vec J$ as a rotation in three
dimensions, and as one would expect, such rotations preserve the
``squared length'' $J^2$ of $\vec J$.  It now follows that the
eigenspaces of $J^2$ are invariant under the errors, and therefore
that the eigenspaces are good places to look for noiseless
subsystems. The eigenvalues of $J^2$ are of the form $j(j+1)$, where
$j$ is the spin quantum number of the corresponding eigenspace.  There
are two eigenspaces, one with spin $j={1\over 2}$ and the other with
spin $j={3\over 2}$.  The figure shows a thought experiment that
involves passing the three-particle system through a type of beam
splitter (BS) or Stern-Gerlach apparatus sensitive to $J^2$.  Using
such a beam splitter, the system of particles can be made to go in one of
two directions depending on $j$. In the figure, if the system's state
is in the the spin-$3\over 2$ subspace, it passes through the beam
splitter; if it is in the spin-$1\over 2$ subspace, the system is
reflected up. It can be shown that the subspace with $j={3\over 2}$ is
four-dimensional and spanned by the states that are symmetric under
particle permutations.  Unfortunately, there is no noiseless subsystem
in this subspace (see Sect.~\ref{sec:symm}). The spin-$1\over 2$
subspace is also four dimensional and spanned by the states in
Eq.~\ref{eq:3spinid}.  The spin-$1\over 2$ property of the subspace
implies that the spin operators $J_u$ act in a way that is
algebraically identical to the way $\sigma_u/2$ acts on a single
spin-$1\over 2$ particle. This property implies the existence of the
syndrome subsystem introduced in the text.  Conventionally, the
spin-$1\over 2$ subspace is thought of as consisting of two orthogonal
two-dimensional subspaces each behaving like a spin-$1\over 2$ with
respect to the $J_u$.  This choice of subspaces is not unique, but by
associating them with two logical states of a noiseless qubit, one can
obtain the subsystem identification of Eq.\ref{eq:3spinid}.  Some care
needs to be taken to ensure that the noiseless qubit operators commute
with the $J_u$, as they should (see Sect.~\ref{sec:symm}). In the
thought experiment, one can imagine unitarily
rotating the system emerging in the upper path to make explicit the
syndrome spin-$1\over 2$ subsystem and the noiseless qubit with which
it must be paired. The result of this rotation is shown.}
\end{herefig}

\pagebreak

\section{Error Models}
\label{sec:error_models}

We have seen several models of physical systems and errors in the
examples of the previous sections.  Most physical systems under
consideration for QIP consist of particles or degrees of freedom that
are spatially localized, a feature reflected in the error models
that are usually investigated. Because we also expect the physically
realized qubits to be localized, the standard error models deal with
quantum errors that act independently on different qubits. Logically
realized qubits, such as those implemented by subsystems
different from the physically obvious ones, may have more complicated
residual error behaviors.

\subsection{The Standard Error Models for Qubits}

The most investigated error model for qubits consists of ``independent,
depolarizing errors''.  This model has the effect of
completely depolarizing each qubit independently with probability $p$
(see Eq.~\ref{eq:cdep}).  For one qubit, the model is the least biased
in the sense that it is symmetric under rotations. As a result, every
state of the qubit is equally affected. Independent depolarizing
errors are considered to be the quantum analogue of the classical
independent bit flip error model.

Depolarizing errors are not typical for physically realized
qubits. However, given the ability to control individual qubits, it is
possible to enforce the depolarizing model (see below).
Consequently, error-correction methods designed to control
depolarizing errors apply to all independent error models.
Nevertheless, it is worth keeping in mind that given detailed
knowledge of the physical errors, a special purpose method is usually
better than one designed for depolarizing errors.
We therefore begin by showing how one can think about arbitrary
error models.

There are several different ways of describing errors affecting a
physical system $\sysfnt{S}$ of interest.  For most situations, in
particular if the initial state of $\sysfnt{S}$ is pure, errors can be
thought of as being the result of coupling to an initially independent
environment for some time. Because of this coupling, the effect of
error can always be represented by the process of adjoining an
environment $\sysfnt{E}$ in some initial state $\kets{0}{E}$ to the
arbitrary state $\kets{\psi}{S}$ of $\sysfnt{S}$, followed by a
unitary coupling evolution $\slb{U}{E\,S}$ acting jointly on
$\sysfnt{E}$ and $\sysfnt{S}$. Symbolically, the process can be
written as the map
\begin{equation}
\kets{\psi}{S}\; \err{\rightarrow}\; \slb{U}{E\,S}\kets{0}{E}\kets{\psi}{S}.
\end{equation}
Choosing an arbitrary orthonormal basis consisting of the states
$\kets{e}{E}$ for the state space of the environment, the process can be
rewritten in the form:
\begin{eqnarray}
\kets{\psi}{S} &\err{\rightarrow} &
  \slb{\idop}{E} \slb{U}{E\,S}\kets{0}{E}\kets{\psi}{S}\nonumber\\
  &=&
  \left(\sum_e\kets{e}{E}\bras{e}{E}\right)\slb{U}{E\,S}\kets{0}{E}\kets{\psi}{S} \nonumber\\ 
   &=&
  \sum_e\kets{e}{E}\Big(\bras{e}{E}\slb{U}{E\,S}\kets{0}{E}\Big)\kets{\psi}{S} \nonumber\\
    &=& \sum_e \kets{e}{E}\slb{A_e}{S}\kets{\psi}{S},
\label{eq:elsum}
\end{eqnarray}
where the last step defines operators $\slb{A_e}{S}$ acting on $\sysfnt{S}$
by $\slb{A_e}{S}=\bras{e}{E}\slb{U}{E\,S}\kets{0}{E}$.  The expression
$\sum_e\kets{e}{E}\slb{A_e}{S}$ is called an ``environment labeled
operator''. The unitarity condition implies that $\sum_e
A_e^\dagger A_e = \idop$ (with system labels omitted).  The environment basis
$\kets{e}{E}$ need not
represent any physically meaningful choice of basis of a real
environment. For the purpose of error analysis, the states $\kets{e}{E}$
are formal states that ``label'' the error operators $A_e$.  One can
use an expression of the form shown in Eq.~\ref{eq:elsum} even when
the $\ket{e}$ are not normalized or orthogonal, keeping in mind
that as a result, the identity implied by the unitarity condition changes.

Note that the state on the right side of Eq.~\ref{eq:elsum}
representing the effect of the errors is correlated with the
environment. This means that after removing (or ``tracing over'') the
environment, the state of $\sysfnt{S}$ is usually mixed.  Instead of
introducing an artificial environment, we can also describe the errors
by using the density operator formalism for mixed states.  Define
$\rho=\ketbras{\psi}{\psi}{S}$. The effect of the errors on the
density matrix $\rho$ is given by the transformation
\begin{equation}
\rho\;\err{\rightarrow}\; \sum_e A_e \rho A_e^\dagger.
\end{equation}
This is the ``operator sum'' formalism~\cite{kraus:qc1983a}.

The two ways of writing the effects of errors can be applied
to the depolarizing-error model for one qubit.
As an environment-labeled operator, depolarization with probability
$p$ can be written as
\begin{equation}
  \sqrt{1-p}\kets{0}{E}\idop
   + {\sqrt{p}\over 2}\Big(
       \kets{1}{E}\idop +
       \kets{x}{E}\sigma_x+
       \kets{y}{E}\sigma_y+
       \kets{z}{E}\sigma_z
     \Big),
\label{eq:depmodel}
\end{equation}
where we introduced five abstract, orthonormal environment states to
label the different events. In this case, one can think of the model
as applying no error with probability $1-p$, or completely
depolarizing the qubit with probability $p$. The latter event is
represented by applying one of $\idop,\sigma_x,\sigma_y$ or $\sigma_z$
with equal probability $p/4$. To be able to think of the model as
randomly applied Pauli matrices, it is crucial that the environment
states labeling the different Pauli matrices be orthogonal. The square
roots of the probabilities appear in the operator because in an
environment-labeled operator, it is necessary to give quantum
amplitudes.  Environment labeled operators are useful primarily
because of their great flexibility and redundancy.

In the operator sum formalism, depolarization with probability $p$
transforms the input density matrix $\rho$ as
\begin{eqnarray}
  \rho &\rightarrow&
     (1-p)\rho +
      {p\over 4}\left(
       \idop\rho\idop +
       \sigma_x\rho\sigma_x +
       \sigma_y\rho\sigma_y +
       \sigma_z\rho\sigma_z
      \right)\nonumber\\ 
    &=&
     (1-3p/4)\rho +
      {p\over 4}\left(
       \sigma_x\rho\sigma_x +
       \sigma_y\rho\sigma_y +
       \sigma_z\rho\sigma_z
      \right).
\end{eqnarray}
Because the operator sum formalism has less redundancy, it is
easier to tell when two error effects are equivalent.

In the remainder of this section, we discuss how one can use active
intervention to simplify the error model. To realize this
simplification, we intentionally randomize the qubit so that the
environment cannot distinguish between the different ``axes'' defined
by the Pauli spin matrices. Here is a simple randomization that
actively converts an arbitrary error model for a qubit into one that
consists of randomly applying Pauli operators according to some
distribution.  The distribution is not necessarily uniform so the new
error model is not yet depolarizing.  Before the errors act, apply a
random Pauli operator $\sigma_u$ ($u=0,x,y,z$, $\sigma_0=\idop$).
After the errors act apply the inverse of that operator,
$\sigma_u^{-1} = \sigma_u$; then ``forget'' which operator was
applied.  This randomization method is called
``twirling''~\cite{bennett:qc1996a}.  To understand twirling, we use
environment labeled operators to demonstrate some of the techniques
useful in this context.  The sequence of actions implementing twirling
can be written as follows (omitting labels for $\sysfnt{S}$):
\begin{equation}
 \begin{array}[b]{rcll}
 \ket{\psi} &\rightarrow& 
      {1\over 2}\sum_u\kets{u}{C}\sigma_u\ket{\psi}
      & \mbox{apply a random $\sigma_{u}$, remembering $u$ with the help of the system $\sysfnt{C}$.}
      \\
  &\err{\rightarrow}&
      \sum_e \kets{e}{E}{1\over 2}\sum_u\kets{u}{C}A_e\sigma_u\ket{\psi}
      &\mbox{errors act.}\\
  &\rightarrow &
      \sum_{e} \kets{e}{E}{1\over 2}\sum_u\kets{u}{C}\sigma_u A_e\sigma_u\ket{\psi}
      &\mbox{apply $\sigma_u = \sigma_u^{-1}$.}\\
  &\rightarrow &
      \sum_{eu} \kets{eu}{EC}{1\over 2}\sigma_u A_e\sigma_u\ket{\psi}
      &\mbox{forget which $u$ was used by absorbing its memory in $\sysfnt{E}$.}
  \end{array}
\label{eq:prand}
\end{equation}
The system $\sysfnt{C}$ that was artificially introduced
to carry the memory of $u$ may be a classical memory because there
is no need for coherence between different $\kets{u}{C}$.

To determine the equivalent random Pauli operator error model, it is
necessary to rewrite the total effect of the procedure using an
environment labeled sum involving orthogonal environment states and
Pauli operators.  To do so, express $A_e$  as a sum of the Pauli
operators, $A_e=\sum_v\alpha_{ev}\sigma_v$, using 
the fact that the $\sigma_v$ are a linear basis for the space of one-qubit
operators. Recall the fact that $\sigma_u$ anticommutes with
$\sigma_v$ if $0\not=u\not=v\not=0$. Thus
$\sigma_u\sigma_v\sigma_u=(-1)^{\langle v,u\rangle}\sigma_v$, where
$\langle v,u\rangle = 1$ if $0\not=u\not=v\not=0$ and $\langle
v,u\rangle = 0$ otherwise.  We can now rewrite the last expression of
Eq.~\ref{eq:prand} as follows:
\begin{eqnarray}
\sum_{eu} \kets{eu}{EC}{1\over 2}\sigma_u A_e\sigma_u\ket{\psi}
  &=&
\sum_{eu} \kets{eu}{EC}{1\over 2}\sigma_u \sum_v\alpha_{ev}\sigma_v\sigma_u\ket{\psi}\nonumber\\
  &=&
\sum_v\left(\sum_{eu}{1\over 2}\alpha_{ev}(-1)^{\langle v,u\rangle}\kets{eu}{EC}\right) \sigma_v\ket{\psi}.
\label{eq:tildevnow}
\end{eqnarray}
It can be checked that the states ${1\over 2}\sum_{u}(-1)^{\langle v,u\rangle}\kets{eu}{EC}$ are orthonormal for different $e$ and $v$.
As a result the
states $\sum_{eu}{1\over 2}\alpha_{ev}(-1)^{\langle v,u\rangle}\kets{eu}{EC}$
are orthogonal for different $v$ and have 
probability (square norm) given by
$p_v=\sum_e|\alpha_{ev}|^2$. Introducing
$\sqrt{p_v}\kets{\tilde v}{EC}=\sum_{eu}{1\over 2}\alpha_{ev}(-1)^{\langle v,u\rangle}\kets{eu}{EC}$, we can write the sum of Eq.~\ref{eq:tildevnow}
as
\begin{equation}
\sum_v\left(\sum_{eu}{1\over 2}\alpha_{ev}(-1)^{\langle v,u\rangle}\kets{eu}{EC}\right) \sigma_v\ket{\psi} = \sum_v\sqrt{p_v}\kets{\tilde v}{EC}\sigma_v\ket{\psi},
\end{equation}
showing that the twirled error model behaves like randomly applied Pauli
matrices with $\sigma_v$ applied with probability $p_v$.  It is a
recommended exercise to reproduce the above argument using the operator sum
formalism.

To obtain the standard depolarizing error model with equal
probabilities for the Pauli matrices, it is necessary to strengthen
the randomization procedure by applying a random member $U$ of the
group generated by the $90^\circ$ rotations around the $x$, $y$ and
$z$ axes before the error and then undoing $U$ by applying $U^{-1}$.

Randomization can be used to transform any one-qubit error model
into the depolarizing error model. This explains why the depolarizing
model is so useful for analyzing error correction techniques
in situations in which errors act independently on different qubits.
However, in many physical situations, the independence
assumptions are not satisfied. For example, errors from common
internal couplings between qubits are generally pairwise correlated to
first order. In addition, the operations required to manipulate the
qubits and to control the encoded information act on pairs at a time,
which tends to spread even single qubit errors.  Still, in all these
cases, the primary error processes are local.  This means that there
usually exists an environment labeled sum expression for the total
error process in which the amplitudes associated with errors acting
simultaneously at $k$ locations in time and space decrease
exponentially with $k$. In such cases, error-correction methods that
handle all or most errors involving sufficiently few qubits are still
applicable.

\subsection{Quantum Error Analysis}
\label{sec:noise_analysis}

One of the most important consequences of the subsystems
interpretation of encoding quantum information in a physical system is
that the encoded quantum information can be error-free even though
errors have severely changed the state of the physical system.  Almost
trivially, any error operator acting only on the syndrome subsystem
has no effect on the quantum information. The goal of error correction
is to actively intervene and maintain the syndrome subsystem in states
where the dominant error operators continue to have little effect on
the information of interest.  An important issue in analyzing error
correction methods is to estimate the residual error in the encoded
information.  A simple example of how that can be done was discussed
for the quantum repetition code.  The same ideas can be applied in
general.  Let $\sysfnt{S}$ be the physical system in which the
information is encoded and $\kets{\psi}{S}$ an initial state
containing such information with the syndrome subsystem appropriately
prepared.  Errors and error-correcting operations modify the
state. The new state can be expressed using environment labeling as
$\sum_e\kets{e}{E}\slb{A_e}{S}\kets{\psi}{S}$.  In view of the
partitioning into information-carrying and syndrome subsystems,
``good'' states $\kets{e}{E}$ are those states for which
$\slb{A_e}{S}$ acts only on the syndrome subsystem given that the
syndrome has been prepared.  The remaining states $\kets{e}{E}$ form the set of
``bad'' states, $\err{\cB}$.  The error probability $p_e$ can be bounded
from above by
\begin{eqnarray}
p_e &\leq &
  \left|\sum_{e\in\err{\cB}}\kets{e}{E}\slb{A_e}{S}\kets{\psi}{S}\right|^2\nonumber\\
  &\leq&
  \left(\sum_{e\in\err{\cB}}|\kets{e}{E}|\; |\slb{A_e}{S}|_1\right)^2,
\label{eq:grossbnd}
\end{eqnarray}
where $|A|_1 = \max_{\phi}\bra{\phi}A\ket{\phi}$, the maximum being taken
over normalized states. The second inequality usually leads to a gross
overestimate but is independent of the encoded information and often
suffices for obtaining good results.  Because the environment-labeled
sum is not unique, a goal of the representation of the errors acting
on the system is to use ``good'' operators to the largest extent
possible. The flexibility of these error-expansions makes them very
useful for analyzing error models in conjunction with error-correction
methods. 

In principle, we can obtain better expressions for $p_e$ by
calculating the density matrix $\rho$ of the state of the subsystem
containing the desired quantum information.  This calculation involves
``tracing over'' the syndrome subsystem. The matrix $\rho$ can then be
compared to the intended state. If the intended state is pure, given
by $\ket{\phi}$, the probability of error is given by
$1-\bra{\phi}\rho\ket{\phi}$, which is the probability that a
measurement that distinguishes between $\ket{\phi}$ and its orthogonal
complement fails to detect $\ket{\phi}$.  The quantity
$\bra{\phi}\rho\ket{\phi}$ is called the ``fidelity'' of the state
$\rho$.

For applications to communication, the goal is to be able to reliably
transmit arbitrary states through a communication channel, 
which may be physical or realized via an encoding/decoding scheme.
It is therefore important to characterize the reliability of the
channel independent of the information transmitted.
Eq.~\ref{eq:grossbnd} can be used to obtain state-independent bounds
on the error probability but does not readily provide a single measure
of reliability.  One way to quantify the reliability is to identify
the error of the channel with the average error $\epsilon_a$ over all
possible input states.  The reliability is then given by the average
fidelity $1-\epsilon_a$ Another elegant way appropriate for QIP
is to use the ``entanglement
fidelity''~\cite{schumacher:qc1996a}. Entanglement fidelity measures
the error when the input is maximally entangled with an identical
``reference'' system. In this process, the reference system is
imagined to be untouched, so that the state of the reference system
together with the output state can be compared to the original
entangled state.  For a one-qubit channel labeled $\sysfnt{S}$, the
reference system is a qubit, which we label with $\sysfnt{R}$.  An
initial, maximally entangled state is
\begin{equation}
\ket{B}={1\over\sqrt{2}}\left(\kets{\bitzero}{R}\kets{\bitzero}{S}+
  \kets{\bitone}{R}\kets{\bitone}{S}\right).
\end{equation}
The reference qubit is assumed to be perfectly isolated and not
affected by any errors.  The final state $\slb{\rho}{R\,S}$ is
compared to $\ket{B}$, which gives the entanglement fidelity according
to the formula $f_e=\bra{B}\slb{\rho}{R\,S}\ket{B}$.  The entanglement
error is $\epsilon_e=1-f_e$. It turns out that this definition does
not depend on the choice of maximally entangled state.  Fortunately,
the entanglement error and the average error $\epsilon_a$ are related
by a linear expression:
\begin{equation}
\epsilon_a= {2\over 3}\epsilon_e.
\end{equation}
For $k$-qubit channels, the constant ${2\over 3}$ is replaced by
$2^k/(2^k+1)$. Experimental measurements of these fidelities do not
require the reference system.  There are simple averaging formulas to
express them in terms of the fidelities for transmitting each of a
sufficiently large set of pure states. An example of the experimental
determination of the entanglement fidelity when the channel is
realized by error-correction is provided in~\cite{knill:qc2001a}.

\section{From Quantum Error Detection to Error Correction}
\label{sec:fromqedtoec}

In the independent depolarizing error model with small probability $p$
of depolarization, the most likely errors are those that affect a
small number of qubits.  That is, if we define the ``weight'' of a
product of Pauli operators to be the number of qubits affected, the
dominant errors are those of small weight. Because the probability of a
non-identity Pauli operator is $3p/4$ (see Eq.~\ref{eq:depmodel}), one
expects about ${3p\over 4} n$ of $n$ qubits to be changed.  As a
result, good error-correcting codes are considered to be those for
which all errors of weight $\leq e \simeq {3p\over 4}n$ can be
corrected.  It is desirable that $e$ have a high ``rate'', which means
that it is a large fraction of the total number of qubits, $n$ (the
``length'' of the code). Combinatorially, good codes are characterized
by a high minimum distance, a concept that arises naturally
in the context of error-detection.

\subsection{Quantum Error Detection}

Let $C$ be a quantum code, that is, a subspace of the state space of a
quantum system. Let $P$ be the operator that projects onto $C$, and
$P^{\perp}=\idop-P$ the one that projects onto the orthogonal
complement. Then the pair $P,P^{\perp}$ is associated with a
measurement that can be used to determine whether a state is in the
code or not. If the given state is $\ket{\psi}$, then the result of
the measurement is $P\ket{\psi}$ with probability $|P\ket{\psi}|^2$
and $P^\perp\ket{\psi}$ otherwise. As in the classical case, an
error-detection scheme consists of preparing the desired state
$\ket{\psi_i}\in C$, transmitting it through, say, a quantum channel,
then measuring whether the state is still in the code, accepting
the state if it is, and rejecting
it otherwise. We say that $C$ detects error operator
$\err{E}$ if states accepted after $\err{E}$ had acted are unchanged
except for an overall scale. Using the projection operators, this is
the statement that for every state $\ket{\psi_i}\in C$,
$P\err{E}\ket{\psi_i} = \lambda_{\err{E}}\ket{\psi_i}$.  Because $P\ket{\psi}$
is in the code for every $\ket{\psi}$, it follows that $P\err{E}
P\ket{\psi}=\lambda_{\err E} P\ket{\psi}$. Therefore, a
characterization of detectability is gven by:
\begin{equation}
\begin{eqthm}{Theorem}
$\err{E}$ is detectable by $C$ if and only if
$P\err{E} P =\lambda_{\err{E}} P$ for some $\lambda_{\err{E}}$.
\end{eqthm}
\end{equation}
It is not difficult to see that a second characterization is given by:
\begin{equation}
\begin{eqthm}{Theorem}
$\err{E}$ is detectable by $C$ if and only if for all
$\ket{\psi},\ket{\phi}\in C$, $\bra{\psi}\err{E}\ket{\phi}
=\lambda_{\err{E}}\braket{\psi}{\phi}$ for some $\lambda_{\err{E}}$.
\end{eqthm}
\end{equation}
A third characterization, which we state without proof, is obtained by
taking the condition for classical detectability in
Thm.~\ref{thm:classdetect} and replacing ``$\not=$'' by ``orthogonal
to'':
\begin{equation}
\begin{eqthm}{Theorem}
$\err{E}$ is detectable by $C$ if and only if for all
$\ket{\phi},\ket{\psi}$ in the code with $\ket{\phi}$ orthogonal to
$\ket{\psi}$, $\err{E}\ket{\phi}$ is orthogonal to $\ket{\psi}$.
\end{eqthm}
\end{equation}

For a given code $C$, the set of detectable errors is closed under
linear combinations. That is, if $\err{E}_1$ and $\err{E}_2$ are both
detectable, then so is $\alpha \err{E}_1+\alpha \err{E}_2$. This
useful property implies that to check detectability, one has to
consider only the elements of a linear basis for the space of errors of
interest.

Consider $n$ qubits with independent depolarizing errors.  A robust
error-detecting code should detect as many of the small weight errors
as possible.  This requirement motivates the definition of ``minimum
distance'': The code $C$ has minimum distance $d$ if the
smallest-weight product of Pauli operators $E$ for which $C$ does not
detect $E$ is $d$.  The notion comes from classical codes for bits,
where a set of code words $C'$ has minimum distance $d$ if the
smallest number of flips required to change one code word in $C'$ into
another one in $C'$ is $d$. For example, the repetition code for three
bits has minimum distance $3$. Note that the minimum distance for the
quantum repetition code is one: Applying $\slb{\sigma_z}{1}$ preserves
the code and changes the sign of $\ket{\bitone\bitone\bitone}$ but not
of $\ket{\bitzero\bitzero\bitzero}$. As a result, $\slb{\sigma_z}{1}$
is not detectable. The notion of minimum distance can be generalized
for error models with specified ``first order'' error
operators~\cite{knill:qc1999b}.  In the case of depolarizing errors,
the first order error operators are single qubit Pauli matrices, which
are the errors of weight one.

\subsection{Quantum Error Correction}

Let $\err{\cE}=\{\err{E}_0=\id,\err{E}_1,\ldots\}$ be the set of
errors that we wish to be able to correct.  When is there a decoding
procedure for the code $C$ such that all errors in $\err{\cE}$ are
corrected?  When such a decoding procedure exists, we say that
$\err{\cE}$ is ``correctable'' (by $C$).  A situation in which
correctability of $\err{\cE}$ is apparent occurs when the errors
$\err{E}_i$ are unitary operators satisfying the condition that
$\err{E}_i C$ are mutually orthogonal subspaces.  The repetition code
has this property for the set of errors consisting of the identity and
Pauli operators acting on a single qubit.  In this situation, the
procedure for decoding is to first make a projective measurement to
determine which of the subspaces $\err{E}_i C$ the state is in, and
then to apply the inverse of the error operator, $\err{E}_i^\dagger$.
This situation is not far from the generic one. One
characterization of correctability is in the following theorem:
\begin{equation}
\begin{eqthm}{Theorem}
$\err{\cE}$ is correctable if and only if there is a
linear transformation of the set $\err{\cE}$ such that the operators
$\err{E}'_i$ in the new set satisfy the following properties: (1)~The
$\err{E}'_iC$ are mutually orthogonal, and (2)~$\err{E}'_i$ restricted to
$C$ is proportional to a restriction to $C$ of a unitary operator.  
\end{eqthm}
\label{thm:ecbyorth}
\end{equation}
To relate this characterization to detectability, note that the two
properties imply that $(\err{E}'_i)^\dagger\err{E}'_j C$ is orthogonal
to $C$ if $i\not=j$, and $(\err{E}'_i)^\dagger \err{E}'_i$ restricted
to $C$ is proportional to the identity on $C$. In other words, the
$(\err{E}'_i)^\dagger\err{E}'_j$ are detectable. This detectability
condition applied to the original error set constitutes a second
characterization of correctability, given in the next theorem:
\begin{equation}
\begin{eqthm}{Theorem}
$\err{\cE}$ is
correctable if and only if the operators in the set
$\err{\cE}^\dagger\err{\cE} = \{\err{E}_1^\dagger\err{E}_2 :
\err{E}_i\in\err{\cE}\}$ are detectable.\end{eqthm}
\label{thm:ecbydet}
\end{equation}
Before explaining the characterizations of correctability, we consider
the situation of $n$ qubits, where the characterization by
detectability~(\ref{thm:ecbydet}) leads to a useful relationship
between minimum distance and correctability of low weight errors:
\begin{equation}
\begin{eqthm}{Theorem}
If a code on $n$ qubits has a minimum distance of at least $2e+1$, then the
set of errors of weight at most $e$ is correctable.
\end{eqthm}
\label{thm:mindist}
\end{equation}
This theorem follows by
observing that the weight of $\err{E}_1^\dagger \err{E}_2$ is at most
the sum of the weights of the $\err{E}_i$.  As a result of this
observation, the problem of finding good ways of correcting all errors
up to a maximum weight reduces to that of constructing codes with
sufficiently high minimum distance. Thus questions such as ``what is the
maximum dimension of a code of minimum distance $d$ on $n$ qubits?''
are of great interest. As in the case of classical coding theory this
problem appears to be very difficult in general.  Answers are known
for small $n$~\cite{calderbank:qc1996b} and there are asymptotic
bounds~\cite{ashikhmin:qc1999a}. Of course, for achieving low error
probabilities, it is not necessary to correct all errors of weight
$\leq e$, just almost all such errors. For example, the concatenated
codes used for fault-tolerant quantum computation achieve this goal
(see Sec.~\ref{sec:fault_tolerance}).

For the remainder of this section we explain the characterizations of
correctability.  Using the conditions for detectability from the
previous section, the condition for correctability in
Thm.~\ref{thm:ecbydet} is equivalent to
\begin{eqnarray}
P\err{E}_i^\dagger\err{E}_j P &=& \lambda_{ij} P
\label{eq:eccond}
\end{eqnarray}
This condition is preserved under a linear change of basis
for $\err{\cE}$. That is, if $A$ is any invertible
matrix with coefficients $a_{ij}$, we can
define new error operators $\err{D}_k=\sum_i  \err{E}_ia_{ik}$.
For the $\err{D}_k$, the left side of Eq.~\ref{eq:eccond} is
\begin{eqnarray}
P\err{D}_k^\dagger\err{D}_l P&=& 
  P\left(\sum_{ij} \bar a_{ik} \err{E}_i^\dagger \err{E}_ja_{jl}\right)  P
\nonumber\\
  &=& \sum_{ij}\bar a_{ik} a_{jl}\lambda_{ij}P 
\nonumber\\
  &=& \left(A^\dagger\Lambda A\right)_{kl} P,
\end{eqnarray}
where $\Lambda$ is the matrix formed from the $\lambda_{ij}$.  Using
the fact that $\Lambda$ is a positive semidefinite matrix (that is, for
all $x$, $x^\dagger \Lambda x \geq 0$ and $\Lambda^\dagger =
\Lambda$), we can choose $A$ such that $A^\dagger\Lambda A$ 
is of the form $\qaop{\idop}{0}{0}{0}$. In this matrix, the upper
left block is the identity operator for some dimension.

An important consequence of invariance under a change of basis of
error operators is that the set of errors correctable by a particular
code and decoding procedure is linearly closed.  Thus, if $\err{E}$
and $\err{D}$ are corrected by the decoding procedure, then so is
$\alpha\err{E}+\beta\err{D}$.  This observation also follows from the
linearity of quantum mechanically implementable operations.

We explain the condition for correctability by using the subsystems
interpretation of decoding procedures.  For simplicity, assume that
$\id\in\err{\cE}$.  To show that correctability of $\err{\cE}$ implies
detectability of all $\err{E}\in\err{\cE}^\dagger\err{\cE}$, suppose
that we have a decoding procedure that recovers the information
encoded in $C$ after any of the errors in $\err{\cE}$ have occurred.  Every
physically realizable decoding procedure can be implemented by first
adding ``ancilla'' quantum systems in a prepared pure state to form a
total system labeled $\sysfnt{T}$, then applying a unitary map $U$
to the state of $\sysfnt{T}$, and finally separating $\sysfnt{T}$
into a pair of systems $\sysfnt{S},\sysfnt{Q}$, where
$\sysfnt{S}$ corresponds to the syndrome subsystem, and $\sysfnt{Q}$ is a
quantum system with the same dimension as the code that carries the
quantum information after decoding.  Denote the state space of the
physical system containing $C$ as $\cH$, and the state space of system
$\sysfnt{X}$ by $\cH_\sysfnt{X}$, where $\sysfnt{X}$ is any one
of the other systems.  Let $V$ be the unitary operator that encodes
information by mapping $\cH_\sysfnt{Q}$ onto $C\subseteq\cH$. We have
the following relationships:
\begin{equation}
\cH_\sysfnt{Q} \stackrel{V}{\leftrightarrow} C\subseteq \cH \subseteq \cH_\sysfnt{T} \stackrel{U}{\leftrightarrow}
\cH_\sysfnt{S}\tensor\cH_\sysfnt{Q}.
\label{eq:speq}
\end{equation}
Here, we used bidirectional arrows ``$\leftrightarrow$'' to emphasize
that the operators $V$ and $U$ can be inverted on their range and
therefore identify the states in their domains with the states in
their ranges. The inclusion $\cH\subseteq\cH_{\sysfnt{T}}$ implicitly
identifies $\cH$ with the subspace determined by the prepared pure
state on the ancillas.  The last state space of Eq.~\ref{eq:speq} is
expressed as a tensor product (``$\tensor$''), which is the state
space of the combined system $\sysfnt{SQ}$.  For states of
$\cH_\sysfnt{Q}$ we write
$\ket{\psi}=\kets{\psi}{Q}\stackrel{V}{\leftrightarrow}\kets{\psi}{L}\in
C$.  Because $\id$ is a correctable error, it must be the case that
$\kets{\psi}{L} \stackrel{U}{\leftrightarrow}
\kets{0}{S}\ket{\psi}\in \cH_\sysfnt{S}\tensor\cH_\sysfnt{Q}$
for some state $\kets{0}{S}$ of the syndrome subsystem.  To establish
this fact, use linearity of the maps.  In general:
\begin{eqnarray}
\kets{\psi}{L} &\rightarrow& \err{E}_i\kets{\psi}{L}\nonumber \\
              &\stackrel{U}{\leftrightarrow}&
               \kets{i}{S}\ket{\psi}
\end{eqnarray}
The $\kets{i}{S}$ need not be normalized or orthogonal. Let $F$ be the
subspace spanned by the $\kets{i}{S}$.  Then $U$ induces an
identification of $F\tensor\cH_{Q}$ with a subspace $\bar C\subseteq
\cH$. This is the desired subsystem identification.
We can then see how the errors act in this identification:
\begin{equation}
\begin{array}{ccc}
   \kets{\psi}{L} &\leftrightarrow& \kets{0}{S}\ket{\psi} \\
          \downarrow        && \\
   \err{E}_{i}\kets{\psi}{L} &\leftrightarrow&
                    \kets{i}{S}\ket{\psi}
\end{array}
\end{equation}
This means that for all $\ket{\psi}$ and $\ket{\phi}$,
\begin{equation}
\bras{\psi}{L}\err{E}_j^\dagger\err{E}_i\kets{\phi}{L}
 = \brakets{j}{i}{S}\braket{\psi}{\phi},
\end{equation}
that is, all errors in $\err{\cE}^\dagger\err{\cE}$ are detectable.

Now, suppose that all errors in $\err{\cE}^\dagger\err{\cE}$ are
detectable. To see that this implies correctability of $\err{\cE}$,
choose a basis for the errors so that $\lambda_{ij} = \delta_{ij}
\lambda_i$ with $\lambda_i=1$ for $i<s$ and $\lambda_i=0$ otherwise.
Define a subsystem identification by
\begin{equation}
\kets{i}{s}\ket{\psi}\stackrel{W}{\rightarrow}
    \err{E}_i\kets{\psi}{L},
\end{equation}
for $0\leq i<s$. By assumption and construction,
$\bras{\psi}{L}\err{E}_j^\dagger\err{E}_i\kets{\psi}{L} =
\delta_{ij}$, which implies that $W$ is unitary (after linear
extension), and so this is a proper identification. For $i\geq s$,
$\err{E}_i\kets{\psi}{L} = 0$, which implies that for states in the
code, these errors have probability $0$.  Therefore, the identification
can be used to successfully correct $\err{\cE}$. 

\section{Constructing Codes}
\label{sec:constructing}

\subsection{Stabilizer Codes}

Most useful quantum codes are based on ``stabilizer''
constructions~\cite{gottesman:qc1996a,calderbank:qc1996a}.  Stabilizer
codes are useful because they make it easy to determine which Pauli-product
errors are detectable and because they can be interpreted as special
types of classical ``linear'' codes. The latter feature makes it possible to
use well-established techniques from the theory of classical
error-correcting codes to construct good quantum codes.

A stabilizer code of length $n$ for $k$ qubits (abbreviated as an
``$[[n,k]]$ code''), is a $2^k$-dimensional subspace of the state
space of $n$ qubits that is characterized by the set of products of
Pauli operators that leave each state in the code invariant.  Such
Pauli operators are said to ``stabilize'' the code.  A simple example
of a stabilizer code is the quantum repetition code introduced in
Sec.~\ref{sec:qrc}. The code's states
$\alpha\ket{\phys{\bitzero\bitzero\bitzero}}+\beta\ket{\phys{\bitone\bitone\bitone}}$
are exactly the states that are unchanged after applying
$\slb{\sigma_z}{1}\slb{\sigma_z}{2}$ or
$\slb{\sigma_z}{1}\slb{\sigma_z}{3}$.

To simplify the notation, we write $I=\idop, X=\sigma_x, Y=\sigma_y$, and
$Z=\sigma_z$.  A product of Pauli operators can then be written as
$ZIXI=\slb{\sigma_z}{1}\slb{\sigma_x}{3}$ (as an example of length
$4$) with the ordering determining which qubit is being acted upon by
the operators in the product.

We can understand the properties of stabilizer codes by working out
the example of the quantum repetition code with the stabilizer formalism. A
stabilizer of the code is $S=\{ZZI,ZIZ\}$. Let $\bar S$ be the set of
Pauli products that are expressible up to a phase as products of
elements of $S$. For the repetition code, $\bar S=\{III, ZZI, ZIZ,
IZZ\}$.  $\bar S$ consists of all Pauli products that stabilize the
code.  The crucial property of $S$ is that its operators commute, that
is, for $A,B\in S$, $AB=BA$.  According to results from linear
algebra, it follows that the state space $\cH$ can be decomposed
into orthogonal subspaces $\cH_\lambda$ such that for $A\in S$ and
$\ket{\psi}\in\cH_\lambda$, $A\ket{\psi} =
\lambda(A)\ket{\psi}$. The $\cH_\lambda$ are the common eigenspaces of
$S$. The stabilizer code $C$ defined by $S$ is the subspace
stabilized by the operators in $S$, which means that it is given by
$\cH_\lambda$ with $\lambda(A)=1$. The subspaces for other
$\lambda(A)$ have equivalent properties and are often included in the
set of stabilizer codes. For the repetition code, the stabilized
subspace is spanned by the logical basis
$\ket{\bitzero\bitzero\bitzero}$ and $\ket{\bitone\bitone\bitone}$.
From the point of view of stabilizers, there are two ways in which a
Pauli product $B$ can be detectable:  (1) If $B\in\bar S$,
because in this case $B$ acts as the identity on the code, and (2)
if $B$ anticommutes with at least one member (say $A$) of $S$.  To
see the second way, let $\ket{\psi}$ be in the code.  Then
$A\left(B\ket{\psi}\right) = \left(AB\right)\ket{\psi}=
-\left(BA\right)\ket{\psi}= -B\left(A\ket{\psi}\right) =
-B\ket{\psi}$. Thus $B\ket{\psi}$ belongs to $\cH_\lambda$ with
$\lambda(A)=-1$. Because this subspace is orthogonal to $C=\cH_1$, $B$ is
detectable.  We define the set of Pauli products that commute with all
members of $S$ as ${\bar S}^\perp$. Thus, $B$ is detectable if either
$B\not\in {\bar S}^\perp$ or $B\in\bar S$. Note that because $\bar S$
consists of commuting operators, $\bar S\subseteq {\bar S}^\perp$.

To construct a stabilizer code that can correct all errors of weight
at most one (a ``quantum one-error-correcting code''), it suffices to
find $S$ with the minimum weight of non-identity members of ${\bar
S}^\perp$ being at least three ($3=2\cdot 1+1$, see Thm.~\ref{thm:mindist}).
In this case we say that ${\bar
S}^\perp$ has minimum distance three.  As an example, we can exhibit
a stabilizer for the famous length-five one-error-correcting code
for one qubit~\cite{bennett:qc1996a,laflamme:qc1996a}:
\begin{equation}
S = \{XZZXI, IXZZX, XIXZZ, ZXIXZ\}.
\end{equation}
As a general rule, it is desirable to exhibit the stabilizer
minimally, which means that no member is the product up to a phase of
some of the other members. In this case, the number of qubits encoded
is $n-|S|$, where $n$ is the length of the code and $|S|$ is the
number of elements of $S$.

The correspondence between stabilizer codes and classical binary codes
is obtained by replacing the symbols $I,X,Y$ and $Z$ in a Pauli product by
$00,01,10$ and $11$, respectively. Thus, the members of the stabilizer can
be thought of as binary vectors of length $2n$.  We use arithmetic
modulo two for sums, inner products and application of a binary
matrix. Because the numbers modulo two ($\mathbb{Z}_2$) form a
mathematical ``field'', the basic properties of vectors spaces and
linear algebra apply to binary vectors and matrices.  Thus, the
stabilizer is minimal in the sense introduced above if the
corresponding binary vectors are independent over
$\mathbb{Z}_2$. Given two binary (column) vectors $x$ and $y$ of length two
associated with Pauli products, the property of anticommuting is
equivalent to $x^T B y=1$, where $B$ is the block diagonal $2n\times
2n$ matrix with $2\times 2$ blocks given by $\qaop{0}{1}{1}{0}$.  This
means that ${\bar S}^\perp$ can be identified with the set of vectors
$x$ such that $x^T B y=0$ for all binary vectors $y$ associated with
the members of $S$. It turns out that the inner product $\langle
x,y\rangle = x^TBy$ arises in the study of classical codes over the
four-element mathematical field $GF(4)$, which can be represented by
the vectors $00,01,10$ and $11$ with addition modulo $2$ and a new
multiplication operation.  This relationship leads to the construction
of many good stabilizer codes~\cite{calderbank:qc1996b}.

\subsection{Conserved quantities, symmetries and noiseless subsystems.}
\label{sec:symm}

Even though a physical system may be exposed to error, some of its
properties are often not affected by the errors.  If these ``conserved
quantities'' can be identified with the defining quantities of qubits
or other information units, error-free storage of information can be
ensured without active intervention.  This is the idea
behind noiseless subsystems.

When do noiseless subsystems exist and how can they be constructed?
The examples discussed in the previous sections show that a noiseless
subsystem may be a subset of physical qubits, as in the trivial
two-qubit example, or it may require a more abstract subsystem
identification, as in the example of the three spin-${1\over 2}$
particles.  As will be explained, in both cases, there are quantities
conserved by the errors that can be used to identify the noiseless
subsystem.

A simple classical example for the use of conserved quantities
consists of two physical bits subject to errors that either flip both
bits or leave them alone. A quantity invariant under this noise model
is the parity $P(s)$ of a state $s$ of the two bits.  The parity
$P(s)$ is defined as the number of $\phys{\bitone}$'s in the bitstring
$s$ reduced modulo $2$:
$P(\phys{\bitzero\bitzero})=P(\phys{\bitone\bitone})=0$ and
$P(\phys{\bitzero\bitone})=P(\phys{\bitone\bitzero})=1$.  Flipping
both bits does not change the value of $P$. Consequently, the two
values of $P$ can be used to identify the two states of a noiseless
bit. The syndrome subsystem can be associated with the value
(nonconserved) of the first physical bit using the function defined
by $F(\phys{\bitzero\mathfrak{b}})=0$,
$F(\phys{\bitone\mathfrak{b}})=1$. The corresponding subsystem
identification is obtained by using the values of $P$ and $F$ as the
states of the syndrome (left) and the noiseless, information-carrying
subsystem (right) according to $\phys{\mathfrak{ab}} \leftrightarrow
F(\phys{\mathfrak{ab}})\cdot P(\phys{\mathfrak{ab}})$.

In quantum systems, conserved quantities are associated with the
presence of symmetries, that is, with operators that commute with all
possible errors. In the trivial two-qubit example,
operators acting only on qubit $\sysfnt{2}$ commute
with the error operators. In particular, if $\err{E}$ is any one
of the errors, $\err{E}\slb{\sigma_u}{2}=
\slb{\sigma_u}{2}\err{E}$, for $u=x,y,z$.
It follows that the expectations of $\slb{\sigma_u}{2}$ are
conserved. That is, if $\rho$ is the initial state (density matrix) of
the two physical qubits and $\rho'$ is the state after the errors
acted, then
$\trace\;{\slb{\sigma_u}{2}\rho'}=\trace\;{\slb{\sigma_u}{2}\rho}$.  Because
the state of qubit $\sysfnt{2}$ is completely characterized by these
expectations, it follows immediately that it is unaffected by the
noise.

The trivial two-qubit example suggests a general strategy for finding
a noiseless qubit: First determine the commutant of the errors, which
is the set of operators that commute with all errors. Then find a
subset of the commutant that is algebraically equivalent to the
operators characterizing a qubit. The equivalence can be formulated as
a one-to-one map $f$ from qubit operators to operators in the
commutant. For the range of $f$ to be algebraically equivalent, $f$
must be linear and satisfy $f(A^\dagger) = f(A)^\dagger$ and
$f(AB)=f(A)f(B)$. Once such an equivalence is found, a fundamental
theorem from the representation theory of finite dimensional operator
algebras implies that a subsystem identification for a noiseless
qubit exists~\cite{knill:qc1999b,viola:qc2000c}.

The strategy can be applied to the example of three spin-${1\over 2}$
particles subject to collective errors. One can determine the
commutant by using the physical properties of spin to find the
conserved quantities associated with operators in the commutant, as
suggested in Fig.~\ref{fig:3spin1/2}. Alternatively, observe that by
definition, this error model is symmetric under permutations of the
particles. Therefore, the actions of these permutations on the state
space form a group $\Pi$ of unitary operators commuting with the
errors.  It is a fact that the commutant of the set of collective
errors consists of the linear combinations of operators in $\Pi$. With
respect to the group $\Pi$, one can immediately determine the space
$V_{3/2}$ of symmetric states, that is, those that are invariant under
the permutations.  It is spanned by
\begin{equation}
\ket{\phys{\uparrow\uparrow\uparrow}},\;\;
{1\over\sqrt{3}}\Big(\ket{\phys{\uparrow\uparrow\downarrow}}+
\ket{\phys{\uparrow\downarrow\uparrow}}+
\ket{\phys{\downarrow\uparrow\uparrow}}\Big),\;\;
{1\over\sqrt{3}}\Big(\ket{\phys{\uparrow\downarrow\downarrow}}+
\ket{\phys{\downarrow\uparrow\downarrow}}+
\ket{\phys{\downarrow\downarrow\uparrow}}\Big),\;\;
\ket{\phys{\downarrow\downarrow\downarrow}}.
\end{equation}
A basic result from the representation theory of groups implies that
the projection onto $V_{3/2}$ is given by $P_{3/2}={1\over
6}\sum_{g\in\Pi}g$.  The orthogonal complement $V_{1/2}$ of $V_{3/2}$
is invariant under $\Pi$ and can be analyzed separately.  With the
subsystem identification of Eq.~\ref{eq:3spinid} already in hand, one
can see that the permutation $\pi_1$ which permutes the spins
according to
$\sysfnt{1}\rightarrow\sysfnt{2}\rightarrow\sysfnt{3}\rightarrow\sysfnt{1}$
acts on the noiseless qubit by applying $Z_{240^\circ}=e^{-i\sigma_z
2\pi/3}$, a $240^\circ$ rotation around the $z$-axis.  Similarly, the
permutation $\pi_2$ which exchanges the last two spins acts as
$\sigma_x$ on the qubit. To make them algebraically equivalent to the
corresponding qubit operators, it is necessary to eliminate their
action on $V_{3/2}$ by projecting onto $V_{1/2}$:
$\pi'_1=(1-P_{3/2})\pi_1$ and $\pi'_2=(1-P_{3/2})\pi_2$.  Sums of
products of $\pi'_1$ and $\pi'_2$ are equivalent to the corresponding
sums of products of $Z_{240^\circ}$ and $\sigma_x$, which generate all
qubit operators. To get the subsystem identification of
Eq.~\ref{eq:3spinid}, one can start with a common eigenstate
$\ket{\psi}$ of $\pi'_1$ (a $z$-rotation on the noiseless qubit) and
$2J_z$ (the syndrome subsystem's $\sigma_z$) with eigenvalues
$e^{-i2\pi/3}$ and $1$, respectively. The choice of eigenvalues
implies that $\ket{\psi} \leftrightarrow
\ket{\phys{\uparrow}}\cdot\ket{\bitzero}$ in the desired
identification.  The other logical states of the syndrome spin-$1\over
2$ and the noiseless qubit can be obtained by applying $\pi'_2$,
$2J_x$ and $\pi'_2 2J_x$ to $\ket{\psi}$, which act by flipping the
states of the qubit or the syndrome spin.  This method for obtaining
the subsystem identification generalizes to other operator
equivalences and error operators.

\section{Fault Tolerant Quantum Communication and Computation}
\label{sec:fault_tolerance}

The utility of information and information processing depends on the
the ability to implement large numbers of information units and
information processing operations.  We say that an implementation of
information processing is scalable if the implementation can realize
arbitrarily many information units and operations without loss of
accuracy and with physical resource overheads that are polynomial
(or ``efficient'') in the number of information units and operations.
Scalable information processing is achieved by implementing
information fault tolerantly.

One of the most important results of the work in quantum
error-correction and fault-tolerant computation is the accuracy
threshold theorem, according to which scalability is possible, in
principle, for quantum information.

\begin{equation}\begin{eqthm}{Theorem}
Assume the requirements for scalable QIP
(see below).  If the error per gate is less than a threshold, then it
is possible to efficiently quantum compute arbitrarily accurately.
\end{eqthm}
\end{equation}

\subsection{Requirements for Scalable QIP}

The value of the threshold accuracy (or error) depends strongly on
which set of requirements is used, in particular, the error model that
is assumed. The requirements are closely related to the basic
requirements for constructing a quantum information
processor~\cite{divincenzo:qc2000a} but have to include explicit
assumptions on the error model and on the temporal and spatial aspects
of the available quantum control.

\noindent{\bf Scalable physical systems:}
It is necessary to have access to physical systems that are able to support
qubits or other basic units of quantum information.  The systems must
be scalable, that is, they must be able to support any number of
independent qubits.
\nocite{viola:qc2000c}

\noindent{\bf State preparation:}
One must be able to prepare any qubit in the standard initial state
$\ket{\bitzero}$.  Any preexisting content is assumed to be lost, as
would happen if, for example, the qubit is first discarded and then
replaced by a prepared one. The condition can be weakened; That is, it is
sufficient that a large fraction of the qubits can be prepared in this
way.

\noindent{\bf Measurement:}
A requirement is the ability to measure any qubit in the logical
basis. Again, it is sufficient that a sufficiently large fraction
of the qubits are measurable.  For solving computational problems
with deterministic answers, the standard projective measurement can be
replaced by weak measurements that return a noisy number whose
expectation is the probability that a qubit is in the state
$\ket{1}$~\cite{knill:qc2001f}.

\noindent{\bf Quantum control:}
One must have the ability to implement a universal set of unitary
quantum gates acting on a small number (usually at most two at a time)
of qubits.  For most accuracy thresholds, it is necessary to be able
to apply the quantum control in parallel to any number of disjoint
pairs of qubits. This parallelism requirement can be weakened if a
nearly noiseless quantum memory is available.  The requirement that it
be possible to apply two-qubit gates to any pair of qubits is
unrealistic given the constraints of three-dimensional space. Work on
how to deal with this problem is ongoing~\cite{aharonov:qc1999a}.  The
universality assumption can be substantially weakened by replacing
some or all unitary quantum gates with operations to prepare special
states or by having additional measurement capabilities. See, for
example~\cite{nielsen:qc2001c} and the references therein.

\noindent{\bf Errors:}
The error probability per gate must be below a threshold and satisfy
independence and locality properties (see
Sec.~\ref{sec:error_models}). The definition of ``gate'' includes the
`no-op'', which is the identity operation implemented over the time
required for a computational step. For the most pessimistic independent,
local error models, the error threshold is above $\sim 10^{-6}$.  For the
independent depolarizing error model, it is believed to be better than
$10^{-4}$~\cite{gottesman:qc2000b}.  For some special error
models, the threshold is substantially higher. For example, for
the independent ``erasure'' error model, where error events are always
detected, the threshold is above $.01$, and for an error model whose
errors are specific, unintentional measurements in the standard basis
of a qubit, the threshold is $1$~\cite{knill:qc2000a,knill:qc2001h}.  
The threshold is also well above $.01$ when the goal is only to
transmit quantum information through noisy quantum
channels~\cite{briegel:qc1998a}.

\subsection{Realizing Fault-Tolerance}

The existing proofs
of the accuracy threshold theorems consist of explicit
instructions for building a scalable quantum information processor and
analyses of its robustness against the assumed error model.  The
instructions for realizing scalable computation are based on the
following simple idea. Suppose that the error rate per operation for
some way of realizing qubits is $p$. We can use these qubits and a
quantum error-correcting code to encode logical qubits for which the
storage error rate is reduced. For example, if a one-error correcting
code is used, the error rate per storage interval for the logical
qubits is expected to be $\leq cp^2$ for some constant $c$. Suppose
that we can show how to implement encoded operations, preparations,
measurement and the subroutines required for error-correction such
that this inequality is now valid for each basic encoded step, perhaps
for a larger constant $C$. Suppose furthermore that the errors for the
encoded information still satisfy the assumed error model.  The newly
defined logical qubits then have an error rate of $\leq Cp^2$, which
is less than $p$ for $p<1/C$. We can use the newly realized qubits as
a foundation for making higher level logical qubits. This results in
multiple levels of encodings. In the next level (level $2$), the error
rate is $\leq C^3p^4$, and after $k$ iterations it is $\leq
C^{2^k-1}p^{2^k}$, a doubly-exponentially decreasing function of $k$.
This procedure is called ``concatenation''
(Fig.~\ref{fig:concatenation}). Because the complexity, particularly
the number of physical qubits needed for each final logical qubit,
grows only singly-exponentially in $k$, the procedure is efficient.
Specifically, to achieve a logical error of $\epsilon$ per operation
requires of the order of $|\log(\epsilon)|^r$ resources per logical
qubit for some finite $r$.  In practice, this simple idea
is still dauntingly complex, but there is hope that for realistic
errors in physical systems and by cleverly trading off different
variations of these techniques, much of the theoretical complexity can
be avoided~\cite{steane:qc1999a}.

\begin{herefig}
\scalebox{.8}{
\begin{picture}(8.5,6.2)(-4.25,0)\Large
\nputgr{-1.5,0}{b}{}{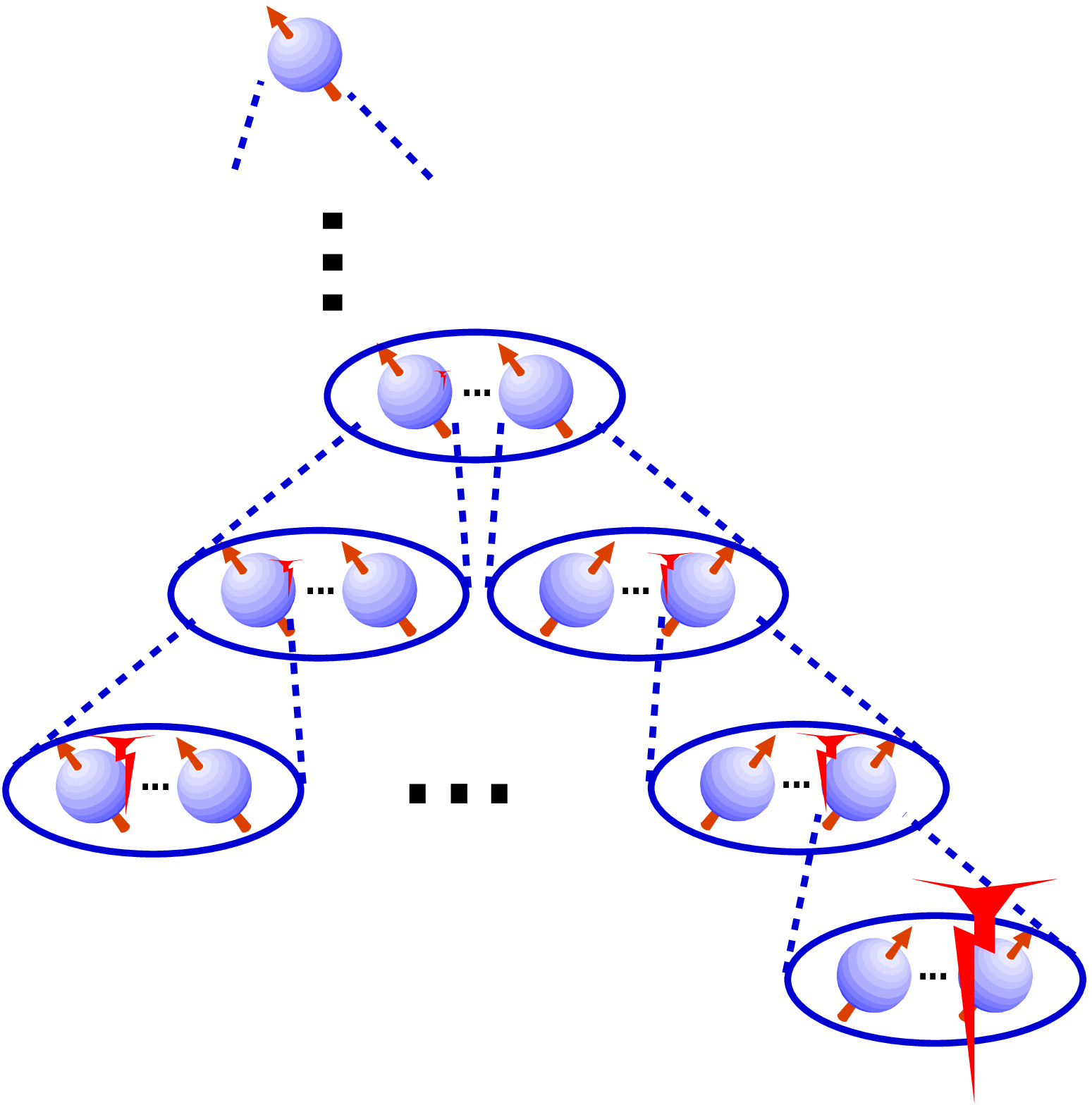}
\nputbox{1.95,6.25}{tc}{Level}
\nputbox{3.35,6.25}{tc}{Error rate}
\nputbox{2.6,6.25}{t}{\rule{2pt}{5.9in}}
\nputbox{2.6,6.0}{c}{\rule{3in}{2pt}}
\nputbox{1.95,.35}{bc}{1}
\nputbox{3.35,.35}{bc}{$p$}
\nputbox{1.95,1.5}{bc}{2}
\nputbox{3.35,1.5}{bc}{$p_2\leq Cp^2$}
\nputbox{1.95,2.65}{c}{3}
\nputbox{2.6,2.65}{l}{\setlength{\arraycolsep}{2pt}
$\left\{\begin{array}{rcl}p_3&\leq& C(Cp^2)^2\\ &=& C^{2^2-1}p^{2^2}\end{array}\right.$}
\nputbox{1.95,3.75}{c}{4}
\nputbox{2.6,3.75}{l}{\setlength{\arraycolsep}{2pt}
$\left\{\begin{array}{rcl}p_4&\leq& C(C^3p^{2^2})^2\\ &=& C^{2^3-1}p^{2^3}\end{array}\right.$}
\nputbox{1.95,4.8}{c}{$\vdots$}
\nputbox{1.95,5.5}{b}{k}
\nputbox{2.8,5.5}{lb}{$\leq C^{2^k-1}p^{2^k}$}
\end{picture}}
\label{fig:concatenation}
\herefigcap{Schematic representation of concatenation.
The bottom level represents qubits realized more-or-less directly in a
physical system. Each next level represents logical qubits defined by
means of subsystems in terms of the previous level's
qubits. More efficient subsystems might represent multiple qubits
in one code block rather than the one qubit per code block shown here.}
\end{herefig}

Many important developments and ideas of quantum information were
ultimately needed to realize encoded operations, preparations,
measurements and error-correction subroutines that behave well with
respect to concatenation. Stabilizer codes provide a particularly nice
setting for implementing many of these techniques.  One reason is that
good stabilizer codes are readily constructed.  Another is that they
enable encoding operations in a way that avoids spreading errors
between the qubits of a single code word~\cite{gottesman:qc1997a}.  In
addition, there are many tricks based on teleportation that can be
used to maintain the syndrome subsystems in acceptably low-error states
and to implement general operations
systematically~\cite{gottesman:qc1999a}.  To learn more about all of
these techniques, see the textbook by Nielsen and
Chuang~\cite{nielsen:qc2001a} and the works of
Gottesman~\cite{gottesman:qc1997a} and
Preskill~\cite{preskill:qc1998a}.

\section{Concluding Remarks}
\label{sec:conclusion}

The advancements in quantum error-correction and fault-tolerant
QIP have shown that in principle scalable
quantum computation is achievable. This is a crucial result because it
suggests that experimental efforts in QIP
will eventually lead to more than a few small scale applications of
quantum information to communication and problems with few
qubits. However, the general techniques for achieving scalability that
are known are difficult to realize. Existing technologies are far from
achieving sufficient accuracy even for just two qubits---at least in terms
of the demands of the usual accuracy threshold theorems.  There is
hope that more optimistic thresholds can be shown to apply if one considers
the specific constraints of a physical device,
better understands the dominant sources of errors, and exploits
tailor-made ways of embedding quantum information into
subsystems. Current work in this area is focused on finding such
methods of quantum error control. These methods include approaches to
error control not covered in this introduction---for example,
techniques for actively turning off the error-inducing environmental
interactions~\cite{viola:qc1998a,viola:qc1999a} and modifications to
controlling quantum systems that eliminate systematic and calibration
errors~\cite{levitt:qc1982a,cummins:qc1999a}.  Further work is also
needed to improve the thresholds for the more pessimistic error models
and for developing more-efficient scalability schemes.

\vspace*{\baselineskip}

\noindent{\bf Acknowledgements}: We thank Nikki Cooper and Ileana Buican
for their extensive encouragement and editorial help.

\vspace*{\baselineskip}

\makeaddress

\bibliographystyle{unsrt}
\bibliography{journalDefs,qc}

\begin{thebibliography}{10}

\bibitem{braunstein:qc2000b}
Special focus issue: Experimental proposals for quantum computation.
\newblock {\em Fort. Phys.}, 48:767--1138, 2000.

\bibitem{steane:qc1995a}
A.~Steane.
\newblock Multiple particle interference and quantum error correction.
\newblock {\em Proc. R. Soc. Lond. A}, 452:2551--2577, 1996.

\bibitem{shor:qc1995b}
P.~W. Shor.
\newblock Scheme for reducing decoherence in quantum computer memory.
\newblock {\em Phys. Rev. A}, 52:2493--2496, 1995.

\bibitem{gottesman:qc1996a}
D.~Gottesman.
\newblock A class of quantum error-correcting codes saturating the quantum
  hamming bound.
\newblock {\em Phys. Rev. A}, 54:1862--1868, 1996.

\bibitem{calderbank:qc1996a}
A.R. Calderbank, E.M. Rains, P.W. Shor, and N.J.A. Sloane.
\newblock Quantum error correction and orthogonal geometry.
\newblock {\em Phys. Rev. Lett.}, 78:405--408, 1997.

\bibitem{calderbank:qc1996b}
A.~R. Calderbank, E.~M. Rains, P.~W. Shor, and N.~J.~A. Sloane.
\newblock Quantum error correction via codes over gf(4).
\newblock {\em IEEE Trans. Inf. Theory}, 44:1369--1387, 1998.

\bibitem{shor:qc1996a}
P.~W. Shor.
\newblock Fault-tolerant quantum computation.
\newblock In {\em Proceedings of the 37th Symposium on the Foundations of
  Computer Science (FOCS)}, pages 56--65, Los Alamitos, California, 1996. IEEE
  press.

\bibitem{kitaev:qc1996a}
A.~Yu. Kitaev.
\newblock Quantum error correction with imperfect gates.
\newblock In O.~Hirota et~al., editor, {\em Quantum Communication and Computing
  and Measurement}, New York, 1997. Plenum.

\bibitem{knill:qc1996b}
E.~Knill and R.~Laflamme.
\newblock Concatenated quantum codes.
\newblock Technical Report LAUR-96-2808, Los Alamos National Laboratory, {\tt
  knill@lanl.gov}, 1996.
\newblock quant-ph/9608012.

\bibitem{aharonov:qc1996a}
D.~Aharonov and M.~Ben-Or.
\newblock Fault-tolerant quantum computation with constant error.
\newblock In {\em Proceedings of the 29th Annual ACM Symposium on the Theory of
  Computation (STOC)}, pages 176--188, New York, New York, 1996. ACM Press.

\bibitem{aharonov:qc1999a}
D.~Aharonov and M.~Ben-Or.
\newblock Fault-tolerant quantum computation with constant error.
\newblock quant-ph/9906129, 1999.

\bibitem{knill:qc1997a}
E.~Knill, R.~Laflamme, and W.~Zurek.
\newblock Resilient quantum computation: Error models and thresholds.
\newblock {\em Proc. R. Soc. Lond. A}, 454:365--384, 1998.

\bibitem{knill:qc1998a}
E.~Knill, R.~Laflamme, and W.~H. Zurek.
\newblock Resilient quantum computation.
\newblock {\em Science}, 279:342--345, 1998.

\bibitem{gottesman:qc1997a}
D.~Gottesman.
\newblock A theory of fault-tolerant quantum computation.
\newblock {\em Phys. Rev. A}, 57:127--137, 1998.

\bibitem{preskill:qc1998a}
J.~Preskill.
\newblock Reliable quantum computers.
\newblock {\em Proc. R. Soc. Lond. A}, 454:385--410, 1998.

\bibitem{knill:qc2001c}
E.~Knill, R.~Laflamme, H.~Barnum, D.~Dalvit, J.~Dziarmaga, J.~Gubernatis,
  L.~Gurvits, G.~Ortiz, L.~Viola, and W.~Zurek.
\newblock Introduction to quantum information processing.
\newblock Technical Report LAUR-01-4761, Los Alamos National Laboratory, 2001.
\newblock To appear in LA Science.

\bibitem{knill:qc2001f}
R.~Laflamme, E.~Knill, D.~Cory, E.~M. Fortunato, T.~Havel, C.~Miquel,
  R.~Martinez, C.~Negrevergne, G.~Ortiz, M.~A. Pravia, S.~Sinha, R.~Somma, and
  L.~Viola.
\newblock Introduction to {NMR} quantum information processing.
\newblock Technical Report LAUR-02-6132, Los Alamos National Laboratory, 2001.
\newblock To appear in LA Science.

\bibitem{kraus:qc1983a}
K.~Kraus.
\newblock {\em States, Effects and Operations: Fundamental Notions of Quantum
  Theory}.
\newblock Lecture Notes in Physics, Vol.\ 190. Springer-Verlag, Berlin, 1983.

\bibitem{bennett:qc1996a}
C.~H. Bennett, D.~P. DiVincenzo, J.~A. Smolin, and W.~K. Wootters.
\newblock Mixed state entanglement and quantum error-correcting codes.
\newblock {\em Phys. Rev. A}, 54:3824--3851, 1996.

\bibitem{schumacher:qc1996a}
B.~Schumacher.
\newblock Sending entanglement through noisy quantum channels.
\newblock {\em Phys. Rev. A}, 54:2614--2628, 1996.

\bibitem{knill:qc2001a}
E.~Knill, R.~Laflamme, R.~Martinez, and C.~Negrevergne.
\newblock Implementation of the five qubit error correction benchmark.
\newblock {\em Phys. Rev. Lett.}, 86:5811--5814, 2001.

\bibitem{knill:qc1999b}
E.~Knill, R.~Laflamme, and L.~Viola.
\newblock Theory of quantum error correction for general noise.
\newblock {\em Phys. Rev. Lett.}, 84:2525--2528, 2000.

\bibitem{ashikhmin:qc1999a}
A.~Ashikhmin and S.~Litsyn.
\newblock Upper bounds on the size of quantum codes.
\newblock {\em IEEE Trans. Inf. Theory}, 45:1206--1216, 1999.

\bibitem{laflamme:qc1996a}
R.~Laflamme, C.~Miquel, J.-P. Paz, and W.~H. Zurek.
\newblock Perfect quantum error-correcting code.
\newblock {\em Phys. Rev. Lett.}, 77:198, 1996.

\bibitem{viola:qc2000c}
L.~Viola, E.~Knill, and R.~Laflamme.
\newblock Constructing qubits in physical systems.
\newblock {\em J. Phys. A}, 34(LAUR-00-5877):7067--7080, 2001.

\bibitem{divincenzo:qc2000a}
D.P. DiVincenzo.
\newblock The physical implementation of quantum computation.
\newblock {\em Fort. Phys.}, 48:771--783, 2000.

\bibitem{nielsen:qc2001c}
M.~A. Nielsen.
\newblock Universal quantum computation using only projective measurement,
  quantum memory, and preparation of the $|0\rangle$ state.
\newblock quant-ph/0108020, 2001.

\bibitem{gottesman:qc2000b}
D.~Gottesman and J.~Preskill.
\newblock Unpublished analysis of the accuracy threshold., 1999.

\bibitem{knill:qc2000a}
E.~Knill, R.~Laflamme, and G.~Milburn.
\newblock Thresholds for linear optics quantum computation.
\newblock Technical Report LAUR-00-3477, Los Alamos National Laboratory, 2000.
\newblock quant-ph/0006120.

\bibitem{knill:qc2001h}
E.~Knill.
\newblock Linear optics quantum computation i--v.
\newblock Tutorial lectures, online at
  \texttt{http://online.itp.ucsb.edu/qinfo01/knill$\{$,1,2,3,4,5$\}/$}, (expand
  curly brackets for six links), 2001.

\bibitem{briegel:qc1998a}
H.-J. Briegel, W.~D\"ur, J.~I. Cirac, and P.~Zoller.
\newblock Quantum repeaters for communication.
\newblock quant-ph/9803056, 1998.

\bibitem{steane:qc1999a}
A.~Steane.
\newblock Efficient fault-tolerant quantum computing.
\newblock {\em Nature}, 399:124--126, 1999.

\bibitem{gottesman:qc1999a}
D.~Gottesman and I.~L. Chuang.
\newblock Demonstrating the viability of universal quantum computation using
  teleportation and single-qubit operations.
\newblock {\em Nature}, 402:390--393, 1999.

\bibitem{nielsen:qc2001a}
M.~A. Nielsen and I.~L. Chuang.
\newblock {\em Quantum Computation and Quantum Information}.
\newblock Cambridge University Press, 2001.

\bibitem{viola:qc1998a}
L.~Viola and S.~Lloyd.
\newblock Dynamical suppression of decoherence in two-state quantum systems.
\newblock {\em Phys. Rev. A}, 58:2733--2744, 1998.

\bibitem{viola:qc1999a}
L.~Viola, E.~Knill, and S.~Lloyd.
\newblock Dynamical decoupling of open quantum systems.
\newblock {\em Phys. Rev. Lett.}, 82:2417--2421, 1999.

\bibitem{levitt:qc1982a}
M.~H. Levitt.
\newblock Symmetrical composite pulse sequences for {NMR} population-inversion
  1. compensation for radiofrequency field inhomogeneity.
\newblock {\em J. Mag. Res.}, 48:234--264, 1982.

\bibitem{cummins:qc1999a}
H.~K. Cummins and J.~A. Jones.
\newblock Use of composite rotations to correct systematic errors in {NMR}
  quantum computation.
\newblock quant-ph/9911072, 1999.

\end{thebibliography}

\section{Glossary}

\begin{description}
\setlength{\itemsep}{0pt}\setlength{\parskip}{0pt}\setlength{\parsep}{0pt}

\item[\textbf{Bit}.]
The basic unit of deterministic information. It is a system
that can be in one of two possible states, $\bitzero$ and $\bitone$.

\item[\textbf{Bit string}.]
A sequence of $\bitzero$'s and $\bitone$'s that represents a state of a
sequence of bits. Bit strings are words in the binary alphabet.

\item[\textbf{Classical information}.]
The type of information based on bits and bit strings and more
generally on words formed from finite alphabets. This is the
information used for communication between people.  Classical
information can refer to deterministic or probabilistic information,
depending on the context.

\item[\textbf{Code}.]
A set of states that can be used to represent information. The set of
states needs to have the properties of the type of information to be
represented. The code is usually a subset of the states of a given
system $\sysfnt{Q}$. It is then a \emph{$\sysfnt{Q}$-code} or a
\emph{code on $\sysfnt{Q}$}.  If information is represented by a state
in the code, $\sysfnt{Q}$ is said to \emph{carry} the information.

\item[\textbf{Code word}.]
A state in a code. The term is primarily used for classical
codes defined on bits or systems with non-binary alphabets.

\item[\textbf{Concatenation}.] An iterative procedure in which
higher-level logical information units are implemented in terms of
lower-level units.

\item[\textbf{Control error}.]
An error due to non-ideal control in applying operations or gates.

\item[\textbf{Communication channel}.]
A means for transmitting information from one place to another.
It can be associated with a physical system in which the information
to be transmitted is stored by the sender. The system
is subsequently conveyed to the receiver, who can then
make use of the information.

\item[\textbf{Correctable error set}.]
For a given code, a set of errors such that there exists an
implementable procedure $R$ that, after any one of these errors
$\err{E}$ acts on a state $x$ in the code, returns the system to the
state: $x=R\err{E}x$. What procedures are implementable depends on the
type of information represented by the system and, if it is a physical
system, its physics.

\item[\textbf{Decoding}.]
The process of transferring information from an encoded form to its
``natural'' form.  In the context of error correction, decoding is
often thought of as consisting of two steps, one which removes the
errors' effects (sometimes called the recovery procedure) and one
that extracts the information (often also called decoding, in a
narrower sense).

\item[\textbf{Depolarizing errors}.]
An error model for qubits in which random Pauli operators are applied
independently to each qubit.

\item[\textbf{Detectable error}.]
For a given code, an error that has no effect on an initial state in
the code if an observation determines that the state is still in the
code. If the state is no longer in the code, the error is said to
have been detected and the state no longer represents valid
information.

\item[\textbf{Deterministic information}.]
The type of information based on bits and bit strings. This is the
same as classical information but explicitly excludes probabilistic
information.

\item[\textbf{Encoding}.]
The process of transferring information from its ``natural'' form to
an encoded form. It requires an identification of the valid states
associated with the information and the states of a code.  The process
acts on an information unit and replaces it with the system whose
state space contains the code.

\item[\textbf{Environment}.]
In the context of information encoded in a physical system, it refers
to other physical systems that may interact with the
information-carrying system.

\item[\textbf{Environmental noise}.]
Noise due to unwanted interactions with the environment.

\item[\textbf{Error}.]
Any unintended effect on the state of a system, particularly in
storing or otherwise processing information.

\item[\textbf{Error basis}.]
A set of state transformations that can be used to represent any
error. For quantum systems, errors can be represented as operators
acting on the system's state space, and an error basis is a maximal,
linearly independent set of such operators.

\item[\textbf{Error control}.]
The term for general procedures that limit the effects of errors
on information represented in noisy, physical systems.

\item[\textbf{Error correction}.]
The process of removing the effects of errors on encoded information.

\item[\textbf{Error-correcting code}.]
A code with additional properties that enable a decoding procedure to
remove the effects of the dominant sources of errors on encoded
information. Any code is error-correcting for some error-model in this
sense. To call a code ``error-correcting'' emphasizes the fact that it
was designed for this purpose.

\item[\textbf{Error model}.]
An explicit description of how and when errors happen in a given
system. Typically, a model is specified as a probability distribution
over error operators.  More general models may need to be considered,
particularly in the context of fault tolerant computation, for which
correlations in time are important.

\item[\textbf{Fault tolerance}.]
A property of encoded information that is being processed with
gates. It means that errors occurring during processing, including
control errors and environmental noise, do not seriously affect the
information of interest.

\item[\textbf{Gate}.]
An operation applied to information for the purpose
of information processing.

\item[\textbf{Hamming distance}.]
The Hamming distance between two binary words (sequences of $\bitzero$
and $\bitone$) is the number of positions in which the two words
disagree.

\item[\textbf{Hilbert space}.]
A $n$-dimensional Hilbert space consists of all complex
$n$-dimensional vectors. A defining operation in a Hilbert space is
the inner product.  If the vectors are thought of as column vectors,
then the inner product $\langle x,y\rangle$ of $x$ and $y$ is obtained
by forming the conjugate transpose $x^\dagger$ of $x$ and calculating
$\langle x,y\rangle=x^\dagger y$. The inner product induces the usual
norm $|x|^2 =
\langle x,x\rangle$.

\item[\textbf{Information}.]
Something that can be recorded, communicated and computed
with. Information is \emph{fungible}, which implies that its meaning
can be identified regardless of the particulars of the physical
realization. Thus, information in one realization (such as ink on a
sheet of paper) can be easily transferred to another (for example,
spoken words). Types of information include deterministic,
probabilistic and quantum information. Each type is characterized by
information units, which are abstract systems whose states represent
the simplest information of this type. These define the ``natural''
representation of the information.  For deterministic information the
unit is the \emph{bit}, whose states are symbolized by $\bitzero$ and
$\bitone$. Information units can be put together to form larger
systems and can be processed with basic operations acting on a small number
of units at a time.

\item[\textbf{Length}.]
For codes on $n$ basic information units, the length of the code is
$n$.

\item[\textbf{Minimum distance}.]
The smallest number of errors that is not detectable by a code.  In
this context, the error model consists of a set of error operators
without specified probabilities.  Typically the concept is used for
codes on $n$ information units and the error model consists of
operators acting on any one of the units. For a classical binary code,
the minimum distance is the smallest Hamming distance between two code
words.

\item[\textbf{Noise}.]
Any unintended effect on the state of a system, particularly an effect
with a stochastic component due to incomplete isolation of the system
from its environment.

\item[\textbf{Operator}.]
A function transforming the states of a system. Operators may be
restricted depending on the system's properties. For example,
operators acting on quantum systems are always assumed to be linear.

\item[\textbf{Pauli operators}.]
The Hermitian matrices $\sigma_x,\sigma_y$ and $\sigma_z$
(Eq.~\ref{eq:paulidef}) acting on qubits. It is often convenient to
consider the identity operator to be included in the set of Pauli
operators.

\item[\textbf{Physical system}.]
A system explicitly associated with a physical device or particle.
The term is used to distinguish between abstract systems used to
define a type of information and specific realizations, which are
subject to environmental noise and errors due to other imperfections.

\item[\textbf{Probabilistic bit.}]
The basic unit of probabilistic information.  It is a system whose
state space consists of all probability distributions over the two
states of a bit. The states can be thought of as describing the
outcome of a biased coin flip before the coin is flipped.

\item[\textbf{Probabilistic information}.]
The type of information obtained when the state
spaces of deterministic information are extended with arbitrary
probability distributions over the deterministic states.
This is the main type of classical information to which
quantum information is compared.

\item[\textbf{Quantum information}.]
The type of information obtained when the state space of
deterministic information is extended with arbitrary superpositions of
deterministic states.  Formally, each deterministic state is
identified with one of an orthonormal basis vector in a Hilbert space
and superpositions are unit-length vectors that are expressible as
complex linear sums of the chosen basis vectors.  Ultimately it is
convenient to extend this state space again by permitting probability
distributions over the quantum states. This is still called
quantum information.

\item[\textbf{Qubit}.]
The basic unit of quantum information. It is the quantum extension of
the deterministic bit; that is, its state space consists of
the unit-length vectors in a two dimensional Hilbert space.

\item[\textbf{Repetition code}.]
The classical, binary repetition code of length $n$ consists of the
two words $\bitzero\bitzero\ldots\bitzero$ and
$\bitone\bitone\ldots\bitone$. For quantum variants of this code 
one applies the superposition principle to obtain the states
consisting of all unit-length complex linear combinations of the two
classical code words.

\item[\textbf{Scalability}.]
A property of physical implementations of information processing that
implies that there are no bounds on accurate information
processing. That is, arbitrarily many information units can be
realized and they can be manipulated for an arbitrarily long amount of
time without loss of accuracy. Furthermore, the realization is
polynomially efficient in terms of the number of information units and
gates used.

\item[\textbf{States}.]
The set of states for a system characterizes the system's behavior and
possible configurations.

\item[\textbf{Subspace}.]
For a Hilbert space, a subspace is a linearly closed subset of the
vector space. The term can be used more generally for a system
$\sysfnt{Q}$ of any information type: A subspace of $\sysfnt{Q}$ or,
more specifically, of the state space of $\sysfnt{Q}$ is a subset of
the state space that preserves the properties of the information type
represented by $\sysfnt{Q}$.

\item[\textbf{Subsystem}.]
A typical example of a subsystem is the first (qu)bit in a system
consisting of two (qu)bits. In general, to obtain a subsystem of system
$\sysfnt{Q}$, one first selects a subset $C$ of
$\sysfnt{Q}$'s state space and then identifies $C$ as the state space
of a pair of systems. Each member of the pair is then a subsystem of
$\sysfnt{Q}$. Restrictions apply depending on the types of information
carried by the system and subsystems.  For example, if $\sysfnt{Q}$ is
quantum and so are the subsystems, then $C$ has to be a linear
subspace and the identification of the subsystems' state space with
$C$ has to be unitary.

\item[\textbf{Subsystem identification}.]
The mapping or transformation that identifies the state space of two
systems with a subset $C$ of states of a system $\sysfnt{Q}$.  In saying that
$\sysfnt{L}$ is a subsystem of $\sysfnt{Q}$, we also introduce a second
subsystem $\sysfnt{S}$ and identify  the state space
of the combined system $\sysfnt{LS}$ with $C$.

\item[\textbf{Syndrome}.]
One of the states of a syndrome subsystem. It is often used more narrowly
for one of a distinguished set of basis states of a syndrome subsystem.

\item[\textbf{Syndrome subsystem}.]
In identifying an information-carrying subsystem in
the context of error-correction, the other member of the
pair of subsystems required for the subsystem identification
is called the syndrome subsystem. The terminology comes
from classical error-correction, in which the syndrome is
used to determine the most likely error that has happened.

\item[\textbf{System}.]
An entity that can be in any of a specified number of states. An
example is a desktop computer whose states are determined by the
contents of its various memories and disks. Another example is a
qubit, which can be thought of as a particle whose state space is
identified with complex, two-dimensional, length-one vectors.
Here, a system is always associated with a type of information,
which in turn determines the properties of the state space.
For example, for quantum information the state space
is a Hilbert space. For deterministic information,
it is a finite set called an alphabet.

\item[\textbf{Twirling}.] A randomization method for ensuring
that errors act like a depolarizing error model. For one qubit, it
involves applying a random Pauli operator before the errors occur and
then undoing the operator by applying its inverse.

\item[\textbf{Unitary operator}.]
A linear operator $U$ on a Hilbert space that preserves
the inner product. That is, for all $x$ and $y$,
$\langle Ux,Uy\rangle=\langle x,y\rangle$.
If $U$ is given in matrix form, then this condition is equivalent
to $U^\dagger U = \idop$.

\item[\textbf{Weight}.]
For a binary word, the weight is the number of $\bitone$'s
in the word. For an error operator acting on $n$ systems
by applying an operator to each one of them, the weight
is the number of non-identity operators applied.

\end{description}

\end{document}